\documentclass[fleqn,usenatbib]{./references/mnras}
\usepackage{import}

\import{./preface/}{preamble.sty}
\def \aj {{\rm {AJ}}}
\def \araa {{\rm {ARA\&A}}}
\def \apj {{\rm {ApJ}}}

\def \apjs {{\rm {ApJS}}}
\def \apjl {{\rm {ApJL}}}

\def \aap {{\rm {A\&A}}}

\def \mnras {{\rm {MNRAS}}}

\def \pasp {{\rm {PASP}}}

\def \nat {{\rm {Nat}}}
\def \kms {{\rm km\,s$^{-1}$}}

\def \sol {{\rm M$_\odot$}}

\def \arcsec {{\rm $^{\prime\prime}$}}
\def \micron{\hbox{$\upmu$m}}

\def \cmsq {{\rm cm$^{-2}$}}
\def \cmcb {{\rm cm$^{-3}$}}
\def \gcmsq{{\rm g\,cm$^{-2}$}}

\usepackage{array}
\newcolumntype{L}[1]{>{\raggedright\let\newline\\\arraybackslash\hspace{0pt}}m{#1}}
\newcolumntype{C}[1]{>{\centering\let\newline\\\arraybackslash\hspace{0pt}}m{#1}}
\newcolumntype{R}[1]{>{\raggedleft\let\newline\\\arraybackslash\hspace{0pt}}m{#1}}

\def \cCtHt {{c-C$_{3}$H$_{2}$}}

\begin{document}
\label{firstpage}

\title[Cluster formation in the Galactic Centre]{Young massive star cluster formation in the Galactic Centre is driven by global gravitational collapse of high-mass molecular clouds}

\author[A.~T.~Barnes et al.]{A.~T.~Barnes,$^{1,2,3}$\thanks{E-mail: ashleybarnes.astro@gmail.com}
 S.~N.~Longmore,$^{1}$ 
 A.~Avison,$^{4,5}$
 Y.~Contreras,$^{6}$ 
 A.~Ginsburg,$^{7}$ \and
 J.~D.~Henshaw,$^{8}$ 
 J.~M.~Rathborne,$^{9}$
 D.~L.~Walker,$^{10,11}$
  J.~Alves,$^{12}$ 
  J.~Bally,$^{13}$ \and
 C.~Battersby,$^{14}$
 M.~T.~Beltr{\'a}n,$^{15}$ 
 H.~Beuther,$^{8}$ 
 G.~Garay,$^{16}$
 L.~Gomez,$^{10}$ 
  J.~Jackson,$^{17}$ \and
 J.~Kainulainen,$^{18}$ 
 J.~M.~D.~Kruijssen,$^{19}$
 X.~Lu,$^{11}$
 E.~A.~C.~Mills,$^{20}$ 
 J.~Ott$^{7}$ and
 T.~Peters$^{21}$
 \\
$^{1}$Astrophysics Research Institute, Liverpool John Moores University, 146 Brownlow Hill, Liverpool L3 5RF, UK\\
$^{2}$Max-Planck-Institut f\"{u}r extraterrestrische Physik, Gie\ss enbachstra\ss e 1, 85748, Garching, Germany \\
$^{3}$Argelander-Institut f\"{u}r Astronomie, Universit\"{a}t Bonn, Auf dem H\"{u}gel 71, 53121, Bonn, Germany \\
$^{4}$UK Atacama Large Millimeter/submillimeter Array Regional Centre Node\\
$^{5}$Jodrell Bank Centre for Astrophysics, The University of Manchester, Manchester, M13 9PL, UK \\
$^{6}$Leiden Observatory, Leiden University, PO Box 9513, NL-2300 RA Leiden, Netherlands \\
$^{7}$National Radio Astronomy Observatory, 1003 Lopezville Rd., Socorro, NM, 87801, USA \\
$^{8}$Max-Planck-Institut f\"{u}r Astronomie, K\"{o}nigstuhl 17, 69117 Heidelberg, Germany \\
$^{9}$CSIRO Astronomy and Space Science, P.O. Box 76, Epping NSW, 1710, Australia\\
$^{10}$Joint ALMA Observatory, Alonso de C{\'o}rdova 3107, Vitacura, Santiago, Chile \\
$^{11}$National Astronomical Observatory of Japan, 2-21-1 Osawa, Mitaka,Tokyo, 181-8588, Japan \\
$^{12}$University of Vienna, Department of Astrophysics, Türkenschanzstra\ss e 17, 1180 Vienna, Austria\\
$^{13}$CASA, University of Colorado, 389-UCB, Boulder, CO, 80309, USA\\
$^{14}$Department of Physics, University of Connecticut, Storrs, CT, 06269 USA\\
$^{15}$INAF-Osservatorio Astrofisico di Arcetri, Largo E. Fermi 5, I-50125 Firenze, Italy\\
$^{16}$Departamento de Astronom{\'i}a, Universidad de Chile, Casilla 36-D, Santiago, Chile\\
$^{17}$ SOFIA Science Center, USRA, NASA Ames Research Center, Mountain View, CA, 94043, USA \\ 
$^{18}$Chalmers University of Technology, Onsala Space Observatory, 439 92 Onsala, Sweden\\
$^{19}$Astronomisches Rechen-Institut, Zentrum f\"{u}r Astronomie der Universit\"{a}t Heidelberg, M\"{o}nchhofstra\ss e 12-14, 69120 Heidelberg, Germany \\
$^{20}$Physics Department, Brandeis University, 415 South Street, Waltham, MA, 02453, USA \\
$^{21}$Max-Planck-Institut f\"{u}r Astrophysik, Karl-Schwarzschild-Stra\ss e 1, D-85748 Garching, Germany
}
\date{Accepted 2019 March 13. Received 2019 March 13; in original form 2018 November 26.}

\pubyear{2018}


\pagerange{\pageref{firstpage}--\pageref{lastpage}}
\maketitle
\begin{abstract}

Young massive clusters (YMCs) are the most compact, high-mass stellar systems still forming at the present day. The precursor clouds to such systems are, however, rare due to their large initial gas mass reservoirs and rapid dispersal timescales due to stellar feedback. Nonetheless, unlike their high-z counterparts, these precursors are resolvable down to the sites of individually forming stars, and hence represent the ideal environments in which to test the current theories of star and cluster formation. Using high angular resolution (1\arcsec\,/\,0.05pc) and sensitivity ALMA observations of two YMC progenitor clouds in the Galactic Centre, we have identified a suite of molecular line transitions -- e.g. c-C$_{3}$H$_{2}$~($7-6$) -- that are believed to be optically thin, and reliably trace the gas structure in the highest density gas on star-forming core scales. We conduct a virial analysis of the identified core and proto-cluster regions, and show that half of the cores (5/10) and both proto-clusters are unstable to gravitational collapse. This is the first kinematic evidence of global gravitational collapse in YMC precursor clouds at such an early evolutionary stage. The implications are that if these clouds are to form YMCs, then they likely do so via the ``conveyor-belt'' mode, whereby stars continually form within dispersed dense gas cores as the cloud undergoes global gravitational collapse. The concurrent contraction of both the cluster-scale gas and embedded (proto)stars ultimately leads to the high (proto)stellar density in YMCs.

\end{abstract}



\section{Introduction}\label{sec:intro}

Studies of the most massive and dense molecular clouds are key in developing our understanding of the extremes of star and stellar cluster formation. The largest clusters currently forming within the Galaxy today are referred to as young massive clusters (or YMCs), which can be characterised as having masses $M_\mathrm{YMC}>10^{4}$\,\sol, ages $<100$\,Myr, radii $R_\mathrm{YMC} < 1$\,pc and being gravitationally bound (as outlined in the review by \citealp{portegies-zwart_2010}). Given these properties, YMCs have been suggested as the current day analogues of the early universe globular clusters (e.g. \citealp{elmegreen_1997, kruijssen_2015a}). 

Molecular clouds with sufficient mass ($M_\mathrm{YMC}\sim10^{5}$\,\sol) to form such clusters are, however, very rare, and only a handful of candidate objects have been currently identified \citep{ginsburg_2012, longmore_2012, urquhart_2013, contreras_2017, jackson_2018}. Nonetheless, investigating the very early stages of YMC evolution, before the onset of star formation, is crucial in understanding how these systems formed (e.g. \citealp{walker_2015, walker_2016}). 

The current theories for cluster formation differ in their predictions for the spatial density distribution of the gas within molecular clouds, just before the onset of star formation, and how this compares to the density distribution of stars within the resultant cluster. In other words, these theories ask: how could a molecular cloud with an observed initial mean density of $\rho_\mathrm{cloud}^\mathrm{inital}$ form a typical YMC with a mean density of $\rho_\mathrm{YMC}^\mathrm{final} \approx 10^{3\,\pm1}$\sol\,pc$^{-3}$ (approximately equivalent to a molecular hydrogen number density within a molecular cloud of $n_\mathrm{H_2} \approx 10^{4\,\pm1}$\,cm$^{-3}$; \citealp{portegies-zwart_2010})? In the most simplistic terms, the models can be described as the following (see review by \citealp{longmore_2014}): 
\begin{itemize}
\item[i)] ``Conveyor-belt'' ($\rho_\mathrm{cloud}^\mathrm{inital} < \rho_\mathrm{YMC}^\mathrm{final}$; $R_\mathrm{cloud}^\mathrm{inital} >> 1$\,pc): the molecular cloud has an initial gas density distribution lower than the stellar distribution of the final YMC (i.e. $\rho_\mathrm{cloud}^\mathrm{inital} < 10^{3\,\pm1}$\sol\,pc$^{-3}$). Star formation can occur throughout the cloud following it hierarchical gas density distribution (e.g. \citealp{larson_1981}). As the system evolves, both the gas and the embedded protostellar population concurrently globally collapse, until all the gas has formed stars or been expelled, and stellar dynamics eventually dominate (e.g. \citealp{girichidis_2012, kruijssen_2012, kruijssen_2012a, zamora-aviles_2012}). The merging of the initially hierarchical structure, imprinted on the protostellar population from the gas, forms a smooth, centrally concentred, bound stellar cluster (e.g. \citealp{fujii_2012, parker_2014}).
\item[ii)] ``In situ'' ($\rho_\mathrm{cloud}^\mathrm{inital}  \approx \rho_\mathrm{YMC}^\mathrm{final}$; $R_\mathrm{cloud}^\mathrm{inital} \approx 1$\,pc): star formation is initially inhibited within the molecular cloud, and the gas alone contracts to reach a density similar to the final YMC stellar density (i.e. $\rho_\mathrm{cloud}^\mathrm{inital} \approx 10^{3\,\pm1}$\sol\,pc$^{-3}$). Stars then form at this higher gas density, and do not have to change their density distribution to reach that of the final YMC stellar density.
\item[iii)] ``Popping'' ($\rho_\mathrm{cloud}^\mathrm{intial}  > \rho_\mathrm{YMC}^\mathrm{final}$; $R_\mathrm{cloud}^\mathrm{inital} << 1$\,pc): as in scenario ii), the molecular cloud collapses with inhibited star formation, but down to an even higher gas density (smaller radius) than the final YMC stellar density (i.e. $\rho_\mathrm{cloud}^\mathrm{inital} > 10^{3\,\pm1}$\sol\,pc$^{-3}$). Star formation then proceeds at this higher gas density. As the stellar population is formed, the cluster exhausts or expels its gas content, hence removing its gravitational influence, and the cluster expands towards its final, lower stellar density distribution (e.g. \citealp{lada_1984, boily_2003, bastian_2006, baumgardt_2007}).
\end{itemize}
However, definitively discriminating between these models is complicated by the scale-free nature of molecular clouds, as imposing arbitrary density (or extinction) thresholds to define clouds can lead to differing interpretations depending on whether ongoing star formation is included within the boundary. Despite these caveats, a relatively simple test to discriminate between the conveyor-belt, in situ, and popping cluster formation scenarios can be conducted by comparing molecular clouds, proto-clusters and clusters at different evolutionary stages. If these star-forming proto-clusters clouds are observed with $\rho_\mathrm{cloud}^\mathrm{initial} > \rho_\mathrm{proto-cluster} > \rho_\mathrm{YMC}^\mathrm{final}$ then these can only form a YMC through the conveyor-belt scenario. 

\begin{figure*}
\centering
\includegraphics[trim = 0mm 0mm 10mm 0mm, clip,angle=0,width=0.98\textwidth]{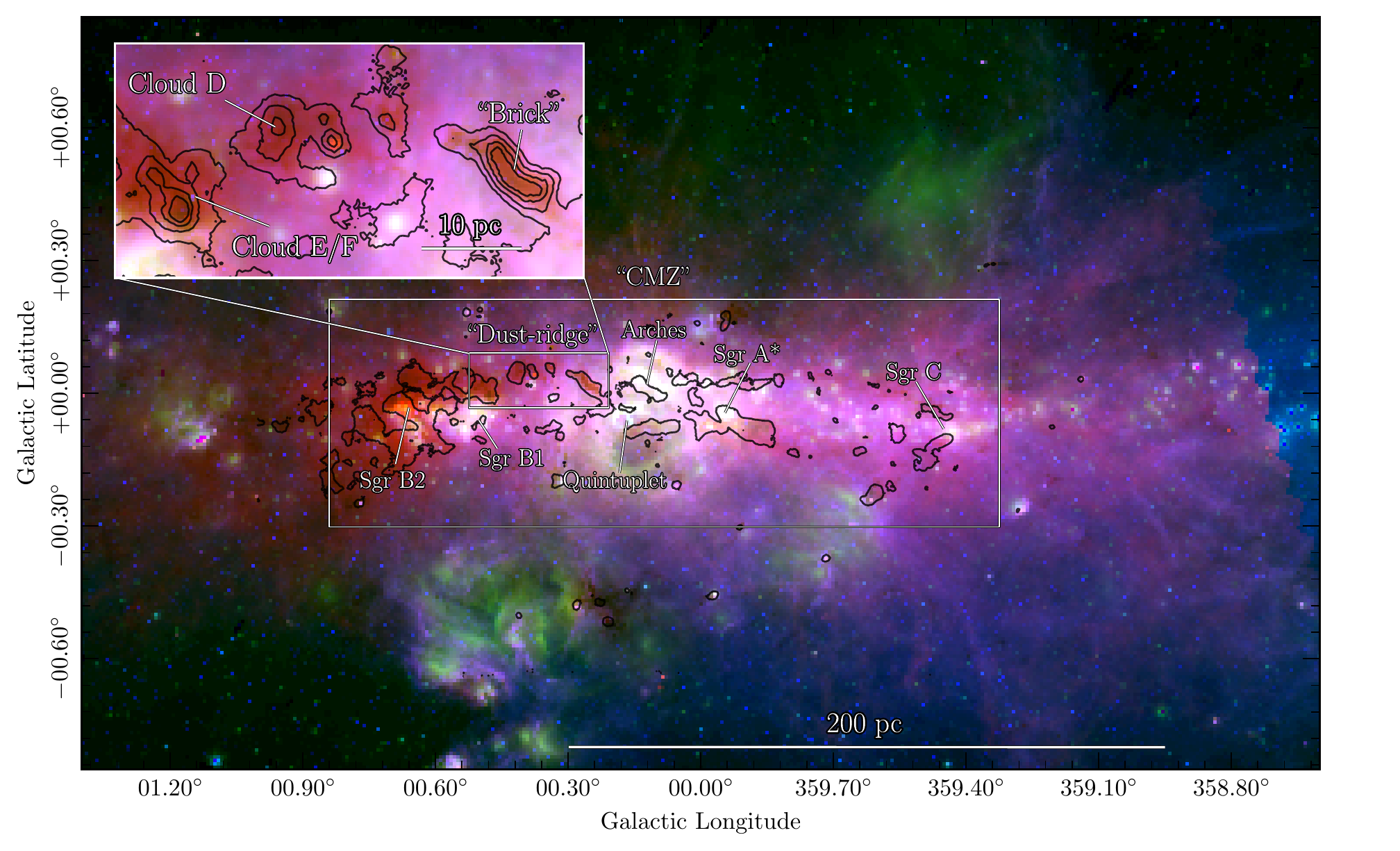}\vspace{-4mm}
\caption{A three colour image of the Galactic Centre. In this image, red is 70\,\micron\ emission from {\it Herschel} Hi-GAL \citep{molinari_2010}, green is 24\,\micron\ emission from {\it Spitzer} MIPSGAL \citep{carey_2009}, and blue is 8\,\micron\ emission from {\it Spitzer} GLIMPSE \citep{churchwell_2009}. Labeled are the sources of interest throughout this region, and shown as rectangles are the approximate regions of the Central Molecular Zone (or CMZ) and the dust-ridge. Observations from the BOLOCAM Galactic Plane Survey (BGPS) is shown in a contour of 0.5 Jy\,beam$^{-1}$ for the main panel, and contours of 0.5, 1, 2, 3, 7, 20 and 50 Jy\,beam$^{-1}$ for the zoom-in panel, which have been chosen to highlight the regions of dense molecular gas ($>$\,10$^{23}$\,\cmsq; \citealp{ginsburg_2013}). Shown in the upper left is a zoom-in of the dust-ridge region, which contains the sources Cloud D and Cloud E/F that are studied in this work. Shown in the lower right of full and zoom-in images are scale-bars representing projected lengths of $\sim$\,200\,pc and $\sim$\,10\,pc, respectively, at a distance of $\sim$\,8\,kpc \citep{reid_2014, grav_collab_2018}.} 
\label{three_col_map}
\end{figure*}

Along these lines, \citet{walker_2016} have conducted an extensive study of YMC progenitors within both the disc of the Galaxy (W49, W51, G010.472+00.026, G350.111+0.089, G351.774-00.537, G352.622-01.077) and central 200\,pc of the Galaxy (G0.253+0.016, Cloud D, E/F), referred to as the Central Molecular Zone (CMZ, see Figure\,\ref{three_col_map}, also see \citealp{walker_2015}). These authors find that quiescent clouds in both environments do not have the densities required to form a YMC (CMZ: Sgr B2 main, north, Arches; Disc: NGC 3603, Trumpler 14, W1), and they only begin to approach high enough densities when they harbour a significant level of star formation (i.e. they have evolved for $>$1\,Myr; e.g. see Sgr B2 in their Figures 7 \& 8). This would suggest that the conveyor-belt scenario for cluster formation is the most common throughout the Galaxy. 

The result that all the observed YMC progenitors have a common formation mechanism, regardless of environment, is somewhat surprising given that the central 200\,pc of the Galaxy has very extreme environmental conditions (e.g. \citealp{kruijssen_2013}). It may even be surprising that massive YMC progenitor clouds can exist within the Galactic Centre at all without rapidly forming stars, as their average densities are factors of a few to several orders of magnitude larger than required for many of the commonly adopted critical densities for star formation, which are typically calibrated for disc environments ($\sim$\,10$^{4}$\,\cmcb; e.g. \citealp{lada_2010, lada_2012}). It has, however, been noted for several decades that despite containing $\sim$\,80\,per cent of the Galaxy's dense molecular gas (2\,-\,6\,$\times10^{7}$\,\sol; \citealp{morris_1996}), the CMZ does not appear to be forming stars at a proportional rate (e.g. \citealp{guesten_1983, caswell_1983, taylor_1993}), with recent estimates at $<$\,10\,per cent of the Galaxy's total star formation rate (e.g. \citealp{longmore_2013, barnes_2017}). Indeed, high-resolution observations show a distinct lack of core condensates within the Galactic Centre molecular clouds \citep{rathborne_2014a, rathborne_2015, walker_2018, kauffmann_2017a, kauffmann_2017b}, in comparison to similar density and age high-mass star forming regions within the disc, which are typically highly fragmented on core scales ($\sim$\,0.1pc; c.f. \citealp{wang_2014, dirienzo_2015, henshaw_2016c, henshaw_2016d, kainulainen_2013b, kainulainen_2017, motte_2018, beuther_2019}). It has been suggested that this lack of star and dense core formation is due to the higher fraction of turbulent gas that is sinusoidally (divergence-free) driven within the the Galactic Centre (e.g. \citealp{rathborne_2014, kruijssen_2014a, federrath_2016, barnes_2017, ginsburg_2018}).
 
In this work we target the young massive cluster progenitor clouds found towards the ``dust-ridge'' region of the CMZ (e.g. \citealp{lis_1999, longmore_2013a}), which is highlighted in the three colour image presented in Figure\,\ref{three_col_map}. The dust-ridge region is composed of several massive (e.g. \citealp{walker_2016}), relatively quiescent (e.g. \citealp{immer_2012, barnes_2017}), and kinematically complex molecular clouds (e.g. \citealp{henshaw_2016, henshaw_2019}), and is thought to have been formed by a recent ($<$Myr) flow of gas into the CMZ from larger Galactic radii ($\sim$\,kpc; e.g. \citealp{sormani_2019}). Specifically, here we present results based on the high-angular resolution, high-sensitivity, high-dynamic-range Atacama Large Millimetre array (ALMA) observations of the dust-ridge clouds ``Cloud D" (G0.412+0.052) and ``Cloud E/F" (G0.489+0.010).\footnote{These clouds were originally referred by \citet{lis_1999} to as Clouds ``D", ``E" and ``F", who separated the structures based on dust continuum emission (i.e. ``D", ``E" and ``F"). However, recent analysis of molecular line observations suggest Clouds E and F may be physically linked (e.g. \citealp{henshaw_2016}). Therefore, as they are covered by the same mosaic in the observations presented in this work, these are henceforth referred to as a single cloud, "Cloud E/F".}. These clouds are massive (gas masses of $\sim$\,$10^{5}$\,\sol), compact (radii of $\sim$1\,pc), and are thought to harbour only the earliest stages of star formation (no prominent \ion{H}{ii} regions; e.g. \citealp{caswell_2010, immer_2012, titmarsh_2016, lu_2019}), and hence represent the ideal candidates to distinguish between the current cluster formation mechanisms.

The ALMA observations presented here will be used to investigate a range of outstanding questions relating to core, star and cluster formation and evolution over a series of future works. In this first paper, we present an overview of the datasets of both clouds (i.e. both continuum and line observations), and focus our analysis to understanding YMC formation. This paper is organised as follows. Section\,\ref{sec:obs} presents the data calibration, reduction and imaging techniques used to obtain both the continuum and line datasets. Section\,\ref{results} presents the column density and moment maps, which are used to identify regions of interest within the clouds, and Section\,\ref{analysis} presents a virial analysis of these regions. Section\,\ref{discussion} presents a discussion on the implication of these results, the critical density for star formation, and the implication of the virial analysis on the different theories of star and stellar cluster formation. A summary of this paper is then presented in Section\,\ref{conclusions}. In the online version, the appendix contains several additional tables and figures.

\section{Observations}\label{sec:obs}

\subsection{ALMA interferometric observations}

\begin{table*}
\caption[Information of ALMA Galactic Centre observations.]{Observations log. Shown are the sources, observation date, array configuration, total on-source integration time, and the sources used for the band pass, flux and phase calibrations.}
\small
\centering
\begin{tabular}{ c c c c c c c }
\hline
Cloud & Date & Array & On-source time & Band Pass & Flux & Phase \\ 
& & configuration$^{a}$ & (hour:min) & calibrator & calibrator & calibrator \\ \hline
D & 26/04/2015 & 12\,m C34-1/(2) & 1:11 & J1733-1304, 1924-2914 & Titan, Neptune & J1744-3116 \\
D & 13/08/2015 & 7\,m ACA & 0:35 & J1733-1304 & Titan & J1744-3116 \\
D & 14/08/2015 & 7\,m ACA & 0:42 & J1733-1304 & J1733-1304 & J1744-3116  \\
D & 15/08/2015 & 7\,m ACA & 0:29 & 1733-1304 & J1733-1304 & J1744-3116 \smallskip \\ 

E/F & 26/04/2015 & 12\,m C34-1/(2) &  2:07 & J1517-242, J1733-1304 & Titan & J1744-3116  \\
E/F & 27/04/2015 & 12\,m C34-1/(2) & 0:46 & J1733-1304 & Titan & J1744-3116 \\
E/F & 18/08/2015 & 7\,m ACA & 1:05 & J1733-1304 & Titan, Neptune & J1744-3116 \\
E/F & 03/09/2015 & 7\,m ACA & 0:28 & J1924-2914 & Titan & J1744-3116 \\
E/F & 04/09/2015 & 7\,m ACA & 0:15 & J1924-2914 & Titan & J1744-3116\\
E/F & 20/09/2015 & 7\,m ACA & 0:22 & J1517-2422 & Titan & J1744-3116 \\ \hline

\end{tabular}
\label{observations log}
	\begin{minipage}{\textwidth} \vspace{1mm}
	\footnotesize
	$a$: The parenthetical numbers are added to indicate that the array configuration is expected to contain sufficient baselines to approximate either configuration (see \url{https://almascience.nrao.edu/observing/prior-cycle-observing-and-configuration-schedule}). \\
	\end{minipage}
\vspace{-5mm}
\end{table*}

\begin{table*}
\caption[Parameters of ALMA observations]{Observational parameters.}
\small
\centering
\begin{tabular}{l c c c}
\hline
Observational parameter  & Cloud D & Cloud E/F \\
\hline
Synthesised beam: \\
\hspace{1cm} major axis, $\theta_{\rm major}$ (\arcsec) & 1.47 &  1.27 \\
\hspace{1cm} minor axis, $\theta_{\rm minor}$ (\arcsec) & 0.90 & 0.90 \\
\hspace{1cm} beam position angle, $\theta_{\rm PA}$ ($\degr$) & -23.2 & 0.0 \smallskip\\
Velocity Resolution, $\Delta \upnu_\mathrm{res}$ (\kms) & 1.25 & 1.25 \\
Continuum rms level, $\sigma_{\rm rms}$ (mJy\,beam$^{-1}$)$^a$ & 0.4 &  0.6 \\
Line rms level, $\sigma_{\rm rms}$$^b$ & 0.1\,K [9\,mJy\,beam$^{-1}$]  & 0.07\,K [4\,mJy\,beam$^{-1}$] \\
\hline
\end{tabular}
\label{parameters_alma}
\begin{minipage}{0.6\textwidth} \vspace{1mm}
\footnotesize
$a$: The rms level determined across the full $\sim$\,8\,GHz bandwidth. \\
$b$: The rms level determined within a single channel. \\
\end{minipage}
\end{table*}


To investigate the early stages of star formation within these regions on proto-stellar core scales, high-angular resolution dust continuum and molecular line observations have been taken with ALMA as part of the Cycle 2 project: 2013.1.00617.S (Principal investigator: S.N. Longmore). The observations made use of the Band 6 receiver, configured to use four spectral windows in dual polarisation centred at 250.5\,GHz, 252.5\,GHz, 265.5\,GHz, and 267.5\,GHz, each with a bandwidth of 1875\,MHz (1920 channels), a channel spacing of 977\,kHz (uniformly regridded to 1.25\,\kms\ in all cubes used throughout this work), and resolution of 1129\,kHz (equivalent to 1.35\,\kms\ at 250\,GHz). The observations were carried out in April, August and September 2015 (see Table\,\ref{observations log}). During these dates, the configurations of 12\,m and 7\,m arrays had projected baseline ranges of $15.0 - 348.5$\,m (configuration C34-1) and $8.9 - 48.9$\,m (ACA), respectively, which, at the average observed frequency of $\sim$\,259\,GHz, gives an combined angular resolution of $\sim$\,1\arcsec\ and a maximum recoverable size scale up to $\sim$\,50\arcsec. At this frequency, the primary beam sizes of the 12\,m and 7\,m dishes are  $\sim$\,25\arcsec\ and  $\sim$\,42\arcsec\, respectively. Given these, we proposed for mosaics containing 100 pointings with the 12\,m array, and 37 pointings with the 7\,m array for Cloud D, and 132 pointings with the 12\,m array, and 47 pointings with the 7\,m array for Cloud E/F. The 12\,m array observations for both clouds were, however, not fully completed. This resulted in the final 12\,m array mosaics containing 65 and 88 pointings for Cloud D and E/F, respectively. The missing pointings can be seen to the upper right side of both clouds, and result in an irregularly shaped coverage (see Figures\,\ref{cont_cloud_d} and \ref{cont_cloud_ef}). The complete observational information regarding the final on-source integration time for each array configuration, the observation date, and the bandpass, phase and amplitude calibrators are given in Table\,\ref{observations log}.

\subsubsection{Calibration}

As a result of the missing pointings, the 12\,m observations were assigned a Quality-Assurance stage 0 ``Semi-Pass'' classification, and were not subject to the pipeline reduction and Quality-Assurance stage 2 stage. The raw data, therefore, had to be manually calibrated. This was done in the Common Astronomy Software Applications package {\sc casa}\footnote{see \url{https://casa.nrao.edu}} version 4.4.0 with assistance from the ALMA support scientist at the UK ALMA Regional Centre.\footnote{see \url{http://www.alma.ac.uk/}} For consistency, we also chose to calibrate the 7\,m array observations. 


As is common practice, after calibration we created rough images of the dataset for checking. Upon comparison with observations made with the Submillimeter Array (SMA) towards these sources \citep{walker_2018}, systematic offsets of 3.6\arcsec\ and 2.2\arcsec\ for Cloud D and Cloud E/F, respectively, between the bright, compact sources were found. This was caused by a known problem with around 80 projects observed as part of Cycles 1 to 3, of which 2013.1.00617 is included. The problem was produced by the ALMA online system, which introduced a small mislabelling of the position of each field. This was due to an inconsistency between the procedure for computing the coordinates that are stored in the field table of the data by the online software, and the procedure for computing the delay propagation and antenna pointing coordinates. The issue only affected programs that intended to either map extended areas around a reference (mosaics) or used offset pointings from a reference position. For such maps, this problem would result in a distortion of the final image, which depends on the distance from the reference position, the coordinates of the reference position and the size of the area mapped (for mosaics). Therefore, in addition to the normal astrometric uncertainty, the positions derived from these images have a systematic error whose magnitude depends on the above factors (see ALMA User Support Ticket ID: 6347). This issue was corrected, and the raw data was again downloaded and reduced following the previously produced scripts. A comparison between the ALMA and SMA observations, when smoothed to a comparable angular resolution of $\sim$\,4\arcsec, showed no obvious offset, and hence this issue was deemed to be resolved. We do note, however, an issue still persists in all {\sc casa} versions at the time of publication, whereby regridding from the International Celestial Reference Frame (the default created by {\sc tclean}) to the Galactic coordinate system with {\sc imregrid} produces a systematic offset of $\sim$\,0.5\arcsec\ across the map (ALMA/CASA User Support Ticket ID: 14182/5379). This issue, however, does not effect the results presented in this work, as all the maps (i.e both lines and continuum) contain the same systematic offset, and hence no significant relative difference.

As with the previous calibration, rough images of the final calibrated dataset were produced to check for abnormalities. Given that no further unexpected issues were then present, the next step was to identify the channel ranges which contain strong line emission (see online appendix). These channels were then masked and a first-order polynomial baseline was fit to the remaining channels using the task {\sc uvcontsub}.\footnote{It is preferable to do this at this stage, rather than post imaging with the {\sc imcontsub} routine.} This task produces a ``model'' continuum dataset, which is subtracted from the original dataset to produce a continuum subtracted dataset. The latter of these is used for molecular line imaging, and the former for continuum imaging. We note that using the model continuum dataset for continuum imaging is, however, not advised in the imaging guidelines. To test this, maps were produced using the continuum ``model'' output and produced when masking the line channels in the whole cube in {\sc tclean}. Comparison of these showed that qualitatively, the flux distributions appear to be very similar. Quantitatively, on scales of up to 5\arcsec\ and 10\arcsec\ the fluxes are in agreement to within 10$\%$ and 20$\%$, respectively. The induced uncertainty, along with the known issues when cleaning in {\sc CASA-4.7.0}\footnote{See the North American ALMA Science Center Software Support Team \& the CASA Team Memo \#117.}, which was also checked and found to cause a flux difference of $\sim$\,5$\%$ compared to images produced in the most recent version of CASA at the time of publication ({\sc CASA-5.4.0}), are accounted for in section\,\ref{uncertainties}. Given that these uncertainties do not change the results presented in this paper, and that it was significantly faster to produce a cleaned image using the continuum model output, and, hence, easier to test and refine the imaging method presented below, the images produced using the continuum models for both clouds are used throughout this work.

The calibrated data sets from the 12m and 7m arrays were weighted and imaged together with clean process {\sc tclean} in {\sc CASA-4.7.0}. This was chosen over the standard {\sc clean} function for its increased functionality and improved stability. For example, testing showed that {\sc clean} would begin to diverge from a solution for a much smaller number of cleaning cycles compared to {\sc tclean}. 

\subsubsection{Imaging}\label{imaging}

Initially, a ``basic'' set of parameters (i.e. a ``Hogbom'' deconvolver and a single run with large iterations) was used in {\sc tclean}. The produced images, however, contained many artefacts and had noise levels significantly higher than the theoretical noise limits. To produce the best image quality (e.g. with minimal side-lobe structure), the cleaning of both the continuum maps and molecular line cubes was done in an iterative process. The data were cleaned down to a given noise level (with a ``multiscale'' deconvolver), and then the resultant image was checked. If required, the mask was then adjusted, and the clean continued down to a lower noise threshold. The steps of this process are:  
\begin{itemize} 
\item[i)] The ``dirty'' image was produced by setting the number of clean cycle iterations to zero. Using this dirty map, an initial mask corresponding to some high multiple of the noise was produced (typically $\sim$10\,$\sigma_{\rm rms}$). The mask was then pruned such that structures smaller than a given multiple of the beam size are removed (typically $\sim$3\,beams), hence removing any noise spikes taken into the mask.

\item[ii)] The initial mask was then applied in the {\sc tclean} function, which effectively informed clean where to find the brightest, and therefore most likely real, structures within the map. This procedure was repeated until a specified threshold was reached. At the first pass of this stage, the threshold should be reasonably high (typically $\sim$\,10\,$\sigma_{\rm rms}$), such that only the bright structures are cleaned, and to make sure clean does not begin to diverge early. With this choice of a high initial threshold, fainter, extended emission remained in the residual image.

\item[iii)] A new mask was then made from the residual image produced in stage ii), which has a lower multiple of the noise and a higher multiple of the beam size for pruning than used in the previous masking stage (typically around a factor of two lower threshold and a factor of two larger in beam size than previously used). This mask encompassed more of the larger scale, lower level emission.

\item[iv)] The mask from step iii) was then used in clean, with the image from step ii) as the starting model. Using the image as the starting model allowed clean to continue from step ii), taking into account the information of the bright emission (i.e. effectively removing it before clean begins), such that clean can focus on the lower level, larger scale emission.  

\end{itemize} 
The steps iii) and iv) were then repeated until an acceptable image was reached, or the deconvolved image began to diverge from a sensible solution (e.g. producing large negative bowls in the image). Cleaning the images via this method of dynamically altering the mask, rather than directly cleaning the image down to a threshold of a given sigma level, enhanced the lower level, diffuse emission, whilst suppressing artefacts commonly seen in interferometric images (e.g. large-scale striping across the image). 

\subsection{ALMA and single dish continuum observations}

\begin{figure*}
\centering
\includegraphics[trim = 0cm 23mm 0cm 32mm, clip,angle=0, width= 0.9 \textwidth]{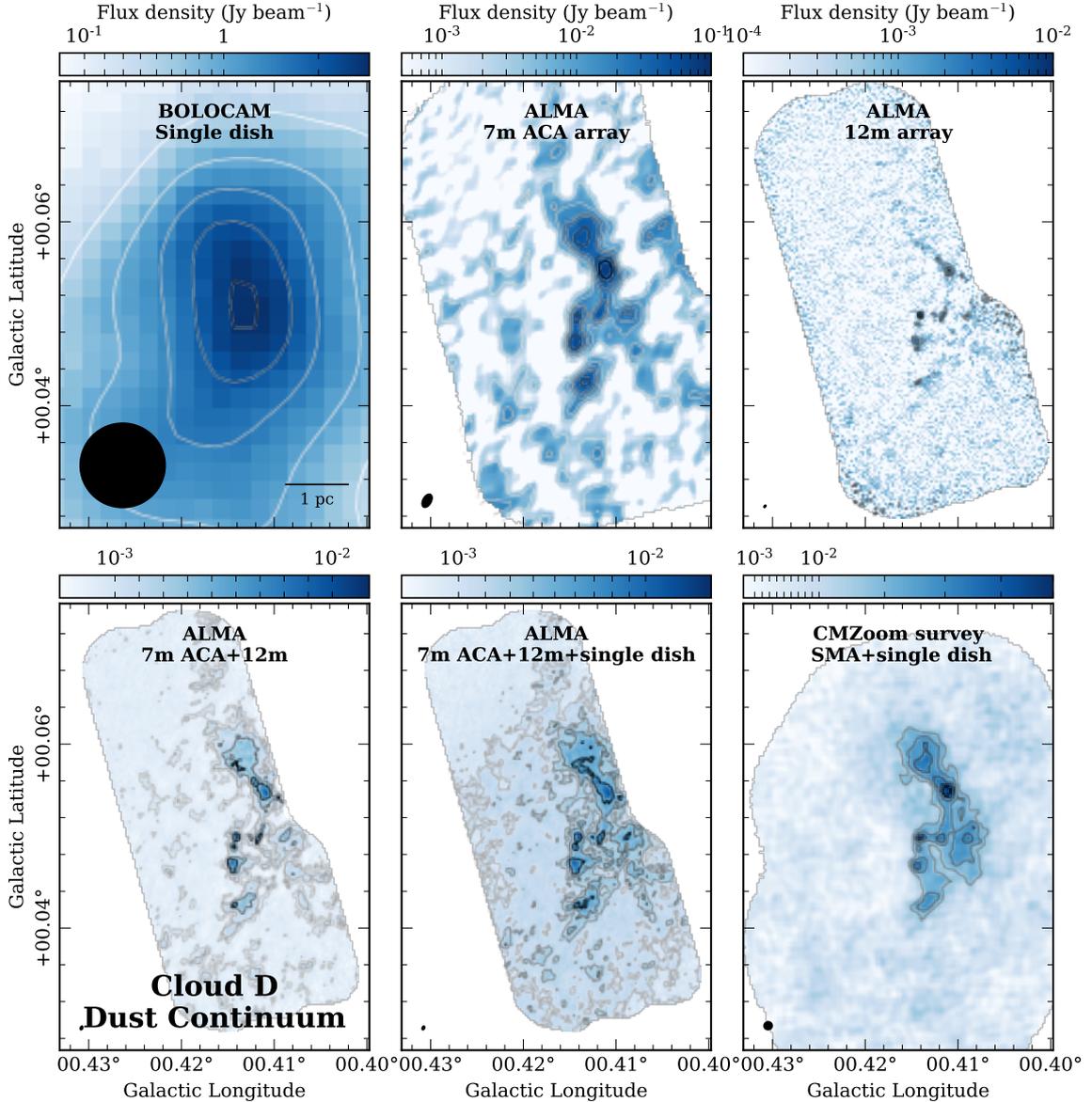}\vspace{-1mm}
\caption{Continuum observations towards Cloud D. The upper left panel shows the single dish observations from the BGPS \citep{ginsburg_2013}, overlaid with contours of [1, 1.5, 2, 2.5] Jy\,beam$^{-1}$. The upper centre panel shows the map produced from the ACA 7m array observations only, overlaid with contours of [3, 9, 15, 20, 30, 40]\,$\sigma_{\rm rms}$, where $\sigma_{\rm rms}$\,$\sim$\,1.4\,mJy\,beam$^{-1}$. The upper right shows the map produced from the 12m array observations only, overlaid with contours of [6,  9, 15]\,$\sigma_{\rm rms}$, where $\sigma_{\rm rms}$\,$\sim$\,0.2\,mJy\,beam$^{-1}$. The lower left panel shows the map produced from the combined 12m and ACA 7m observations, overlaid with contours of [3, 6, 15, 25]\,$\sigma_{\rm rms}$, where $\sigma_{\rm rms}$\,$\sim$\,0.2\,mJy\,beam$^{-1}$. The lower centre panel shows the combined (or ``feathered'') 7m ACA, 12m, and single dish map, overlaid with contours of [3, 6, 9, 15]\,$\sigma_{\rm rms}$, where $\sigma_{\rm rms}$\,$\sim$\,0.4\,mJy\,beam$^{-1}$. The lower right panel shows the SMA map for comparison, overlaid with contours of [3, 4, 5, 6, 7]\,$\sigma_{\rm rms}$, where $\sigma_{\rm rms}$\,$\sim$\,6\,mJy\,beam$^{-1}$ \citep{walker_2018}. Contours for each panel have been chosen to best highlight the structure in the map, with colours from white to black showing increasing levels. Shown in the lower left of each panel is the beam size for each set of observations. Shown in the lower right of the upper left panel is a scale bar for reference.} 
\label{cont_cloud_d}
\end{figure*}

\begin{figure*}
\centering
\includegraphics[trim =  0cm 45mm 0cm 52mm, clip, angle=0, width= 0.9 \textwidth]{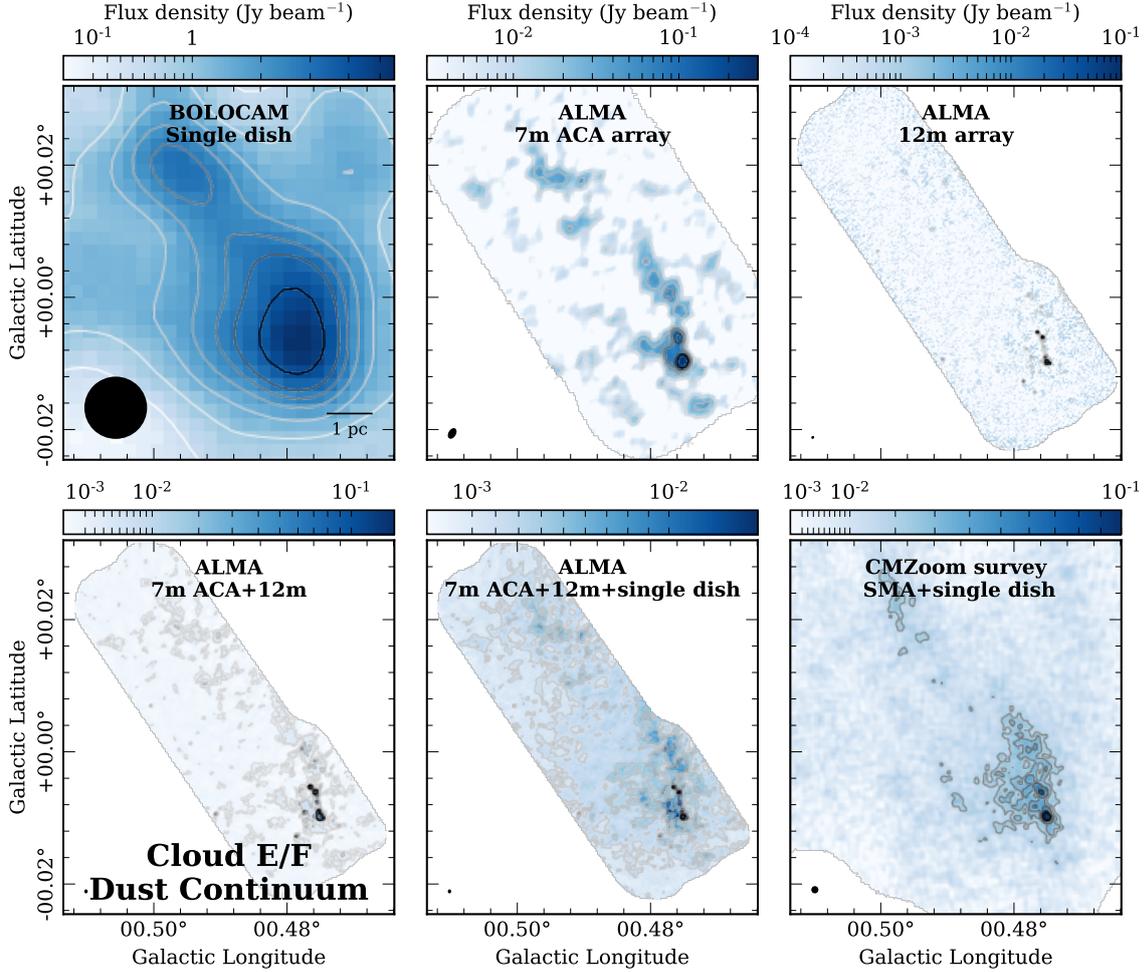}\vspace{-1mm}
\caption{Continuum observations towards Cloud E/F. The upper left panel shows the single dish observations from the BGPS \citep{ginsburg_2013}, overlaid with contours of [1, 1.5, 2, 2.5, 3, 4] Jy\,beam$^{-1}$. The upper centre panel shows the map produced from the ACA 7m array observations only, overlaid with contours of [3, 9, 15, 20, 30, 40]\,$\sigma_{\rm rms}$, where $\sigma_{\rm rms}$\,$\sim$\,3\,mJy\,beam$^{-1}$. The upper right shows the map produced from the 12m array observations only, overlaid with contours of [6,  9, 15, 30, 50, 70]\,$\sigma_{\rm rms}$, where $\sigma_{\rm rms}$\,$\sim$\,0.2\,mJy\,beam$^{-1}$. The lower left panel shows the map produced from the combined 12m and ACA 7m observations, overlaid with contours of [3, 6, 15, 30, 50, 70]\,$\sigma_{\rm rms}$, where $\sigma_{\rm rms}$\,$\sim$\,0.3\,mJy\,beam$^{-1}$. The lower centre panel shows the combined (or ``feathered'') 7m ACA, 12m, and single dish map, overlaid with contours of [3, 6, 9, 15, 30, 50, 70]\,$\sigma_{\rm rms}$, where $\sigma_{\rm rms}$\,$\sim$\,0.6\,mJy\,beam$^{-1}$. The lower right panel shows the SMA map for comparison, overlaid with contours of [3, 4, 5, 6, 9, 12]\,$\sigma_{\rm rms}$, where $\sigma_{\rm rms}$\,$\sim$\,7\,mJy\,beam$^{-1}$ \citep{walker_2018}. Contours for each panel have been chosen to best highlight the structure in the map, with colours from white to black showing increasing levels. Shown in the lower left of each panel is the beam size for each set of observations. Shown in the lower right of the upper left panel is a scale bar for reference.}
\label{cont_cloud_ef}
\end{figure*}

Single dish continuum observations taken with the 10.4-meter diameter Caltech Submillimeter Observatory (CSO), as part of the BOLOCAM Galactic Plane Survey (BGPS; \citealp{ginsburg_2013}), were used to estimate the zero-spacing (i.e. the missing {\it uv}-coverage of the interferometric observations). The BGPS is a publicly available,\footnote{\url{http://irsa.ipac.caltech.edu/data/BOLOCAM_GPS/}} 1.1 mm survey of dust emission in the Northern Galactic plane, covering longitudes -$10^{\circ} < l < 90^{\circ}$ and latitudes $|b|$ $< 0.5^{\circ}$ with a typical rms sensitivity of $30-100$\,mJy in a $\sim$\,33\arcsec\ beam. These observations were chosen as they closely match the frequency and coverage of the ALMA observations, whilst having a moderate cross-over between CSO dish size (10.4\,m) and the smallest baseline of the ALMA observations (8.9\,m). Crossover in dish size is important for the combination of the single dish and interferometric observations, such that the absolute flux scaling of the images can be determined (i.e. so that the flux in the single dish image is conserved).

The single dish observations had to be modified before combination. Firstly, as the BOLOCAM observations are at a slightly different frequency to the ALMA dish observations, the flux was scaled in accordance with,
\begin{equation}
\centering
\frac{F_{\rm ALMA}}{F_{\rm BOLOCAM}}  = \left( \frac{\nu_{\rm ALMA}}{\nu_{\rm BOLOCAM}} \right) ^{\alpha_\nu} \approx \left( \frac{259}{272} \right) ^{3.75} \approx 0.8,
\end{equation}
where $F$ (units of Jy beam$^{-1}$) and $\nu$ (units of GHz) are the continuum intensities and approximate central frequencies of the ALMA and BOLOCAM observations, which are denoted in the subscript. \citet{ginsburg_2013} found that the spectral index from the BOLOCAM to higher frequency {\it Herschel} observations is approximately $\alpha_\nu \sim 3.75$, which is consistent with typical dust emissivity index measurements in the range $1.5 < \beta (= \alpha_\nu - 2) < 2.5$ (e.g. \citealp{paradis_2010}). The BOLOCAM image was then regridded and cropped to the same pixel grid and coverage of the ALMA observations. Additionally, before the combination procedure, the ALMA image was corrected for the primary beam response, which has the effect of enhancing emission towards the edge of the mosaic, where the antenna response (or sensitivity) is lower. 

We used the ``feathering'' technique to combine the prepared BOLOCAM image and ALMA image. Feathering works by taking the Fourier transforms of both images, summing them with a weighting factor applied to each image, and taking the inverse transform to produce a combined image (see \citealp{cotton_2017}). The weighting factor is applied during this procedure such that the combined image has a total flux comparable to the single dish observations. To conduct this procedure, we used the {\sc feather} function from {\sc CASA}-version 4.7.0 with the default parameter set (i.e. effective dish size, single dish scaling and low pass filtering of the single dish observations). As a consistency check, we compared the resultant combined images for both Cloud D and E/F to the single dish BOLOCAM observations (accounting for the frequency difference) and found that the total flux within the mapped region was conserved. Shown in Figures\,\ref{cont_cloud_d} and \ref{cont_cloud_ef} are the BGPS single dish only, and 7m ACA array, 12m array, the combined 7m ACA and 12m array and combined 7m ACA, 12m and single-dish maps towards Clouds D and E/F, respectively. 

The final combined continuum map for Cloud D has an angular beam major axis size, minor axis size, and position angle of $\theta_{\rm major}$: 1.47\arcsec, $\theta_{\rm minor}$: 0.90\arcsec, and $\theta_{\rm PA}$: -23.2$\degr$, respectively, with a 1\,$\sigma_{\rm rms}$ sensitivity of $\sim$\,0.4\,mJy\,beam$^{-1}$. The final combined map for Cloud E/F has $\theta_{\rm major}$: 1.27\arcsec, $\theta_{\rm minor}$: 0.90\arcsec, and $\theta_{\rm PA}$: -0.0$\degr$, and a 1\,$\sigma_{\rm rms}$ sensitivity of $\sim$\,0.6\,mJy\,beam$^{-1}$. These values are summarised in Table\,\ref{parameters_alma}.

\subsection{ALMA line observations}

Since these are the first observations of Clouds D and E/F with this spectral coverage the first step was to identify the detected molecular lines. To do so, dirty images of the whole continuum subtracted data cube where produced for a selection of sub-regions throughout both clouds, i.e. step i) in the process presented in the previous section. These positions were chosen to include both the peaks and the more diffuse continuum emission, and thereby to eliminate any potential bias (see \citealp{rathborne_2015}). To identify the molecular transitions potentially responsible for any emission peak observed above a 3\,$\sigma_{\rm rms}$ threshold, the frequency was firstly adjusted to the source velocities of $\sim$\,20\,\kms\ and $\sim$\,30\,\kms\ for Cloud D and Cloud E/F, respectively \citep{henshaw_2016}, and then compared to the rest frequency of the lines within the Splatalogue spectral line database.\footnote{\url{http://www.splatalogue.net}} In some cases multiple line transitions were present in the database with frequencies in agreement with the observed line emission. To choose between these, we took into account several criteria: whether any transitions from the given molecules had already been observed; for the case of rare isotopologues, whether any transitions from the main isotopologue had already been observed; the expected intensity (either CDMS/JPL or Lovas/AST, as listed by Splatalogue); and whether the upper state energy of the line falls within a reasonable range of 10 - 200\,K (i.e. similar to the highest gas temperature within these cloud, as determined by \citealp{walker_2018}).

The list of molecules detected within the clouds is presented in Table\,\ref{molecules}, and the full information regarding the individual transitions is presented in the online appendix. In some cases, due to many lines being very close in frequency, the selection criteria listed above did not produce a definitive line identification. These cases are highlighted in the notes column of the online table. At the end of online table the frequencies of lines that were detected yet not identified are given, which have been adjusted for the assumed source velocity such that they represent the rest frequency of the associated transition. For comparison, in this online table the molecular transitions that have been detected within the ``Brick'' molecular cloud are presented, which have been identified using complementary Band 6 ALMA observations (Contreras et al. in prep)\footnote{Project: 2012.1.00133.S (Principal investigator: G. Garay)}.

After the cleaning procedure, the final line cubes were converted from flux density to brightness temperature units assuming the standard conversion in the Rayleigh-Jeans limit, which for Cloud D and E/F is approximately 11\,K (Jy beam$^{-1}$)$^{-1}$ and 18\,K (Jy beam$^{-1}$)$^{-1}$, respectively. We note that, unlike the continuum, the cubes are not primary beam corrected, and therefore have uniform $\sigma_{\rm rms}$ levels of $\sim$\,0.1\,K ($\sim$\,9\,mJy\,beam$^{-1}$) and $\sim$\,0.07\,K ($\sim$\,4\,mJy\,beam$^{-1}$) for Cloud D and E/F, respectively (see Table\,\ref{parameters_alma}).

\section{Results}\label{results}

\subsection{Column density analysis}\label{column_density_section}

\begin{figure*}
\centering
\includegraphics[trim = 0mm 0mm 0mm 0mm, clip, angle=0, width=0.95\textwidth]{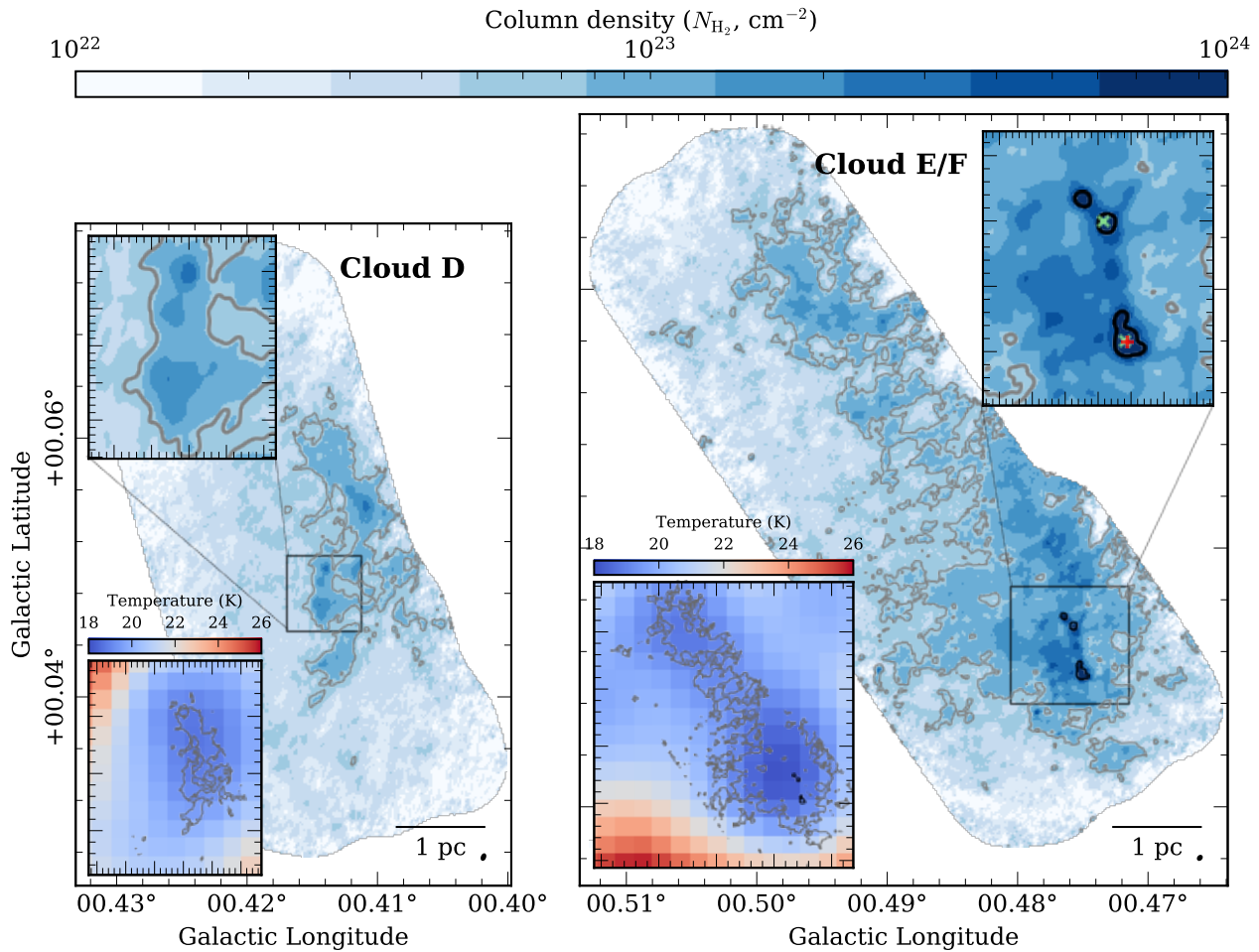}\vspace{-4mm}
\caption{The column density of molecular hydrogen determined towards Cloud D [left panel] and Cloud E/F [right panel]. These have been plotted on the same discretised colour-scale, and angular size scale for ease of comparison. The grey and black contours overlaid on both panels represent column density levels of $N_{\rm H_{2}} = [67, 670] \times10^{21}$\,cm$^{-2}$, respectively (one and two orders of magnitude higher than the star formation relation from \citealp{lada_2012}, respectively). For clarity, a zoom-in of the core region within Cloud E/F is shown in the upper right of the right panel, overlaid with are red $+$ and green $\times$ symbols marking the position of the H$_2$O and CH$_3$OH Class II maser emission, respectively \citep{caswell_2010, titmarsh_2016, lu_2019}. Shown in the lower left of the main panels are maps of the {\it Herschel} derived dust temperature across each region (\citealp{barnes_2017}; Battersby et al. in prep), which were used in the calculation of the column density. The overlaid contours are identical to those in the main panel.}
\label{column_density}
\end{figure*}

The column density distribution and inferred mass above a given threshold are key parameters for many models used to predict the rate of star formation within molecular clouds (e.g. see \citealp{padoan_2014}). The simplest and most commonly used of these is the empirical scaling relation from \citet{lada_2010, lada_2012} (also see \citealp{kainulainen_2014}). The authors suggest that the rate at which stars form increases linearly with increasing mass above a molecular hydrogen column density of $N_{\rm H_{2}} \geq 6.7\times10^{21}$\,cm$^{-2}$. To investigate how this applies to the clouds studied here, the column density of hydrogen has been determined from the dust continuum emission assuming the observed flux-density, ${\rm S}_{\rm \nu}$ (Jy\,beam$^{-1}$),  at a given frequency, ${\rm \nu}$, is well described by the standard equation of radiative transfer, assuming no background contribution. This is given as,
\begin{equation}
S_\nu = \frac{2 h \nu^{3} }{ c^2 } \frac{1}{{\rm exp} (h \nu / k_{\rm B} T) -1}  [1 - {\rm exp }(- \uptau_{\rm \nu})],
\label{eq:column}
\end{equation}
where $T$ is the dust temperature and $c$, $h$ and $k_{\rm B}$ are the speed of light, the Planck constant and Boltzmann constant, respectively. The opacity is defined as,
\begin{equation}
\uptau_\nu = f_{\rm gd}\,\mu_\mathrm{H_2} m_\mathrm{H} \kappa_\nu N_{\rm H_{2}},
\label{eq:opacity}
\end{equation}
where $f_{\rm gd}$ is the gas-to-dust ratio (typically assumed to be 100), $\mu_\mathrm{H_2}$ is the mean molecular weight per hydrogen molecule (2.8\,a.m.u; see e.g. \citealp{kauffmann_2008}), and the dust opacity, $\kappa_{\nu}$, is given as $\kappa_{\nu}$\,=\,$\kappa_0 \left( \nu / \nu_{0} \right)^\beta$ at $\nu_{0}$\,=\,230\,GHz, with the linear absorption coefficient $\kappa_0$\,=\,0.9\,cm$^{2}$\,g$^{-1}$ \citep{ossenkopf_1994} and an index of $\beta$\,=\,1.75 \citep{battersby_2011}. 

To obtain an estimate of the dust temperature we use far-infrared dust continuum emission, as observed with the {\it Herschel} space observatory. Following the method outlined in \citet{battersby_2011}, both the molecular hydrogen column density and dust temperature at each position is determined by fitting the spectral energy distribution from 70$-$500\,\micron\ with a modified black body function (see \citealp{barnes_2017} for a full outline of the procedure). The dust temperature maps across both clouds are shown in the lower left inset panels of Figure\,\ref{column_density}. These maps are interpolated onto the higher angular resolution ALMA pixel grid, and then used to calculate the column density at each position using equations\,\ref{eq:column} and \ref{eq:opacity}. The final column density maps for both clouds are presented in the main panels of Figure\,\ref{column_density}. These have been plotted using the same angular size scale, and a discretised colour-scale has been chosen such that the different density regimes present in both clouds can be easily compared, as shown by the colour bar above the shown panels.

Figure\,\ref{column_density} clearly shows that the vast majority of the observed gas within both of these clouds sits well above the star formation threshold from \citet{lada_2012}. Overlaid on both panels are contours of factors of 10 and 100 times the \citet{lada_2012} threshold, i.e. $N_{\rm H_{2}} = [67, 670] \times10^{21}$\,cm$^{-2}$ (shown as grey and black contours). These contours show that both clouds have typical column densities around an order of magnitude larger than is typically expected for quiescent clouds in the solar neighbourhood. We find that the peak column density towards the centre of Cloud D is 3.1$\times$\,10$^{23}$\cmsq, and towards the south of Cloud E/F it is 3.7$\times$\,10$^{24}$\cmsq. 

The peaks towards the south of Cloud E/F corresponds to previously identified 24\,\micron\ and 70\,\micron\ point sources, and H$_2$O and CH$_3$OH maser emission sources \citep{churchwell_2009, molinari_2010, caswell_2010, titmarsh_2016, lu_2019}, which are thought to pinpoint potential sites of high-mass star formation. For clarity, zoom-ins of these regions are also shown in Figure\,\ref{column_density}, with the position of the H$_2$O and CH$_3$OH maser emission sources within Cloud E/F shown by the green and red crosses, respectively. 


The highest contour of $N_{\rm H_{2}} = 6.7\times10^{23}$\,cm$^{-2}$ (i.e. two orders of magnitude higher than the \citealp{lada_2010} and \citealp{lada_2012} threshold), is only observed towards this region containing the only known signature of star formation across both clouds (see zoom-in region shown in the right panel of Figure\,\ref{column_density}). This contour level is close to an alternative threshold for the formation of high-mass stars, as determined by \citet{krumholz_2008}. These authors suggest that only clouds with a density of $\geq$\,1\,\gcmsq\ (i.e. $N({\rm H}_{2})\,\geq\,6\times\,10^{23}$\,\cmsq) can avoid fragmentation and, therefore, form high-mass stars. Indeed, the mass contained within this contour level could be up to $\sim$\,600\,\sol\ (see section\,\ref{virial_cores}), and is, therefore, capable of forming one or several high-mass stars (assuming a typical star formation efficiency for a core of $\sim$\,25 per cent; e.g. \citealp{enoch_2008}). 

\subsection{Moment map analysis}\label{moment_map_analysis}

\begin{table}
\caption{The molecules and the number of detected transitions for the Galactic Centre clouds (see table in online version for full details on the transitions). In this table, the molecules with cyclic, trans and gauche isomers have been grouped, and only the constituent chemical formula is shown.}
\centering
\small
\begin{tabular}{c c c c}
\hline
 Number  &  \multicolumn{2}{c}{Detected molecule} & Number of \\ 
 of atoms & Main & Isotopologue  & transitions detected \\ 
\hline
2 \\
 & SO & & 2 \smallskip \\
3 \\
& SO$_2$ & & 2\\
& HCN & & 1 \\
& HCO$^+$ & & 1\smallskip \\
4 \\
& HNCO & & 1 \\
&  & HDCO & 1\smallskip  \\
5 \\
& CH$_2$NH & & 3 \\
& HCOOH & & 1 \\
& HCCCH & & 1\smallskip \\
6 \\
& CH$_3$OH & & 30 \\
&  & $^{13}$CH$_3$OH & 4 \\
& CH$_3$SH & & 1 \\
& & $^{13}$CH$_3$CN & 8 \\
& NH$_2$CHO & & 2 \smallskip \\
7 \\
& CH$_3$NH$_2$ & & 1 \\
& CH$_3$CHO & & 5 \smallskip \\
8 \\
& CH$_3$OCHO & & 1 \smallskip\\
9 \\
& CH$_3$OCH$_3$ & & 3 \\
& CH$_3$CH$_2$OH & & 2 \\
& CH$_3$CH$_2$CN & & 8 \\ \hline
\end{tabular}
\label{molecules}
\end{table}

\begin{figure*}
\centering
\includegraphics[trim = 25mm 2mm 25mm 0mm, clip, angle=0, width=1\textwidth]{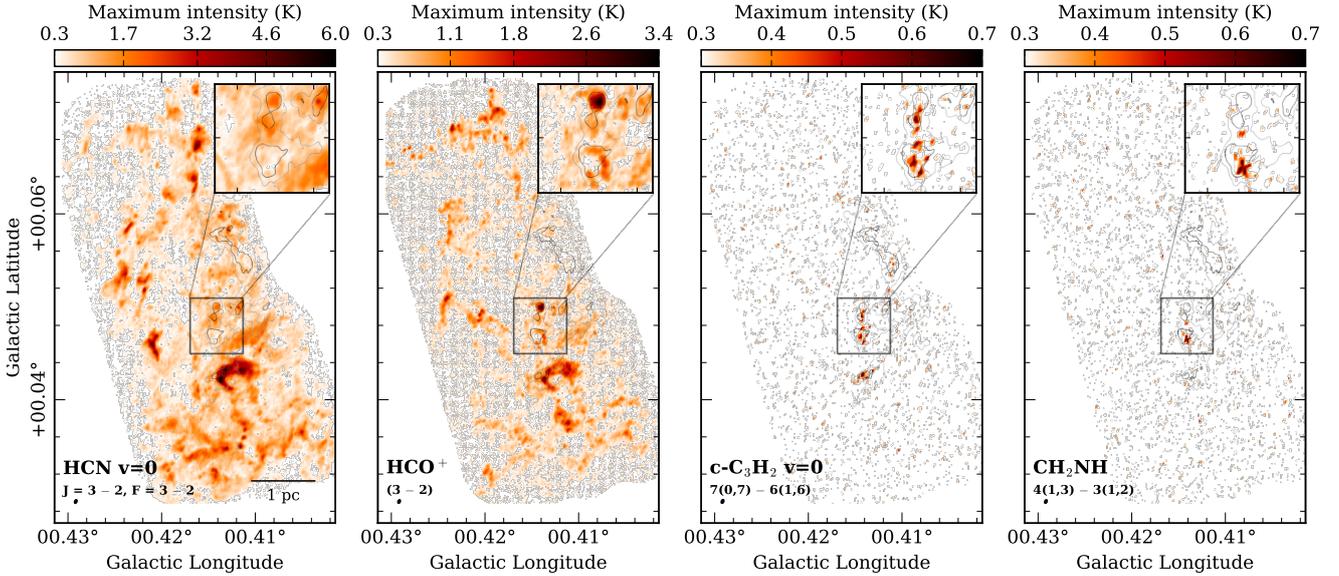}\vspace{-6mm}
\caption{Maximum intensity moment maps of all the molecular transitions detected towards Cloud D [labelled in the lower left of each panel]. Contours overlaid on each map are of the continuum emission, which is shown in levels of [9, 15]\,$\sigma_{\rm rms}$, where $\sigma_{\rm rms}$\,$\sim$\,0.4\,mJy\,beam$^{-1}$. Contours for each panel have been chosen to best highlight the structure in the map, with colours from grey to black showing increasing levels. A zoom-in of the core region is shown in the upper right of the right panels. Shown in the lower left of each panel is the beam size for each set of observations. Shown in the lower right of the left panel is a scale bar for reference.}
\label{moment_maps_d}
\end{figure*}

\begin{figure*}
\centering
\includegraphics[trim = 4mm 12mm 1mm 12mm, clip, angle=0, width=1\textwidth]{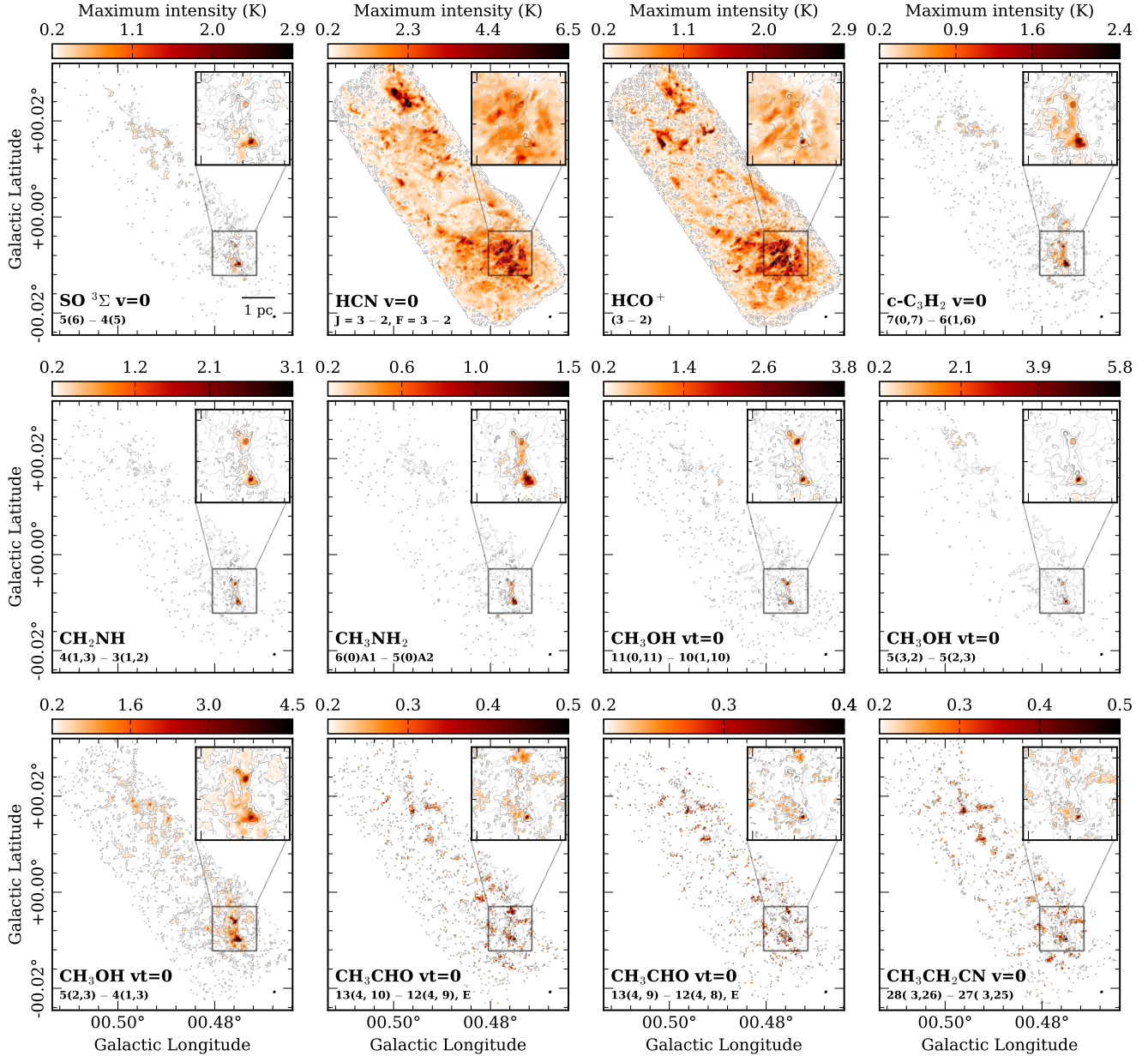}\vspace{-6mm}
\caption{Maximum intensity moment maps of all the ``extended'' and ``moderately extended'' molecular transitions detected towards Cloud E/F [labelled in the lower left of each panel]. Contours overlaid on each map are of the continuum emission, which is shown in levels of [7, 15, 50]\,$\sigma_{\rm rms}$, where $\sigma_{\rm rms}$\,$\sim$\,0.6\,mJy\,beam$^{-1}$. Contours for each panel have been chosen to best highlight the structure in the map, with colours from grey to black showing increasing levels. A zoom-in of the core region is shown in the upper right of each panel. Shown in the lower right of each panel is the beam size for each set of observations. Shown in the lower right of the upper left panel is a scale bar for reference.}
\label{moment_maps_ef}
\end{figure*}

In the previous section, we established that star formation only appears to be present within a very small region of one of the two clouds, despite both having average column densities of around an order of magnitude higher than has been suggested for the onset of star formation \citep{lada_2010, lada_2012}. To investigate how these environmental conditions affects the molecular line emission, this section provides an overview of all the detected transitions and presents a simple moment map analysis to investigate their distribution throughout both clouds.

The identified molecules, ordered by increasing molecular weight, are summarised in Table\,\ref{molecules}. The main molecular configuration and the isotopologues have been sorted into separate columns, and the number of transitions detected for each is shown in the final column (also see table in online version). 

To conduct the moment map analysis, firstly velocity ranges that contain all the emission above a 3$\sigma_{\rm rms}$ threshold were identified for each line. Typically these velocity ranges were between  $15.0 - 30.0$\,\kms\ and $25.0 - 35.0$\,\kms\ for Cloud D and Cloud E/F, respectively. However, the more extended molecules HCO$^+$ and HCN required larger velocity ranges of  -$100 - 130$\,\kms. The moment maps were then created using the {\sc spectral cube} package for {\sc python}.\footnote{\url{https://spectral-cube.readthedocs.io/en/latest/}.}, after masking emission below a 3\,$\sigma_{\rm rms}$ threshold within the cubes. The results of this analysis for all the identified molecular lines can be found in the online appendix of this work.

To highlight the variation in the spatial distribution of the emission from the identified molecules, shown in Figures\,\ref{moment_maps_d} and \ref{moment_maps_ef} are the maximum intensity moments (i.e. the voxels with the highest intensity within the chosen velocity range). In Figure\,\ref{moment_maps_d} we show all the lines detected towards Cloud D. Given the large number of molecular transitions identified within Cloud E/F, only several molecules, which are discussed below, are presented in Figure\,\ref{moment_maps_ef} (see online appendix for all maps).

Emission from the observed hydrogen cyanide, HCN~($3-2$), and formylium, HCO$^+$~($3-2$), transitions cover the largest velocity range and most extended spatial coverage of all the identified molecular lines. This is expected given that they have similar formation conditions, and have similar theoretical excitation densities (i.e. ``critical density'') of around $\sim$\,10$^{6-7}$\,\cmcb.\footnote{Einstein coefficients listed as $A_{\rm ji}\sim$\,8\,$\times$\,$10^{-4}$\,s$^{-1}$ and $A_{\rm ji}\sim$\,1\,$\times$\,$10^{-3}$\,s$^{-1}$ for HCN~($3-2$) and HCO$^+$~($3-2$), respectively. The critical density is approximated as $n_{\rm crit} = A_{\rm ji} \,/\, \left< \sigma_{\rm cross} v_{\rm therm} \right>$, where $\left< \sigma_{\rm cross} v_{\rm therm} \right> \,\approx 10^{-10}$\,\cmcb s$^{-1}$.} However, despite their high critical densities, the HCN and HCO$^+$ maps show little correspondence to the features identifiable within the dust continuum maps (compare to overlaid contours on Figures\,\ref{moment_maps_d} and \ref{moment_maps_ef}), as is typically observed within Galactic Disc star forming regions (e.g. \citealp{kauffmann_2017}). Indeed, \citet{rathborne_2015} found a similar result for the majority of the molecular line transitions identified in ALMA Band 3 observations towards the Brick molecular cloud (see Figure\,\ref{three_col_map}). There are physical mechanisms that may explain this difference in molecular line to dust continuum morphology, such as different spatial distributions of the gas and dust, excitation conditions, and optical depth effects, as well as observational reasons, such as missing spatial scales within the interferometric images. We can not, however, differentiate between these explanations as we are currently missing scales larger than $\sim$\,30\arcsec, due to the lack of single-dish observations for these transitions. 

The methanimine, CH$_2$NH~($4-3$), and cyclopropenylidene, c-HCCCH~($7-6$) (henceforth, \cCtHt), molecules have also been observed within both clouds, both of which also appear to trace similarly compact regions within the clouds. Within Cloud E/F the CH$_2$NH and \cCtHt\ emission is limited to regions towards the north and south of the cloud (i.e. lowest continuum contour on Figure\,\ref{moment_maps_ef}). Within Cloud D these are much fainter, albeit still significant given the 3\,$\sigma_{\rm rms}$ detection threshold of $\sim$\,0.3\,K (see Table\,\ref{parameters_alma}), and are only observed towards the peak in dust continuum emission towards the centre of the mapped region (i.e. highest contour on Figure\,\ref{moment_maps_d}). The critical excitation density of the CH$_2$NH~($4-3$) and \cCtHt~($7-6$) transitions is similar to that of HCN~($3-2$) and HCO$^+$~($3-2$) ($\sim$\,10$^{6-7}$\,\cmcb),\footnote{Einstein coefficients listed as $A_{\rm ji}\sim$\,2\,$\times$\,$10^{-4}$\,s$^{-1}$ and $A_{\rm ji}\sim$\,9\,$\times$\,$10^{-4}$\,s$^{-1}$ for CH$_2$NH~($4-3$) and \cCtHt~($7-6$), respectively.} however the former appear to be selectively tracing only the densest region within the clouds (i.e. above a contour of $\sim2\times10^{23}$\,cm$^{-2}$). This could be an effect of the chemistry within the region, whereby CH$_2$NH and \cCtHt\ are less abundant than HCN or HCO$^+$, and, therefore, more suitable to pinpoint the density peaks. Moreover, we find that the spectra of the CH$_2$NH and \cCtHt\ lines towards both clouds are simple, typically showing only single Gaussian features which have velocities coincident with strong dips in the more complex HCN and HCO$^+$ spectra. These dips are likely a result of the high optical depth of the HCN and HCO$^+$  transitions, whereby the emission is self-absorbed at the mean centroid velocity of the region. This would point to the CH$_2$NH and \cCtHt\ being both optically thin, likely due to their low abundances, which would make them the ideal observational tools to probe the compact, dense gas dynamics of high-mass molecular clouds within the Galactic Centre.

The remaining molecular line transitions shown in Figure\,\ref{moment_maps_ef} have a similar spatial morphology to that of the CH$_2$NH and \cCtHt\ lines, which may indicate that these also trace a similar density regime within Galactic Centre clouds. It is interesting to then consider why these are exclusively observed within Cloud E/F and not within Cloud D. Indeed, the remaining transitions are generally not as bright as CH$_2$NH and \cCtHt\, but still would be above the 3\,$\sigma_{\rm rms}$ detection threshold ($\sim$\,0.3\,K) if they were proportionally as bright, relative to the overall column density of Cloud D, as seen in Cloud E/F. Given that Cloud E/F already contains signs of embedded star formation, which Cloud D does not, it is, therefore, likely that this cloud is more evolved both physically and chemically. Indeed, molecules such as sulphur monoxide (SO) are thought to originate from regions that harbour embedded star formation (e.g. due to strong shocks). It could be that the remaining molecules that are observed within Cloud E/F do not have high enough abundances, or the correct excitation conditions (e.g. high enough density), within Cloud D to emit strongly within the observed frequency range. This would be in agreement with the single dish observations from \citet{jones_2012}, which show complex molecular species, that are not observed in Cloud D, appear to trace Cloud E/F, as well as the evolved, actively star-forming source Sgr B2 (e.g. CH$_3$CCH). These results would then suggest that the molecular line detections within the clouds are linked to their different evolutionary stage.
\section{Analysis}\label{analysis}

\begin{figure*}
\centering
\includegraphics[trim = 0mm 0mm 0mm 0mm, clip,  angle=0, width=0.84\textwidth]{{./../figures/virial/c-HCCCH.251.314_cloud_d-new}.pdf}
\vspace{-2mm}

\includegraphics[trim =  8mm 3mm 10mm 4mm, clip, angle=0, width=0.256\textwidth]{{./../figures/virial/c-HCCCH_d_pub}.pdf}
\includegraphics[trim = 14mm 3mm 10mm 4mm, clip, angle=0, width=0.24\textwidth]{{./../figures/virial/c-HCCCH_d2a_pub}.pdf}
\includegraphics[trim = 14mm 3mm 10mm 4mm, clip, angle=0, width=0.24\textwidth]{{./../figures/virial/c-HCCCH_d4a_pub}.pdf}
\includegraphics[trim = 14mm 3mm 10mm 4mm, clip, angle=0, width=0.24\textwidth]{{./../figures/virial/c-HCCCH_d4b_pub}.pdf}\\ \vspace{-2mm}

\caption{[upper two rows] Dust continuum (measured with ALMA and the SMA) and moment maps of the \cCtHt\ v=0 7(0,7) -- 6(1,6) transition towards the Cloud D proto-cluster region (as labelled in each panel). Overlaid on upper row left panel are the ALMA continuum contours of [6, 9, 15, 20, 25]\,$\sigma_{\rm rms}$, where $\sigma_{\rm rms}$\,$\sim$\,0.4\,mJy\,beam$^{-1}$. Overlaid on the second row left panel are SMA continuum image are contours of [3, 4, 5, 6]\,$\sigma_{\rm rms}$, where $\sigma_{\rm rms}$\,$\sim$\,6\,mJy\,beam$^{-1}$. Overlaid on the remaining map panels are maximum intensity contours of [0.35,  0.42,  0.49,  0.56]\,K. All contours increase in colours from white to grey to black. Shown are the cores d2$_{\rm a}$, d4$_{\rm a}$ and e4$_{\rm b}$ that have been identified within this region, and the average region covering all these cores, referred to as proto-cluster (region d). [lower row] The average spectra of the \cCtHt\ v=0 7(0,7) -- 6(1,6) transition taken from each of the regions. For reference, overlaid on each spectrum are vertical and horizontal dotted lines, which show the systemic velocity of the as proto-cluster region, and the $\sigma_{\rm rms}$ level of $\sim$\,0.1\,K. Also shown are the profiles of the Gaussian fits to each spectrum (fit parameters are given in Table\,\ref{cores_table_d}).} 
\label{core_map_d}
\end{figure*}

\begin{figure*}
\centering
\includegraphics[trim = 0mm 0mm 0mm 0mm, clip,  angle=0, width=0.84\textwidth]{{./../figures/virial/c-HCCCH.251.314_cloud_ef-new}.pdf}
\vspace{-2mm}

\includegraphics[trim = 8mm 15mm 10mm 2mm, clip, angle=0, width=0.256\textwidth]{{./../figures/virial/c-HCCCH.251.314_e_pub}.pdf}
\includegraphics[trim = 14mm 15mm 10mm 2mm, clip, angle=0, width=0.24\textwidth]{{./../figures/virial/c-HCCCH.251.314_e1a_pub}.pdf}
\includegraphics[trim = 14mm 15mm 10mm 2mm, clip, angle=0, width=0.24\textwidth]{{./../figures/virial/c-HCCCH.251.314_e1b_pub}.pdf}
\includegraphics[trim = 14mm 15mm 10mm 2mm, clip, angle=0, width=0.24\textwidth]{{./../figures/virial/c-HCCCH.251.314_e1c_pub}.pdf}\\

\includegraphics[trim =  8mm 3mm 10mm 4mm, clip, angle=0, width=0.256\textwidth]{{./../figures/virial/c-HCCCH.251.314_e1d_pub}.pdf}
\includegraphics[trim = 14mm 3mm 10mm 4mm, clip, angle=0, width=0.24\textwidth]{{./../figures/virial/c-HCCCH.251.314_e1e_pub}.pdf}
\includegraphics[trim = 14mm 3mm 10mm 4mm, clip, angle=0, width=0.24\textwidth]{{./../figures/virial/c-HCCCH.251.314_e2a_pub}.pdf}
\includegraphics[trim = 14mm 3mm 10mm 4mm, clip, angle=0, width=0.24\textwidth]{{./../figures/virial/c-HCCCH.251.314_e2b_pub}.pdf}\vspace{-2mm}

\caption{[upper two rows] Dust continuum (measured with ALMA and the SMA) and moment maps of the \cCtHt\ v=0 7(0,7) -- 6(1,6) transition towards the Cloud E/F proto-cluster region (as labelled in each panel). Overlaid on upper row left panel are the ALMA continuum contours of [10, 15, 20, 30, 50, 70, 90]\,$\sigma_{\rm rms}$, where $\sigma_{\rm rms}$\,$\sim$\,0.6\,mJy\,beam$^{-1}$. Overlaid on the second row left panel are SMA continuum image are contours of [3, 4, 5, 6, 9, 12]\,$\sigma_{\rm rms}$, where $\sigma_{\rm rms}$\,$\sim$\,7\,mJy\,beam$^{-1}$. Overlaid on the remaining map panels are maximum intensity contours of [0.21,  0.42,  0.63,  0.84,  1.05,  1.4 ,  2.1]\,K. All contours increase in colours from white to grey to black. Labeled are the individual cores e1$_{\rm a}$, e1$_{\rm b}$, e1$_{\rm c}$, e1$_{\rm d}$, e1$_{\rm e}$, e2$_{\rm a}$, e2$_{\rm b}$ that have been identified within this region, and the average region covering all these cores, referred to as the proto-cluster (region e). [lower two rows] The average spectra of the \cCtHt\ v=0 7(0,7) -- 6(1,6) transition taken from each of the regions. For reference, overlaid on each spectrum are vertical and horizontal dotted lines, which show the systemic velocity of the proto-cluster region, and the $\sigma_{\rm rms}$ level of $\sim$\,0.07\,K. Also shown are the profiles of the Gaussian fits to each spectrum (fit parameters are given in Table\,\ref{cores_table_ef}).} 
\label{core_map_ef}
\end{figure*}

So far the focus of this work has been on the general dust and molecular line properties of two molecular clouds that reside within the Galactic Centre. We will now focus on an investigation of how the gas dynamics can limit core and star formation by assessing the virial state of the densest gas within these clouds. 

\subsection{Defining the core and proto-cluster regions}\label{virial_cores}

The column density maps presented in Figure\,\ref{column_density} show a variety of dense gas structures, which we attempt to characterise here before beginning the virial analysis. We find that the column density peaks towards the centre of Cloud D, with a value of $\sim$\,3$\times$\,10$^{23}$\cmsq, and that this position is also coincident with emission from the high density tracing molecular lines (\cCtHt\ and CH$_2$NH). The column density peaks within the south of Cloud E/F, with an order of magnitude higher value than observed in Cloud D ($\sim$\,4$\times$\,10$^{24}$\cmsq). This region also contains much more molecular line emission than Cloud D, and both a 24\,\micron\ and 70\,\micron\ point source, and H$_2$O and CH$_3$OH maser emission \citep{churchwell_2009, molinari_2010, caswell_2010}.\footnote{We note that, while the south of Cloud E/F is the only region with any observed star formation indicators, these indicators used are not complete and there could be more star formation activity that is not yet detected. It is feasible that bipolar outflows from any currently unknown protostars could be detected within our HCO$^+$ and HCN observations, yet no unambiguous outflow signatures (e.g line-wings) were detected across the mapped regions.}

Recently, a large survey with SMA has also observed both Clouds D and E/F. The aim of this survey is to uncover the dense gas properties across the entire Galactic Centre, down to scales of around $\sim$\,0.1\,pc ({\it CMZoom} survey; see \citealp{battersby_2017, battersby_2018}). The preliminary results of this survey have already highlighted these regions as has being particularly dense \citep{walker_2018}. These authors show that these regions contain several continuum cores, ``d2'', ``d4'', ``e1'' and ``e2'', with masses ranging between $50 - 400$\,\sol, respectively. It is clear ALMA now resolves these regions into many smaller cores (see Figures\,\ref{cont_cloud_d}, \ref{cont_cloud_ef} and \ref{column_density}). We make use of the {\sc astrodendro} package in python to inspect the complex structure observed within our maps (see e.g. \citealp{goodman_2009}). The highest identified structures within the dendrogram hierarchy (i.e. the "leaves") were then used as a basis to define the elliptical cores within each cloud. We identify three cores within Cloud D, and seven cores within Cloud E/F. We assign a similar nomenclature to these as \citet{walker_2018}, using the subscripts separating the original core fragments (i.e. d2$_{\rm a}$, d4$_{\rm a}$,  d4$_{\rm b}$, e1$_{\rm a}$, e1$_{\rm b}$, e1$_{\rm c}$, e1$_{\rm d}$, e1$_{\rm e}$, e2$_{\rm a}$, e2$_{\rm b}$). These cores are shown on the continuum maps shown in the upper left panels of Figures\,\ref{core_map_d} and \ref{core_map_ef}. Additionally, we identify the larger structure that contains these cores as a ``proto-cluster region'', defined as an ellipse approximately covering the column density contour of $\sim2\times10^{23}$\,cm$^{-2}$ (discussed further in section\,\ref{virial_protocluster}). In the following section, we investigate the dynamics and stability for star formation within these high-density gas regions.

\subsection{Virial state of the cores}\label{virial_cores}

The virial parameter, $\alpha_{\rm vir}$, is the simplest and most commonly used quantity to describe relative importance of the kinetic, $E_{\rm kin}$, and gravitational potential energy, $E_{\rm pot}$, of a parcel of gas. In the idealised case of a spherical core of uniform density supported by only kinetic energy (i.e. no magnetic fields), the virial parameter takes the form (e.g. \citealp{bertoldi_1992}), 
\begin{equation}
\alpha_{\rm vir} = a \frac{2 E_{\rm kin}}{\left| E_{\rm pot} \right|} = a \frac{5 \sigma_\mathrm{line}^{2} R}{G M},
\label{virial}
\end{equation}
where $R$ is the radius of the core, $M$ is the total mass of the core, $\sigma_\mathrm{line}$ is the line-of-sight velocity dispersion of a molecular line, the $G$ is gravitational constant. The factor $a$ accounts for systems with non-homogeneous and non-spherical density distributions, and for a wide range of cloud shapes and density gradients takes a value of $a = 2\,\pm\,1$ \citep{kauffmann_2013}. In the above framework, a value of $\alpha_{\rm vir} < 2$ indicates the cloud is sub-virial and should collapse, whereas for a value of $\alpha_{\rm vir} > 2$ the cloud is super-virial and should expand. The cloud is stable when $\alpha_{\rm vir} = 2$.

The core masses for both clouds have been calculated from the column density map shown in Figure\,\ref{column_density}, for which we use the {\it Herschel} derived dust temperatures, and are presented in Tables\,\ref{cores_table_d} and \ref{cores_table_ef}. \citet{walker_2018} found higher gas temperatures of 86\,K towards core d2, and $>$150\,K towards core e1, which would lower the mass estimates by a factor of several to an order of magnitude. Estimates of the core masses using these higher temperatures for the cores d2$_{\rm a}$ and e1$_{\rm a}$ found here are shown in Tables\,\ref{cores_table_d} and \ref{cores_table_ef}, respectively. We apply this higher temperature to only the core e1$_{\rm a}$ within core e, as it is associated with both an infrared point source and maser emission, and hence is likely heated by an embedded protostar. Whereas, the remaining cores, e1$_{\rm b - e}$, appear to be devoid of star formation, and are therefore expected to have temperatures more similar to those derived using the {\it Herschel} observations (see Tables\,\ref{cores_table_d} and \ref{cores_table_ef}).

The line-of-sight velocity dispersion for use in equation\,\ref{virial} is best derived from a molecular line that reliably probes both the entire core region and the individual cores. We have, therefore, chosen the \cCtHt\ v=0 7(0,7) -- 6(1,6) transition, since it is readily detected in both the cores and in proto-cluster regions, and also has relatively simple spectral profiles (i.e. single velocity components, no sign of broad line wings from outflows). The moment maps for \cCtHt, along with the spectrum averaged across each core is shown in Figures\,\ref{core_map_d} and \ref{core_map_ef}. 

The {\sc python} package {\sc pyspeckit}\footnote{Version: 0.1.20, https://pyspeckit.bitbucket.io} was used to a fit a Gaussian profile to the emission above the $\sigma_{\rm rms}$ level of each spectra. Given the simplicity of the observed profiles, the fitting procedure provided robust fits for a range of initial guesses for the peak brightness temperature, centroid velocity and velocity dispersion required by the {\sc pyspeckit.specfit} package. The results are shown on Figures\,\ref{core_map_d} and \,\ref{core_map_ef} and the fit parameters are given in Tables\,\ref{cores_table_d} and \ref{cores_table_ef}. Given that the measured velocity dispersion, $\sigma_\mathrm{obs}$ is similar in magnitude to the velocity-resolution, $\Delta \upnu_\mathrm{res}$ = 1.25\,\kms\, the velocity-resolution has to be removed in quadrature before the velocity dispersion can be used in the virial equation,  
\begin{equation}
\sigma_\mathrm{line} ^2 = \sigma_\mathrm{obs} ^2 - \frac{\Delta \upnu_\mathrm{res}^2} {8 \,\mathrm{ln} 2}.
\label{velocity_disp}
\end{equation}
The values of the velocity dispersion presented in Tables\,\ref{cores_table_d} and \ref{cores_table_ef} have the velocity-resolution subtracted (i.e. $\sigma_\mathrm{line}$). 

All the necessary variables in equation\,\ref{virial} have now been derived, allowing the virial parameter for each core to be calculated. These are shown in Tables\,\ref{cores_table_d} and \ref{cores_table_ef}, where the values in the parenthesis for cores d2$_{\rm a}$ and e1$_{\rm a}$ have been calculated using the higher temperatures determined by \citet{walker_2018}. When including these higher virial estimates, we find that half (5/10) of the cores have the correct criteria for being gravitationally bound and susceptible to collapse ($\alpha_{\rm vir}\leq2$).


\begin{table*}
\caption{The properties of the cores and proto-cluster within Cloud D (see Figure\,\ref{cont_cloud_d}). Shown is the measured radius of the major and minor axis of the ellipse used to define the cores, effective radius when assuming a spherical geometry, integrated continuum flux, the mean {\it Herschel} dust temperature, gas mass, number density, and the free-fall time. Also shown are the results from the Gaussian fitting procedure of the \cCtHt\ molecule towards each core, of the peak brightness temperature, centroid velocity, velocity dispersion. Shown in the second to last row is the estimated Mach number, $\mathcal M\,=\,\sigma_\mathrm{line} / c_s$, where the gas sound speed is equal to $c_s = (k_{B} T / \mu_\mathrm{P} m_\mathrm{H} )^{0.5}$, where the mean molecular weight per free particle $\mu_\mathrm{P} = 2.33$ (see e.g. \citealp{kauffmann_2008}).  Lastly, the virial parameter is given in the final row.}
\centering
\small
\begin{tabular}{l c c c c c c c c c c}
\hline
\multirow{2}{*}{Property} & Proto-cluster & \multicolumn{3}{c}{Core} \\

& d & d2$_{\rm a}$ & d4$_{\rm a}$ & d4$_{\rm b}$ \\ 
\hline

Minor radius, $R_{\rm minor}$ (\arcsec) & 9.7 & 3.1 & 1.8 & 2.0\\ 
Major radius, $R_{\rm major}$ (\arcsec) & 4.8 & 2.1 & 1.7 & 1.5\\ 
Effective radius, $R_{\rm eff}$ (pc) & 0.3 & 0.11 & 0.07 & 0.07\\ 
Integrated flux (Jy) & 0.5 & 0.11 & 0.06 & 0.04\\ 
Dust temperature (K) & 19.2 & 19.3 & 19.0 & 19.1\\ 
Mass, $M$ (\sol) & 559 & 119 (20)$^{a}$ & 63 & 46\\ 
Density, $n_{\rm H_2}$ ($10^{5}$ \cmcb) & 0.7 & 3.4 (0.6)$^{a}$ & 5.9 & 4.3\\ 
Free-fall time, $t_{\rm ff}$ ($10^{4}$ yr) & 11.4 & 5.3 (12.8)$^{a}$ & 4.0 & 4.7\\ 
Peak brightness temperature, $T_{\rm B}$  (K) & 0.1 & 0.2 & 0.2 & 0.2\\ 
Centroid velocity, $V_{\rm LSR}$ (\kms) & 26.2 & 27.8 & 24.5 & 24.6\\ 
Velocity dispersion, $\sigma$ (\kms) & 1.7 & 1.4 & 1.0 & 1.5\\ 
Mach number, ${\mathcal M}$ & 6.4 & 5.3 (2.5)$^{a}$ & 3.7 & 5.6\\ 
Virial parameter, $\alpha_{\rm vir}$ & 1.7 & 2.0 (11.5)$^{a}$ & 1.2 & 3.9\\ 

\hline 
\end{tabular}
\label{cores_table_d}

	\begin{minipage}{\textwidth} \vspace{1mm}
	\footnotesize
	$a$: The parameters shown in parentheses have been calculated using the higher gas temperature estimate of 86\,K determined by \citet{walker_2018}. \\
	\end{minipage}

\vspace{-5mm}
\end{table*}

\begin{table*}
\caption{The properties for cores and proto-cluster within Cloud E/F, identical to Table\,\ref{cores_table_d} (see Figure\,\ref{cont_cloud_ef}).}
\centering
\small
\begin{tabular}{l c c c c c c c c c c}
\hline
\multirow{2}{*}{Property} & Proto-cluster & \multicolumn{6}{c}{Core} \\

&  e & e1$_{\rm a}$ & e1$_{\rm b}$ & e1$_{\rm c}$ & e1$_{\rm d}$ & e1$_{\rm e}$ & e2$_{\rm a}$ &  e2$_{\rm b}$  \\ 
\hline

Minor radius, $R_{\rm minor}$ (\arcsec) & 11.4 & 1.4 & 0.9 & 0.7 & 1.2 & 1.3 & 1.2 & 1.0\\ 
Major radius, $R_{\rm major}$ (\arcsec) & 8.8 & 1.0 & 0.8 & 0.6 & 0.9 & 1.0 & 1.1 & 0.9\\ 
Effective radius, $R_{\rm eff}$ (pc) & 0.42 & 0.05 & 0.04 & 0.03 & 0.04 & 0.05 & 0.05 & 0.04\\ 
Integrated flux (Jy) & 2.58 & 0.27 & 0.06 & 0.02 & 0.04 & 0.05 & 0.13 & 0.08\\ 
Dust temperature (K) & 18.6 & 18.9 & 18.8 & 18.8 & 18.6 & 18.6 & 18.4 & 18.3\\ 
Mass, $M$ (\sol) & 2993 & 311 (28)$^{a}$ & 72 & 19 & 49 & 59 & 148 & 89 \\ 
Density, $n_{\rm H_2}$ ($10^{5}$ \cmcb) & 1.4 & 83.2 (7.7)$^{a}$ & 52.9 & 34.3 & 20.9 & 20.0 & 46.1 & 53.4 \\ 
Free-fall time, $t_{\rm ff}$ ($10^{4}$ yr) & 8.2 & 1.1 (3.5)$^{a}$ & 1.3 & 1.7 & 2.1 & 2.2 & 1.4 & 1.3 \\ 
Peak brightness temperature, $T_{\rm B}$  (K) & 0.2 & 0.9 & 0.9 & 0.7 & 0.5 & 0.6 & 0.5 & 0.3\\ 
Centroid velocity, $V_{\rm LSR}$ (\kms) & 29.9 & 31.2 & 29.0 & 29.4 & 29.4 & 29.8 & 28.1 & 31.1\\ 
Velocity dispersion, $\sigma$ (\kms) & 1.7 & 1.1 & 2.0 & 2.1 & 1.3 & 1.2 & 1.2 & 2.0\\ 
Mach number, ${\mathcal M}$ & 6.6 & 4.4 (1.6)$^{a}$ & 7.5 & 7.9 & 5.2 & 4.6 & 4.6 & 7.7 \\ 
Virial parameter, $\alpha_{\rm vir}$ & 0.5 & 0.2 (2.7)$^{a}$ & 2.2 & 6.9 & 1.8 & 1.3 & 0.5 & 2.0 \\

\hline 
\end{tabular}

	\begin{minipage}{\textwidth} \vspace{1mm}
	\footnotesize
	$a$: The parameters shown in parentheses have been calculated using the higher gas temperature estimate of 150\,K determined by \citet{walker_2018}. \\
	\end{minipage}

\vspace{-3mm}
\label{cores_table_ef}
\end{table*}

%

\subsection{Virial state of the proto-clusters}\label{virial_protocluster}

Along with assessing the dynamics of the individual cores, it is interesting to consider if these regions collectively could go on to form a part of an Arches or Quintuplet-like Galactic Centre YMC \citep{espinoza_2009, harfst_2010}. Henceforth, we will refer to the larger "core d" and "core e" regions, which contain the smaller scale cores (see Figures\,\ref{core_map_d} and \ref{core_map_ef}), as proto-clusters. We assume that these cores will form the central part of the final cluster; i.e. in order to reach a YMC mass, we expect that stars will also form from the global gas reservoir (caveats discussed in section\,\ref{uncertainties}).  

As a simple investigation into the proto-cluster dynamics, we determine the virial state of the proto-clusters using the method of the previous section. The spectrum for these regions and the Gaussian fit are shown in Figure\,\ref{core_map_d} and \ref{core_map_ef}, and the mass, parameters of fit and the determined virial parameter are shown in Tables\,\ref{cores_table_d} and \ref{cores_table_ef}. As before, we estimate the virial parameter using both the  {\it Herschel} derived dust temperature and the higher temperatures determined by \citet{walker_2018} (as shown in parentheses in Tables\,\ref{cores_table_d} and \ref{cores_table_ef}). We find virial parameters of $\alpha_{\rm vir} (\sigma_\mathrm{line}) = 1.7 \pm 1.1$ for Cloud D, and $\alpha_{\rm vir} (\sigma_\mathrm{line}) = 0.5 \pm 0.3$ for Cloud E/F (quoted uncertainties of 65 per cent; see section\,\ref{uncertainties}). These values suggest that the proto-clusters are susceptible to gravitational collapse.

An alternative analysis would be to determine the relative motions of the cores themselves to determine the relative velocity dispersion, $\sigma_\mathrm{rel}$, rather than taking the velocity dispersion from the average spectrum of the proto-cluster region. To do this, we calculate the difference of all the core centroid velocities from the centroid velocity of the proto-cluster, and define the relative velocity dispersion, as the standard deviation of core centroid velocities. We find that the relative velocity dispersion for the cores in Cloud D is $\sigma_\mathrm{rel} = 1.5$\,\kms\ and for Cloud E/F is $\sigma_\mathrm{rel} =  1.0$\,\kms, which are both lower than the values previously determined when taking the average spectrum across the cores ($\sim$\,1.7\,\kms). To calculate the virial parameter, we use the relative velocity dispersion with the effective radius and mass measurements (see Tables\,\ref{cores_table_d} and \ref{cores_table_ef}). We find that the virial parameter calculated using this method for Cloud D is $\alpha_{\rm vir} (\sigma_\mathrm{rel}) = 1.5 \pm 1.0$ and for Cloud E/F is $\alpha_{\rm vir} (\sigma_\mathrm{rel}) = 0.2 \pm 0.1$ (uncertainties of 65 per cent; see section\,\ref{uncertainties}). Even with the large uncertainties on these estimates virial parameter, which could be up to factors of several, we find that when using the relative velocities the proto-cluster both regions appear to be gravitationally bound ($\alpha_{\rm vir}\leq2$).

\section{Discussion}\label{discussion}

The observations presented in this work have allowed us to determine several physical properties of the so-called Galactic Centre dust-ridge clouds (i.e those that harbour only the early stages of star formation), on scales that have not been previously investigated. In this section, we discuss how these observations advance our understanding of star and cluster formation mechanisms within this extreme environment.

\subsection{The mass distribution for young mass cluster formation}\label{mass_distribution}

It has been previously discussed that the density distributions of progenitor YMCs can hold clues regarding their formation mechanisms. For example, by comparing the mass surface density profiles of several YMC progenitors throughout the Galaxy, \citet{walker_2016} showed that the conveyor-belt formation scenario appears to be the most likely formation mechanism across all environments (also see \citealp{walker_2015}). Here we can use the higher resolution observations presented in this work, along with more recent observations of other Galactic Centre star-forming regions, to test this result on much smaller spatial scales than previously possible.

\begin{figure}
\centering
\includegraphics[trim = 5mm 5mm 5mm 3mm, clip,  angle=0, width=0.95\columnwidth]{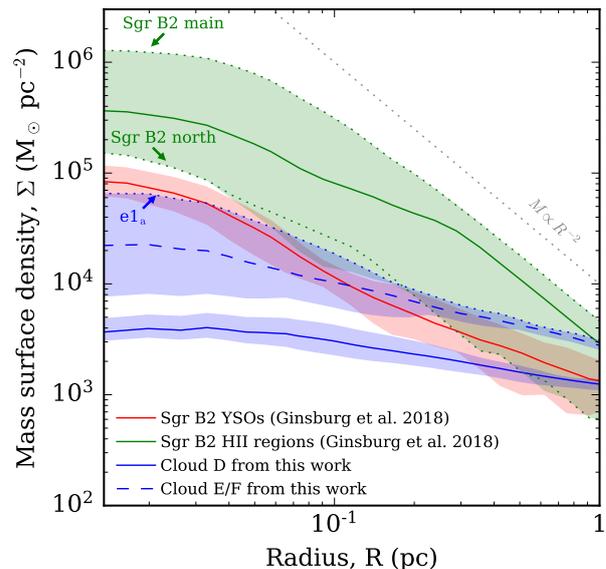}\vspace{-3mm}
\caption{Plot of the mass surface density as a function of radius for several progenitor YMCs within the Galactic Centre. Shown as blue lines are the dust-ridge molecular clouds observed in this work, determined from the column density maps presented in Figure\,\ref{column_density}. The shaded blue region represents the uncertainty produced by varying the aperture centre between the identified cores. The red and green lines show the mass distribution of the young stellar objects (YSOs) and \ion{H}{II} regions identified within Sgr B2 from recent ALMA observations \citep{ginsburg_2018}. Based on the detection limits of these observations, we assume that the unresolved YSOs and \ion{H}{II} regions have masses of 12\,\sol\ and 45.5\,\sol, respectively, which are corrected by factors of 0.14 and 0.09 to account for the limited initial mass sampling (see \citealp{ginsburg_2018}). The dotted blue line shows the profile centred at the the densest observed core region. The dotted green lines highlight well know regions within Sgr B2, as labelled. Also shown as a grey dashed line is the $M\propto R^{-2}$ relation.} 
\label{mass_radius_plot}
\end{figure}

To create the mass surface density profiles for the dust-ridge clouds, we measure the masses at increasing radii from various regions within the mapped regions. For this we use the column density maps (see Figure\,\ref{column_density}), and define the aperture centres as the cores identified in section\,\ref{analysis}. Figure\,\ref{mass_radius_plot} shows the average profiles from apertures within both clouds as solid and dashed blue lines, and the extrema as the shaded blue regions. Also shown in this Figure are the mass surface density profiles of embedded sources within an actively star-forming Galactic Centre cloud, Sagittarius B2 (henceforth Sgr B2), that have been recently identified using ALMA observations \citep{ginsburg_2018}. These authors define the identified embedded sources as young stellar objects (or YSOs), and \ion{H}{II} regions if they are associated with known radio sources (e.g. from \citealp{depree_1998}), and assign them masses according to the following. To be detectable, the \ion{H}{II} regions are assumed to be illuminated by B0 or earlier stars that have masses above 20\,\sol. Each source is, therefore, individually assigned the average mass above this limit, assuming some initial mass function, which they calculate to be 45.5\,\sol. The remaining YSO sources are thought to be proto-stars with masses of $8-20$\,\sol\ that are embedded within warm envelopes. These are individually assigned the average within this mass range of 12\,\sol, again assuming some initial mass function distribution. These values can then be corrected by factors of 14\% for the YSOs and 9\% for the \ion{H}{II} regions to account for the limited initial mass sampling (see \citealp{ginsburg_2018}). We use these corrected masses, i.e. assuming each source can be thought of as an individually embedded sub-cluster, and the source positions to create the mass surface density profiles shown in Figure\,\ref{mass_radius_plot}. Again we vary the aperture centres to create the variation shown as the shaded regions, where the solid lines are the mean values from the variation. 

Figure\,\ref{mass_radius_plot} shows that the gas within Cloud D is the least centrally concentrated on all scales, having a shallow profile from $\sim$\,1\,pc down to $\sim$\,0.01\,pc scales. Cloud E/F has a factor of several higher mass surface density at all measured radii, and also on average a flatter mass surface density profile than in Sgr B2. The upper limit of the shaded region for Cloud E/F, which is taken from the core e1$_{\rm a}$ region (shown as the dashed line in the Figure), however, does show a steepening at around $\sim$\,0.1\,pc that levels off at $\sim$\,0.04\,pc. Interestingly, this core profile appears to be similar to the Sgr B2 YSO profile within the $\sim$\,0.03\,$-$\,0.1\,pc range (i.e. staying within the shaded region for the Sgr B2 YSOs). The Sgr B2 YSO profile does, however, have on average a steeper profile than the clouds on larger scales. The Sgr B2 \ion{H}{II} regions then have an even steeper profile, which typically has factors of several higher mass surface density values than both the Sgr B2 YSO and core e1$_{\rm a}$ profiles. Also shown as dashed green lines on the Figure are the profiles centred on Sgr B2 main ($l$\,=\,0.667\degr, $b$\,=\,-0.035\degr) and Sgr B2 north ($l$\,=\,0.677\degr, $b$\,=\,-0.028\degr), the former representing the densest concentration of protostars within the Sgr B2 region. We then find that the cores within the dust-ridge clouds are on average significantly less dense than the protostars identified within Sgr B2 (approximately two orders of magnitude less dense when comparing to Sgr B2 main), which is in agreement with the results of (\citealp{walker_2015, walker_2016}). 

In summary, and returning to the YMC formation scenarios discussed in section\,\ref{sec:intro}, Figure\,\ref{mass_radius_plot} shows a progression from low-concentration to high-concentration that is correlated with the degree of star formation activity. If one assumes that each of the regions investigated here will end up as a similarly-concentrated central cluster, and that they, therefore, represent different steps of the same evolutionary sequence, this result would be consistent with the conveyor-belt mode of YMC formation (i.e. the concurrent collapse of the gas and stellar objects within a molecular cloud to form a condensed stellar cluster).

\subsection{The critical density for star formation}\label{disc:critial_density}

On a global ($\sim$\,100\,pc) scale, the Galactic Centre is underproducing stars given its mass of dense molecular gas when based on relations calibrated for local environments (e.g. \citealp{longmore_2013, barnes_2017}). This result would suggest that the density at which star formation proceeds within this environment is higher than that observed within the disc. For example, a critical density for star formation of $\sim$\,10$^4$\,\cmcb\ has been suggested for disc star-forming regions \citep{lada_2010, lada_2012}, whereas critical densities as high as $\sim$\,10$^7$\,\cmcb\ have recently been proposed for the Galactic Centre \citep{rathborne_2014, kruijssen_2014a, federrath_2016, kauffmann_2017b, ginsburg_2018}. As discussed in the review by \citet{longmore_2014}, this higher critical density for star formation within the Galactic Centre makes this the ideal environment for the formation of YMC progenitors, because more massive gravitationally bound molecular clouds can form before they begin to form a significant amount of stars. This is key as once stars form their stellar feedback quickly disperses the cloud, and hence limits the contained mass within the resultant stellar cluster.  

In order to see how the observations presented in this work fit into this picture, in the left panel of Figure\,\ref{radius_plot} we plot the mass and radius of the cores identified within Cloud D in blue, and Cloud E/F in red, with those found by \citet{walker_2018} and \citet{lu_2019} using the SMA, and for the larger scales we plot the clouds from \citet{walker_2015}. Here we make the distinction between the cores being the smallest scale structures we identify, and the proto-clusters being the whole core d and e regions (see Figures\,\ref{core_map_d} and \ref{core_map_ef}). Overlaid as diagonal lines are the densities, and for reference we highlight the critical number density for star formation for the Galactic disc \citep{lada_2010, lada_2012, kainulainen_2014} and Galactic Centre \citep{kruijssen_2014a, kauffmann_2017b, krumholz_2017}. This plot shows the Galactic Centre clouds have values between these critical star formation density limits on both the $\sim$\,1\,pc and $<$\,1\,pc scales. We find that the mean density on the smallest scales, for the core regions, within Cloud D and E/F is 5\,$\times$\,10$^{5}$\,\cmcb, and 5\,$\times$\,10$^{6}$\,\cmcb\ (see Table\,\ref{cores_table_ef}), with only the highest mass cores within Cloud E/F approaching the higher Galactic Centre critical density (density within core e1$_\mathrm{a}$ is 8\,$\times$\,10$^{5}$\,\cmcb). Given that we know the Cloud E/F core region contains signs of the very early stages of star formation, this would imply a slightly lower than previously suggested critical density for star formation within the Galactic Centre of $\sim$\,5\,$\times$\,10$^{6}$\,\cmcb. Nonetheless, this is still much larger than the critical density within the Galactic Disc, and hence shows that the density above which stars form should vary across the Milky Way (i.e. $\sim$\,10$^{4}$\,\cmcb\ in the Galactic disc, $\sim$\,10$^{6-7}$\,\cmcb\ in Galactic Centre; e.g. also see \citealp{rathborne_2014}).

The possibility of an environmentally dependent critical density for star formation was recently tested within the  actively star-forming CMZ proto-YMC Sgr B2, by comparing young stellar objects identified with ALMA to the high spatial resolution dust continuum observations, produced by combined {\it Herschel} space observatory observations (SPIRE) with Caltech Submillimeter Observatory (SHARC; for 350\,mm) and James Clerk Maxwell Telescope (SCUBA; for 450\,mm) observations \citep{ginsburg_2018}. These authors find that there are no YSOs below a column density of 10$^{23}$\cmsq, and that half the YSOs reside at $N_{\rm H_{2}}>$10$^{24}$\cmsq\ (i.e. several orders of magnitude higher than the \citealp{lada_2012} threshold). These results are in agreement with those presented here, whereby we only find cores that show the early signs of star formation (i.e. masers; see Figure\,\ref{column_density}) above a column density of $N_{\rm H_{2}}\,\sim\,$10$^{24}$\,\cmsq. Together, these results, therefore, provide further evidence for a higher column density threshold for star formation within the extreme environment of the Galactic Centre.

\begin{figure*}
\centering
\includegraphics[trim = 19mm 1mm 26mm 10mm, clip,  angle=0, width=1\textwidth]{{./../figures/virial/radius_plot}.pdf}\vspace{-6mm}
\caption{[left panel] The gas mass determined for each of the core regions as a function of radius. Blue points are the cores identified in this work, purple and gold are CMZ cores identified using the SMA by \citet{walker_2018} and \citet{lu_2019}, respectively, and dust-ridge clouds from \citet{walker_2015}. The mass and virial ratio determine using the higher temperature estimates for the cores e1$_{\rm a}$ and d2$_{\rm a}$ are show (see text). The x-axis error bars represents an uncertainty of 20 per cent on the radius for the source defined in this work, and 10 per cent for the sources defined by the other authors. Larger uncertainties have been chosen for this work to incorporate the by-eye estimate of the source boundaries. However, this uncertainty should only serve as an estimate, as it is in practise difficult to definitively determine source boundaries due to the hierarchical structure of the interstellar medium. The y-axis error bars represent an uncertainty of 60 and 65 per cent on the mass and virial ratio, respectively (see section\,\ref{uncertainties}). Overlaid are diagonal grey lines that show the corresponding molecular hydrogen number density. Highlighted are the critical density thresholds for star formation, which appear to accurately predict the star formation rate for Galactic Disc star-forming regions, \citep{lada_2010, lada_2012, kainulainen_2014} and the higher critical density that has been determined for the Galactic Centre \citep{kruijssen_2014a, kauffmann_2017b}. [right panel] The virial parameter of the core regions as a function of radius, as calculated using the velocity dispersion determined from the \cCtHt\ v=0 7(0,7) -- 6(1,6) transition fits. Data taken from the same sources as the left panel are also included here for comparison, using identical symbols and colours. The light and dark grey shaded regions show $\alpha_{\rm vir} < 2$ and $\alpha_{\rm vir} < 1$, which represents the regimes of gravitational instability assuming a typical and constant cloud density distribution, respectively \citep{kauffmann_2013}.}
\label{radius_plot}
\end{figure*}

\subsection{Investigating the virial state}\label{discussion_virial}

Up until this study, previous molecular line observations towards Galactic Centre molecular clouds have not been able to identify a molecule that is well correlated with the dust continuum emission (e.g. \citealp{rathborne_2015, lu_2017}). Here, for the first time, we have been able to identify several molecular lines that are selectively tracing both the quiescent and actively star-forming dense gas within the Galactic Centre, and that appear to be optically thin and have relatively simple line profiles (i.e. no inflow or outflow signatures). In section\,\ref{virial_cores} we use these tools to reliably assess the virial state of the cores within the clouds, the results of which are summarised in the right panel of Figure\,\ref{radius_plot}. We find that half of the identified cores have $\alpha_{\rm vir}\leq2$, and are, therefore, within the limits of gravitational collapse for a non-homogeneous and non-spherical parcel of gas (see equation\,\ref{virial}). Given the spatial scales of the ALMA observations ($\sim$\,0.05pc), we would expect these core regions to form single stars, or small bound stellar clusters, on the order of a few to a few tens of solar masses assuming a star formation efficiency of a few tens of per cent. 

In section\,\ref{virial_protocluster} we determined the virial parameter from the relative core velocities, the results of which are also presented on Figure\,\ref{radius_plot}. On this plot the $\sigma_{\rm rel}$ label identifies the markers where the relative velocity dispersion has been used, as opposed to the velocity dispersion determined from the line-width of the region ($\sigma_{\rm line}$). We find that the proto-cluster regions within both clouds, appear to be unstable to gravitational collapse, and most significantly for the proto-cluster region within Cloud E/F ($\alpha_{\rm vir}$\,$\sim$\,0.1). 

We find that the virial ratios for the cores within Cloud E/F, and the proto-cluster region as a whole, are typically lower than those for Cloud D. This would be representative of the more evolved state of Cloud E/F, as highlighted by the higher quantity of observed star formation tracers (see Figure\,\ref{column_density}). This could be explained by a scenario, where the cores have condensed individually and become more gravitationally bound over time, and in addition, the proto-cluster region, as a whole, has globally collapsed to become more centrally concentrated. This would be consistent with a conveyor-belt scenario for cluster formation in the Galactic Centre.

\subsection{Sources of uncertainty}\label{uncertainties}

\subsubsection{Uncertainty propagation on mass and virial estimates}

We adopt a higher than typically assumed uncertainty of 20\,per cent in the absolute flux scale of the ALMA observations, which accounts for the additional uncertainties induced by using the continuum model in the clean process, and the know issue when cleaning in {\sc CASA-4.7.0} (see North American ALMA Science Center Software Support Team \& the CASA Team Memo \#117). In section\,\ref{column_density_section} we assumed a gas-to-dust ratio of 100, but several authors have shown that there is a gradient of decreasing gas-to-dust ratio with decreasing galactocentric radius \citep{schlegal_1998, watson_2011}, a trend which has also been observed in other star-forming galaxies \citep{sandstrom_2013}. Assuming that the gas-to-dust ratio is inversely proportional to the metallicity, the gas-to-dust ratio within the central kpc of the Galaxy would be $\sim$\,50 (e.g. \citealp{sodroski_1995}). In light of this, we estimate the uncertainty on the gas-to-dust ratio to be $\sim$\,50\,per cent. Following \citet{sanhueza_2017}, we assume an uncertainty of 30\,per cent dust opacity, and following \citet{lu_2019} we assume an uncertainty in the distance of $\pm$\,100\,pc (1.2\,per cent). These uncertainties in the dust opacity, the gas-to-dust ratio, dust emission fluxes, and the distance propagate to give an uncertainty of $\sim$\,60 per cent in masses. We estimate that the measured angular sizes have uncertainties of 20\,per cent as a result of the by-eye identification of the cores, whilst the fitting errors of the line widths are of the order 5\,per cent. These propagate to give an uncertainty of $\sim$\,65 per cent on the virial parameter. These uncertainties are represented as error bars in Figure\,\ref{mass_radius_plot}. These do not, however, include the additional sources of uncertainty discussed below, which are typically larger in magnitude yet more difficult to estimate.

\subsubsection{Additional uncertainties}

Firstly, there are several sources of uncertainty stemming from our assumptions of the physical properties of the cloud. The most significant of these is the use of the {\it Herschel} derived dust temperatures to calculate the mass (and column density) across both of the mapped regions (see Figure\,\ref{column_density}). Whilst these temperature maps are accurate over the large spatial scales probed by the {\it Herschel} observations ($\sim$\,30\arcsec), it is very likely that temperature deviations are present on the smaller scales probed by the ALMA observations presented here ($\sim$\,1\arcsec). It is not possible to quantify how these deviations will affect our results, but it is worth noting that the largest expected variations have been accounted for by using the SMA formaldehyde derived temperature measurements (i.e. for cores d2$_{\rm a}$ and e1$_{\rm a}$; see \citealp{walker_2018}).

There are also several uncertainties produced in the virial parameter by our assumed physical source properties. We currently do not know the strength of the magnetic field within the clouds investigated here, which could provide support against gravitational collapse (e.g. \citealp{girichidis_2018}), and thereby increase the calculated virial parameters (see \citealp{kauffmann_2013}). Indeed, magnetic fields over an order of magnitude higher than typically observed within disc molecular clouds have been recently found within the Brick molecular cloud ($\sim$~5000$\,\upmu$G; \citealp{pillai_2015}), although it is not clear if these are sufficient to affect the star formation given the high densities and turbulent conditions within the Galactic Centre \citep{kruijssen_2014a, federrath_2016, barnes_2017}. Shear or tidal forces would also provide support against gravitational collapse, yet have not been accounted for in the virial analysis presented here. These are thought to play a more significant role within the dust-ridge clouds than is typically seen within Galactic disc clouds, due to their association with an orbital stream within the Galactic Centre that can impart strong shear forces (e.g. \citealp{kruijssen_2014a, sormani_2018, kruijssen_2019}). Another mechanism that we have not accounted for in the virial analysis is the effect of external pressure, which, again, is elevated within the Galactic Centre; two to three orders of magnitude greater than typically found in the Galactic disc (see \citealp{walker_2018}). This external pressure would have a similar effect to magnetic fields and shear forces, and cause the regions to have to appear super-virial, because it would equilibrate with the external force. So the kinetic energy density would be higher than its potential energy. It is difficult to assess on what scales these two physical effects would predominantly act, and how that would change the results and interpretations presented in this work.   
 
Additional uncertainties arise from observational and data processing limitations. Firstly, despite being some of the highest spatial resolution ($\sim$\,0.04\,pc), largest ($>>$\,1\,pc), and highest sensitivity observations of entire molecular clouds within the Galactic Centre, the data presented here still have limitations. It is likely that the identified cores fragment further on spatial scales much smaller than the beam size, and that narrower linewidths may be found with better spectral resolution (see \citealp{kauffmann_2017a}). Secondly, in the virial analysis we have used the combined single-dish and ALMA image to calculate the masses, yet the ALMA only image for the \cCtHt\ molecular line. Using the ALMA only continuum image would lower the mass estimates on the individual core scales of $\sim$\,0.05\,pc by $\sim$\,10\,per cent, and on the large core e region scale of  $\sim$\,0.5\,pc by $\sim$\,50\,per cent. Thirdly, we have not conducted a background subtraction when determining the core masses, hence attributing all the mass along the line of sight through the cloud to the cores. We have investigated this effect by subtracting different continuum contour levels before determining the mass, and we find that this does not significantly affect the results of this work.

Finally, an additional caveat arises from our assumption that the clouds studied here will go on to form YMCs. This is a justifiable assumption, given that the two Galactic Centre clouds studied here harbour the ideal initial conditions for YMC formation (i.e. gas masses of $\sim\,10^{5}$\,\sol, radii $\,\sim1$\,pc, $\alpha_{\rm vir} < 1$, and, crucially, little-to-no signs of ongoing star formation; \citealp{longmore_2014}). Future simulations including the effects of star formation on molecular clouds entering the Galactic Centre may be able to further constrain the likelihood that similar clouds to those observed here will form bound YMCs (e.g. based on \citealp{kruijssen_2015, sormani_2018, kruijssen_2019, dale_2019sub}).

\section{Conclusions}\label{conclusions}

We have investigated the mass and density distribution, along with the dynamical state of two molecular clouds within the Galactic Centre. These were chosen to be high-mass ($\sim$\,10$^{5}$\,\sol\ of gas), compact (radius of $\sim$\,1\,pc), and globally gravitationally bound, and hence represent good candidate precursors to the most massive clusters currently forming in the universe today -- called young massive clusters ($\sim$\,10$^4$\,\sol\ of stars). Furthermore, these clouds are known to harbour only the earliest stages of star formation, a necessary condition to distinguish between the theories of cluster formation.

We present high-angular resolution ($\sim$\,1\arcsec), high-sensitivity continuum ($\sim$\,1\,mJy\,beam$^{-1}$) and molecular line ($\sim$\,0.1\,K) ALMA band 6 observations of these clouds. We use the continuum observations to identify the compact, high-density core regions within the clouds, and derive their physical properties. The current mass surface density profiles of the clouds are one to two orders of magnitude below comparable actively star-forming molecular clouds and young massive clusters present within the Galactic Centre. Furthermore, we find evidence for a higher threshold density for star formation within the Galactic Centre of $\sim$\,10$^{6-7}$\,\cmcb, orders of magnitude higher than is estimated for molecular clouds within the Galactic disc. This result has important implications for how we understand and characterise star and cluster formation within other extreme environments, such as within starburst and high redshift galaxies.

In agreement with previous molecular line observations of the Galactic Centre, we find that hydrogen cyanide, HCN~($3-2$), and formylium, HCO$^{+}$~($3-2$), molecular transitions are extended across the clouds, and are spatially uncorrelated with these dust continuum cores (e.g. \citealp{rathborne_2015}). However, uniquely, we do find a suite of optically thin molecular lines, such as methanimine, CH$_2$NH~($4-3$), and cyclopropenylidene, \cCtHt~($7-6$), that selectively trace the quiescent and actively star-forming dense gas within the Galactic Centre. 

The \cCtHt line is used to conduct a virial analysis of the identified core regions. We find that half (5/10) are within the limits of gravitational collapse for a non-homogeneous and non-spherical parcel of gas ($\alpha_{\rm vir}\leq2$). We also investigate the ``proto-cluster'' dynamics by using the ensemble of cores within each cloud. To do so, we use the standard deviation of the relative centroid velocities of cores as a proxy for the velocity dispersion, and calculate the virial parameter. We find that both clouds contain proto-clusters that are sub-virial, and, therefore, if not additionally supported, would also gravitationally collapse ($\alpha_{\rm vir}\leq2$).

Given that we know that star formation has very recently begun within these clouds, these results favour a conveyor-belt scenario for cluster formation. In this scenario, the molecular cloud has an initial density distribution lower than the stellar distribution of the final YMC, and star formation can occur throughout the cloud following its hierarchical density distribution. As the system evolves, both the gas and the embedded protostellar population concurrently globally collapse, until all the gas has formed stars or been expelled, and the final YMC density is reached.

\section*{ACKNOWLEDGEMENTS}

We would like to thank the anonymous referee for their constructive feedback which helped improve the paper. ATB would like to acknowledge the funding provided by Liverpool John Moores University, the Max Planck Institute for Extraterrestrial Physics and funding from the European Union's Horizon 2020 research and innovation programme (grant agreement No 726384). HB acknowledges support from the European Research Council under the Horizon 2020 Framework Program via the ERC Consolidator Grant CSF-648505. JK has received funding from the European Union's Horizon 2020 research and innovation programme under grant agreement No 639459 (PROMISE). JMDK gratefully acknowledges funding from the German Research Foundation (DFG) in the form of an Emmy Noether Research Group (grant number KR4801/1-1) and from the European Research Council (ERC) under the European Union's Horizon 2020 research and innovation programme via the ERC Starting Grant MUSTANG (grant agreement number 714907). JJ would like to acknowledge that this research was conducted in part at the SOFIA Science Center, which is operated by the Universities Space Research Association under contract NNA17BF53C with the National Aeronautics and Space Administration. This paper makes use of the following ALMA data: ADS/JAO.ALMA\#2013.1.00617.S. ALMA is a partnership of ESO (representing its member states), NSF (USA) and NINS (Japan), together with NRC (Canada), MOST and ASIAA (Taiwan), and KASI (Republic of Korea), in cooperation with the Republic of Chile. The Joint ALMA Observatory is operated by ESO, AUI/NRAO and NAOJ.
\bibliographystyle{./references/mnras}
\bibliography{./references/references}
\label{lastpage}

\appendix
\section{Online supplementary material}\label{appendixA}

Table\,\ref{line_idents1} shows the molecular line transitions detected within both clouds. For referencing the large number of molecular transitions, they are grouped according to column density above which emission is observed. These are categorised as follows: 
1) ``extended'': for molecules that are seen above a column density of $\sim1\times10^{22}$\,cm$^{-2}$\,cm$^{-2}$, such that they are seen across the entire mapped region of $2-8$\,pc; 
2) ``moderately extended'': for molecules that are seen above a column density of $\sim2\times10^{23}$\,cm$^{-2}$, such that they are extended across a $1-2$\,pc region; 
3) ``compact'': for molecules that are seen above a column density of $\sim6\times10^{23}$\,cm$^{-2}$\,cm$^{-2}$, such that they are trace only <$1$\,pc regions. 
These categories for all molecular line transitions within both clouds are shown in the table from the online appendix.

\begin{landscape}
\begin{figure}
\vspace{3cm}
\centering
\includegraphics[trim = 0.5cm 0mm 0.6cm 0mm, clip,angle=0,width = 1.34\textwidth]{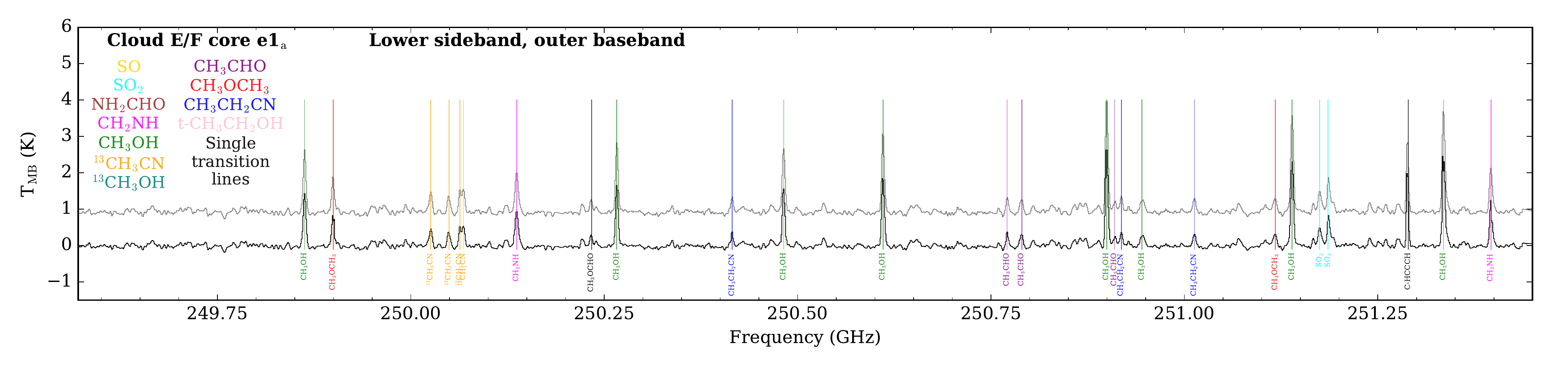}
\includegraphics[trim = 0.47cm 0mm 0.65cm 0mm, clip,angle=0,width = 1.34\textwidth]{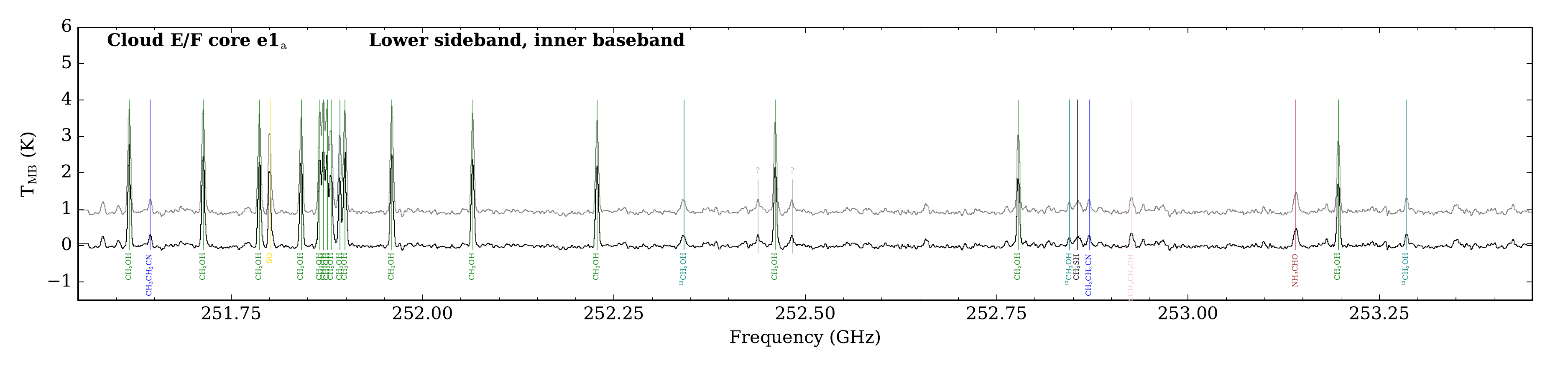}
\caption{The lower sideband, outer [upper panel] and inner [lower panel] basebands, spectra towards the Cloud E/F core e1$_{\rm a}$ region (see Figure\,\ref{core_map_ef}), both with and without the continuum level subtracted [lower and upper line, respectively]. Labeled are the identified molecular transitions, which have been coloured according to their molecular species (see legend in upper left of upper panel). The question marks show unidentified species [shown in grey].}
\label{spectra_core_1_l}
\vfill
\end{figure}
\end{landscape}

\begin{landscape}
\begin{figure}
\vspace{3cm}
\centering
\includegraphics[trim = 0.5cm 0mm 0.6cm 0mm, clip,angle=0,width = 1.34\textwidth]{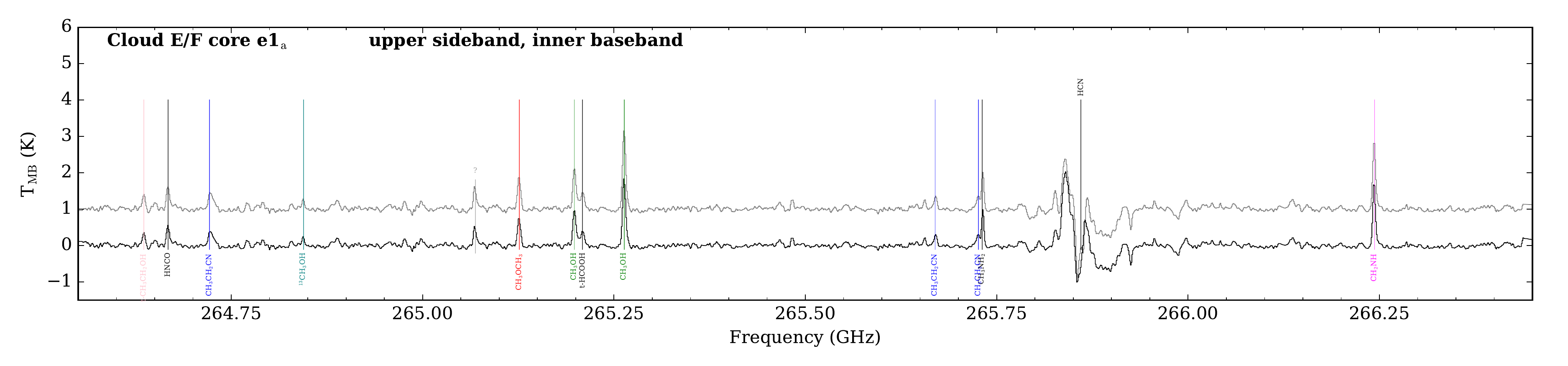}
\includegraphics[trim = 0.47cm 0mm 0.65cm 0mm, clip,angle=0,width = 1.34\textwidth]{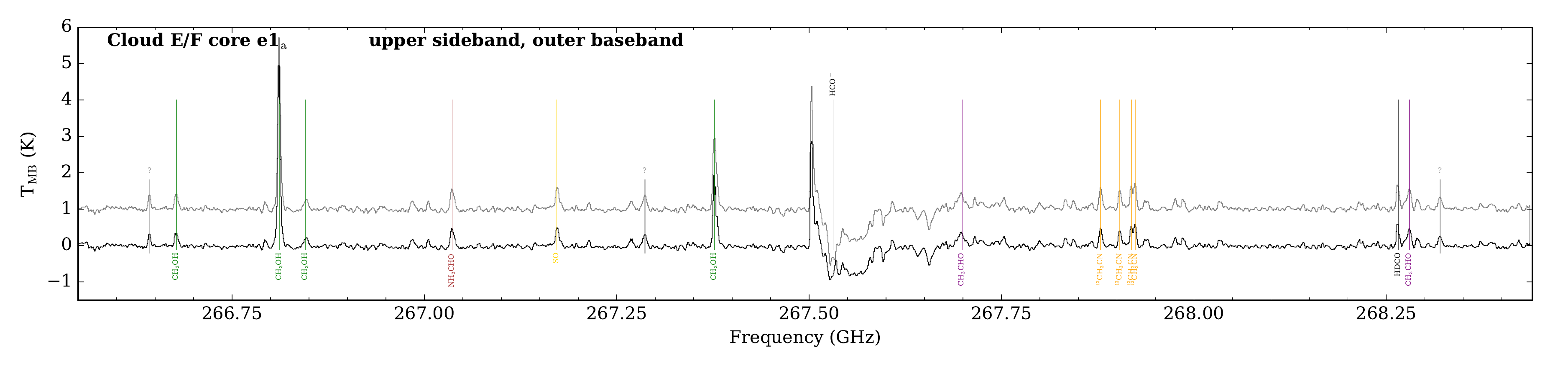}
\caption{The upper sideband, inner [upper panel] and outer [lower panel] basebands, spectra towards the Cloud E/F core e1$_{\rm a}$ region (see Figure\,\ref{core_map_ef}), both with and without the continuum level subtracted [lower and upper line, respectively]. Labeled are the identified molecular transitions, which have been coloured according to their molecular species (see Figure\,\ref{spectra_core_1_l}). The question marks show unidentified species [shown in grey].}
\label{spectra_core_1_u}
\vfill
\end{figure}

\end{landscape}

\begin{table*}
\caption{Table of the molecules identified in Cloud D and E/F . Also show are those identified within the Brick using a similar spectral set-up with ALMA (Contreras et al. in prep). Columns show the molecule name, the transition, the frequency, the upper energy level of the transition, the extent of the emission within each of the clouds, which have been abbreviated (`c' = core, `me' = moderately extended, and `e' = extended), and additional notes where the symbol key is given below the table. Shown at the end of the table are the rest frequencies of the unidentified line transitions, which have been adjusted for the approximate core velocity of $\sim$\,29\,\kms.}
\centering
\small
\begin{tabular}{p{2.2cm} p{4.8cm} L{1.3cm} L{1cm} C{1cm} C{1.2cm} C{1 cm} C{1 cm}}
\hline
Molecule & Transition & Rest &  Upper  & Cloud D & Cloud E/F & Brick & notes \\
 &  & frequency &  energy  & & & & \\
&  &  (GHz) &  (K) & &  & \\ \hline

SO $^3$$\Sigma$ v=0 & 5(6) $-$ 4(5) & 251.83 & 51 & - & me & me &  \\
SO $^3$$\Sigma$ v=0 & 3(4) $-$ 4(3) & 267.20 & 29 & - & c & - &  \\
SO$_2$ v=0 & 13(1,13) $-$ 12(0,12) & 251.20 & 82 & - & c & - &  \\
SO$_2$ v=0 & 8(3,5) $-$ 8(2,6) & 251.21 & 55 & - & c & - &  \\
HCN v=0 & J = 3 $-$ 2, F = 3 $-$ 2 & 265.89 & 26 & e & e & e &  \\
HCO$^+$ & (3 $-$ 2) & 267.56 & 26 & e & e & e &  \\
HDCO & 4(1, 3) $-$ 3(1, 2) & 268.29 & 40 & - & c & - & * \\
HNCO v=0 & 12(1,11) $-$ 11(1,10) & 264.69 & 126 & - & c & - &  \\
t-HCOOH & 3(2, 2) $-$ 3(0, 3) & 265.24 & 19 & - & c & - &  \\
c-C$_3$H$_2$ v=0 & 7(0,7) $-$ 6(1,6) & 251.31 & 51 & c & me & me & ** \\
NH$_2$CHO & 12(2,11) $-$ 11(2,10) & 253.17 & 91 & - & c & - &  \\
NH$_2$CHO & 13(0,13) $-$ 12(0,12) & 267.06 & 91 & - & c & - &  \\
CH$_2$NH & 7(1,6) $-$ 7(0,7) & 250.16 & 97 & - & c & - &  \\
CH$_2$NH & 6(0,6) $-$ 5(1,5) & 251.42 & 64 & - & c & - &  \\
CH$_2$NH & 4(1,3) $-$ 3(1,2) & 266.27 & 40 & c & me & me &  \\
CH$_3$NH$_2$ & 6(0)A1 $-$ 5(0)A2 & 265.76 & 45 & - & me & me &  \\
CH$_3$SH v=0 & 10(3) $-$ 9(3) E & 252.88 & 107 & - & c & - & * \\
CH$_3$OH vt=0 & 14(3,11) $-$ 14(2,12) & 249.89 & 293 & - & c & - &  \\
CH$_3$OH vt=0 & 13(3,10) $-$ 13(2,11) & 250.29 & 261 & - & c & - &  \\
CH$_3$OH vt=0 & 11(0,11) $-$ 10(1,10) & 250.51 & 153 & - & me & - &  \\
CH$_3$OH vt=0 & 12(3,9) $-$ 12(2,10) & 250.64 & 231 & - & c & - &  \\
CH$_3$OH vt=0 & 11(3,8) $-$ 11(2,9) & 250.92 & 203 & - & c & - &  \\
CH$_3$OH vt=0 & 10(3,7) $-$ 10(2,8) & 251.16 & 177 & - & c & - &  \\
CH$_3$OH vt=0 & 9(3,6) $-$ 9(2,7) & 251.36 & 154 & - & c & - &  \\
CH$_3$OH vt=0 & 7(3,4) $-$ 7(2,5) & 251.64 & 115 & - & c & - &  \\
CH$_3$OH vt=0 & 6(3,3) $-$ 6(2,4) & 251.74 & 99 & - & c & - &  \\
CH$_3$OH vt=0 & 5(3,2) $-$ 5(2,3) & 251.81 & 85 & - & me & - &  \\
CH$_3$OH vt=0 & 4(3,1) $-$ 4(2,2) & 251.87 & 73 & - & c & - &  \\
CH$_3$OH vt=0 & 5(3,3) $-$ 5(2,4) & 251.89 & 85 & - & c & - &  \\
CH$_3$OH vt=0 & 6(3,4) $-$ 6(2,5) & 251.90 & 99 & - & c & - &  \\
CH$_3$OH vt=0 & 4(3,2) $-$ 4(2,3) & 251.90 & 73 & - & c & - &  \\
CH$_3$OH vt=0 & 3(3,0) $-$ 3(2,1) & 251.91 & 64 & - & c & - &  \\
CH$_3$OH vt=0 & 3(3,1) $-$ 3(2,2) & 251.92 & 64 & - & c & - &  \\
CH$_3$OH vt=0 & 7(3,5) $-$ 7(2,6) & 251.92 & 115 & - & c & - &  \\
CH$_3$OH vt=0 & 8(3,6) $-$ 8(2,7) & 251.98 & 133 & - & c & - &  \\
CH$_3$OH vt=0 & 9(3,7) $-$ 9(2,8) & 252.09 & 154 & - & c & - &  \\
CH$_3$OH vt=0 & 10(3,8) $-$ 10(2,9) & 252.25 & 177 & - & c & - &  \\
CH$_3$OH vt=0 & 11(3,9) $-$ 11(2,10) & 252.49 & 203 & - & c & - &  \\
CH$_3$OH vt=0 & 12(3,10) $-$ 12(2,11) & 252.80 & 231 & - & c & - &  \\
CH$_3$OH vt=0 & 13(3,11) $-$ 13(2,12) & 253.22 & 261 & - & c & - &  \\
CH$_3$OH vt=0 & 6(1,5) $-$ 5(2,3) & 265.29 & 70 & - & c & me &  \\
CH$_3$OH vt=0 & 23(3,21) $-$ 23(2,22) & 266.70 & 690 & - & c & - &  \\
CH$_3$OH vt=0 & 5(2,3) $-$ 4(1,3) & 266.84 & 57 & - & me & me &  \\
CH$_3$OH vt=0 & 9(0,9) $-$ 8(1,7) & 267.40 & 117 & - & c & me & * \\
CH$_3$OH vt=1 & 17(3) $-$ 18(4), E1 & 250.97 & 771 & - & c & - &  \\
CH$_3$OH vt=1 & 5(1,4) $-$ 6(2,5) & 265.22 & 360 & - & c & - &  \\
CH$_3$OH vt=1 & 14(6,8) $-$ 15(5,11) & 266.87 & 711 & - & c & - &  \\
$^{13}$CH$_3$CN & 14(3) $-$ 13(3), F = 13 -- 12 & 250.05 & 154 & - & c & - &  \\
$^{13}$CH$_3$CN & 14(2) $-$ 13(2), F = 13 -- 12 & 250.07 & 119 & - & c & - &  \\
$^{13}$CH$_3$CN & 14(1) $-$ 13(1), F= 14 $-$ 13 & 250.09 & 97 & - & c & - &  \\
$^{13}$CH$_3$CN & 14(0) $-$ 13(0), F = 13 -- 12 & 250.09 & 90 & - & c & - &  \\
$^{13}$CH$_3$CN & 15(3) $-$ 14(3), F = 14 $-$ 15 & 267.91 & 167 & - & c & - &  \\
$^{13}$CH$_3$CN & 15(2) $-$ 14(2), F = 15 $-$ 14 & 267.93 & 132 & - & c & - &  \\
$^{13}$CH$_3$CN & 15(1) $-$ 14(1) & 267.95 & 110 & - & c & - &  \\

\hline
\end{tabular}

\begin{minipage}{\textwidth}
\smallskip
*: A confused line, and, therefore, the chosen transition should be taken with caution.\\
**: Confusion with the  c-HCCCH v=0 7(1,7) $-$ 6(0,6) transition, which has an identical rest frequency.'\\
\end{minipage}

\label{line_idents1}
\end{table*}

\begin{table*}
 \contcaption{}
\centering
\small
\begin{tabular}{p{2.2cm} p{4.8cm} L{1.3cm} L{1cm} C{1cm} C{1.2cm} C{1 cm} C{1 cm}}
\hline
Molecule & Transition & Rest &  Upper  & Cloud D & Cloud E/F & Brick & notes \\
 &  & frequency &  energy  & & & & \\
&  &  (GHz) &  (K) & &  & \\ \hline

$^{13}$CH$_3$CN & 15(0) $-$ 14(0), F = 16 $-$ 15 & 267.95 & 103 & - & c & - &  \\
$^{13}$CH$_3$OH vt=0 & 15(4,11) $-$ 16(3,13) & 252.37 & 368 & - & c & - & * \\
$^{13}$CH$_3$OH vt=0 & 15(3,12) $-$ 15(2,13) & 252.87 & 322 & - & c & - &  \\
$^{13}$CH$_3$OH vt=0 & 14(3,11) $-$ 14(2,12) & 253.31 & 288 & - & c & - &  \\
$^{13}$CH$_3$OH vt=0 & 4(-2,3) $-$ 3(2,1) & 264.87 & 49 & - & c & - & * \\
CH$_3$CHO vt=0 & 13(4, 10) $-$ 12(4, 9), E & 250.80 & 75 & - & me & me & * \\
CH$_3$CHO vt=0 & 13(4, 9) $-$ 12(4, 8), E & 250.81 & 75 & - & me & - & * \\
CH$_3$CHO vt=0 & 13(3,11) $-$ 12(3,10) A & 250.93 & 105 & - & c & - &  \\
CH$_3$CHO vt=0 & 15(0,15) $-$ 14(1,14) A & 267.73 & 109 & - & c & - &  \\
CH$_3$CHO vt=0 & 14(2, 13) $-$ 13(2, 12) E & 268.31 & 106 & - & c & - &  \\
CH$_3$OCH$_3$ & 15(1,14) $-$ 14(2,13) EE & 249.92 & 113 & - & c & - &  \\
CH$_3$OCH$_3$ & 21(5,16) $-$ 21(4,17) EE & 251.14 & 246 & - & c & - &  \\
CH$_3$OCH$_3$ & 11(2,10) $-$ 10(1,9) EE & 265.15 & 65 & - & c & - &  \\
CH$_3$OCHO v=0 & 20 (3,17) $-$ 19 (3,16) A & 250.26 & 134 & - & c & - &  \\
CH$_3$CH$_2$CN v=0 & 30(0,31) $-$ 30(0,30) & 265.75 & 206 & - & c & - &  \\
CH$_3$CH$_2$CN v=0 & 28( 3,26) $-$ 27( 3,25) & 250.44 & 120 & - & me & me & * \\
CH$_3$CH$_2$CN v=0 & 28(8,21) $-$ 27(8,20) & 250.94 & 246 & - & c & - &  \\
CH$_3$CH$_2$CN v=0 & 28(7,22) $-$ 27(7,21) & 251.04 & 229 & - & c & - &  \\
CH$_3$CH$_2$CN v=0 & 28(4, 25) $-$ 27(4,24) & 251.67 & 193 & - & c & - &  \\
CH$_3$CH$_2$CN v=0 & 28(4,24) $-$ 27(4,23) & 252.90 & 193 & - & c & - &  \\
CH$_3$CH$_2$CN v=0 & 30 (1,29) $-$ 29(1,28) & 264.75 & 202 & - & c & - &  \\
CH$_3$CH$_2$CN v=0 & 31(1,31) $-$ 30(1,30) & 265.70 & 206 & - & c & - &  \\
t-CH$_3$CH$_2$OH & 4(4, 1) $-$ 3(3, 0) & 252.95 & 28 & - & c & me & * \\
t-CH$_3$CH$_2$OH & 16(0,16) $-$ 15(1,15) & 264.66 & 110 & - & c & - & * \\

? & ? & 252.44 & \dots & - & c & -\\
? & ? & 252.51 & \dots & - & c & -\\
? & ? & 265.09  & \dots & - & c & -\\
? & ? & 266.67 & \dots & - & c & -\\
? & ? & 267.31 & \dots & - & c & -\\
? & ? & 268.35 & \dots & - & c & -\\

\hline
\end{tabular}
\label{line_idents2}
\end{table*}

\begin{figure*}
\centering

\includegraphics[trim =  7mm 12mm 5mm 0mm, clip,angle=0,width=0.97\textwidth]{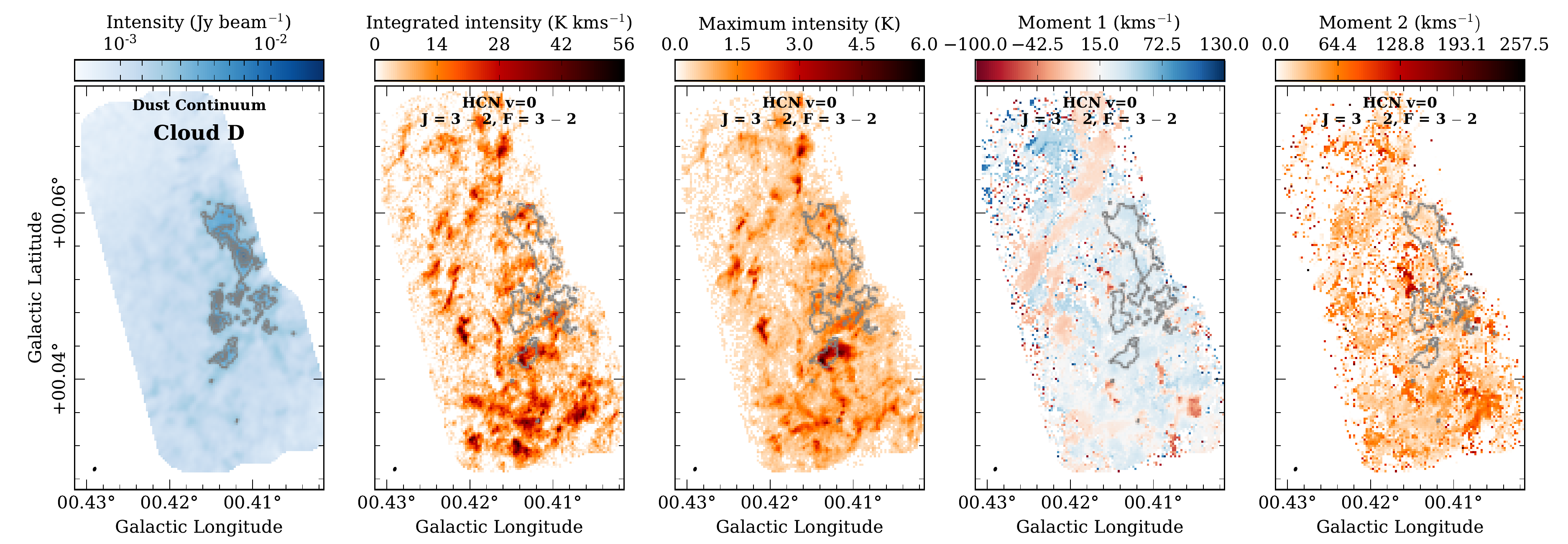}
\includegraphics[trim =  7mm 12mm 5mm 8mm, clip,angle=0,width=0.97\textwidth]{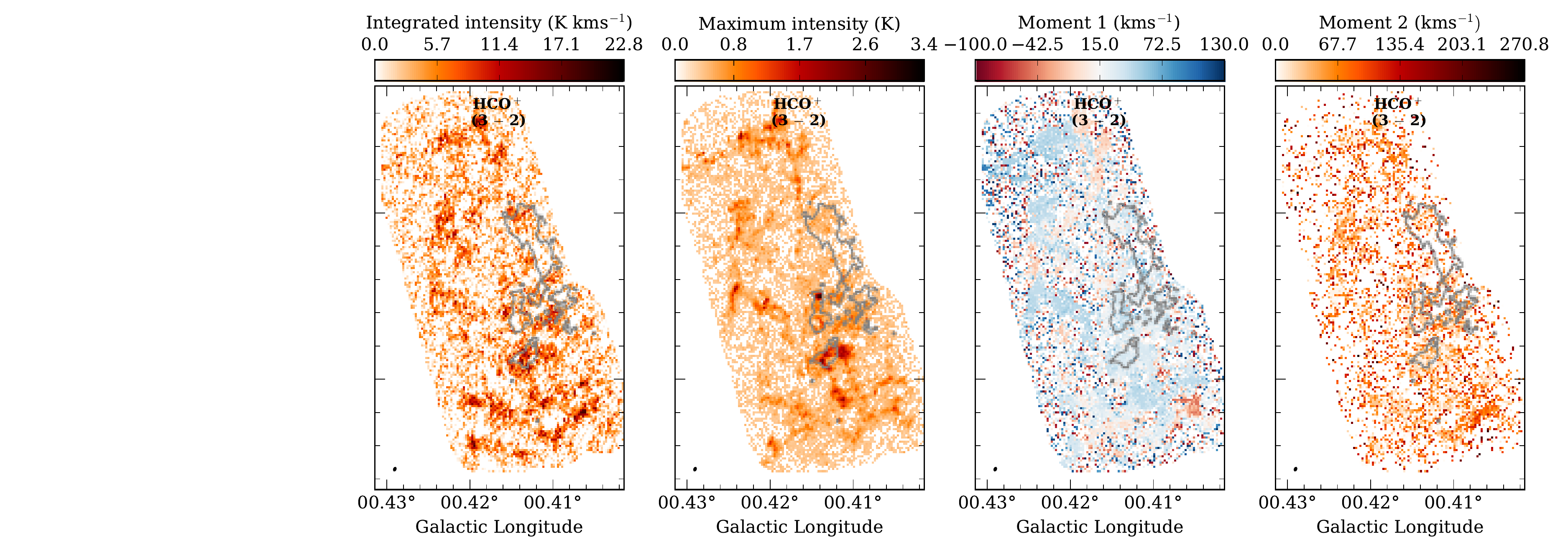}
\includegraphics[trim =  7mm 12mm 5mm 8mm, clip,angle=0,width=0.97\textwidth]{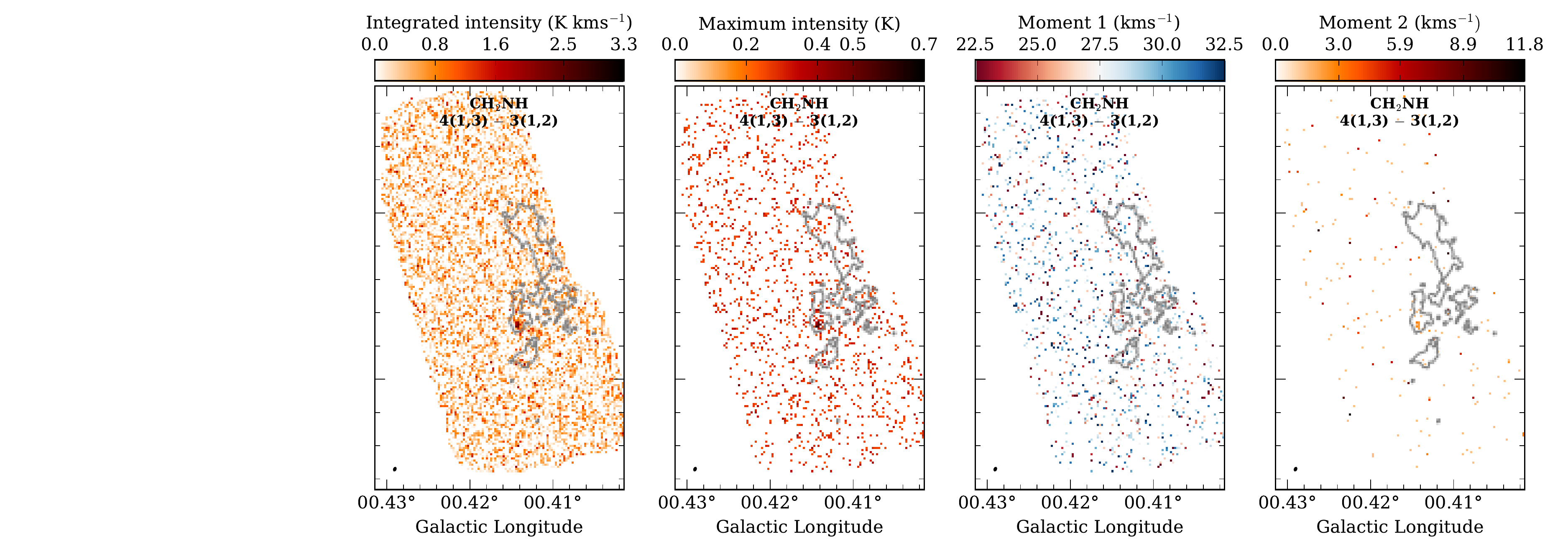}
\includegraphics[trim =  7mm 0mm 5mm 8mm, clip,angle=0,width=0.97\textwidth]{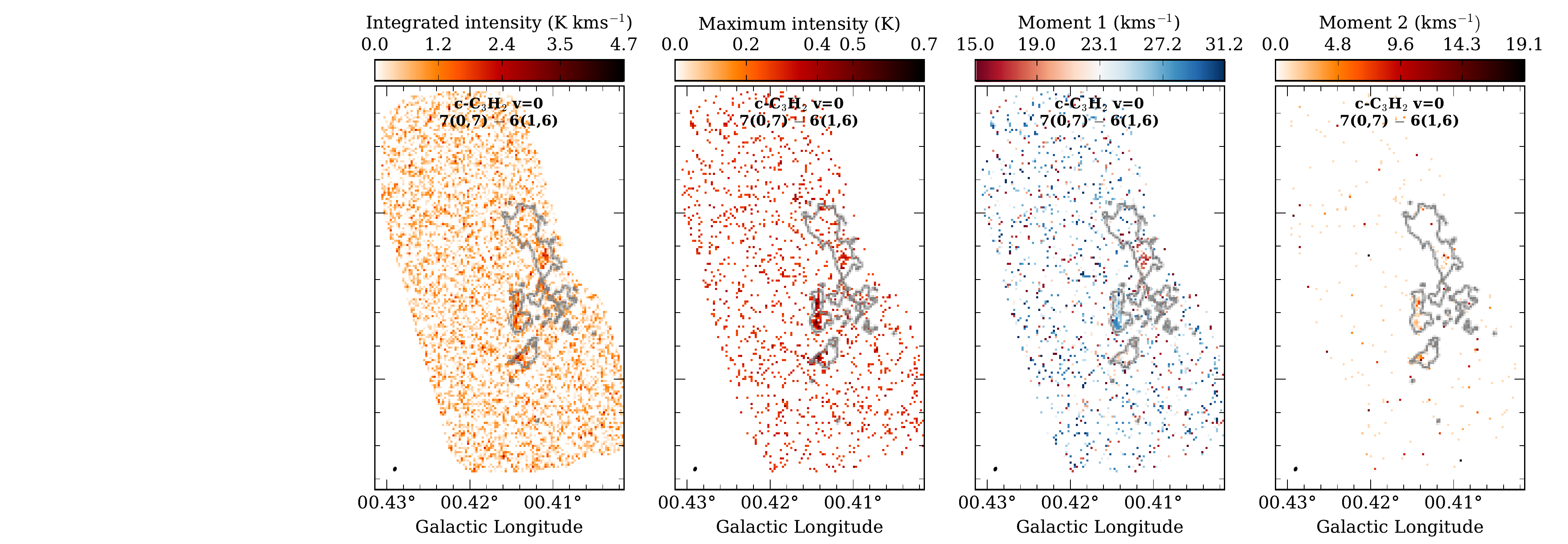}

\caption{Moment maps of the molecular transitions identified towards Cloud D (see table\,\ref{line_idents1}). The analysis for the different molecular transitions are presented in each row, with the molecule labeled at the top of each map. Shown in the upper left is the combined 12m, 7m and single dish continuum map, and then from left to right are moment maps of the integrated intensity, peak intensity, intensity weighted centroid velocity, and intensity weighted velocity dispersion for each molecule. Contours on the upper left panel are of the continuum shown in levels of [4, 6, 8, 15, 20, 30]\,$\sigma_{\rm rms}$, where $\sigma_{\rm rms}$\,$\sim$\,0.4\,mJy\,beam$^{-1}$. The lowest of these contours is repeated on each of the moment maps.}
\label{}
\end{figure*}

\begin{figure*}
\centering
\includegraphics[trim = 0cm 21mm 0cm 10mm, clip,angle=0,width=1\textwidth]{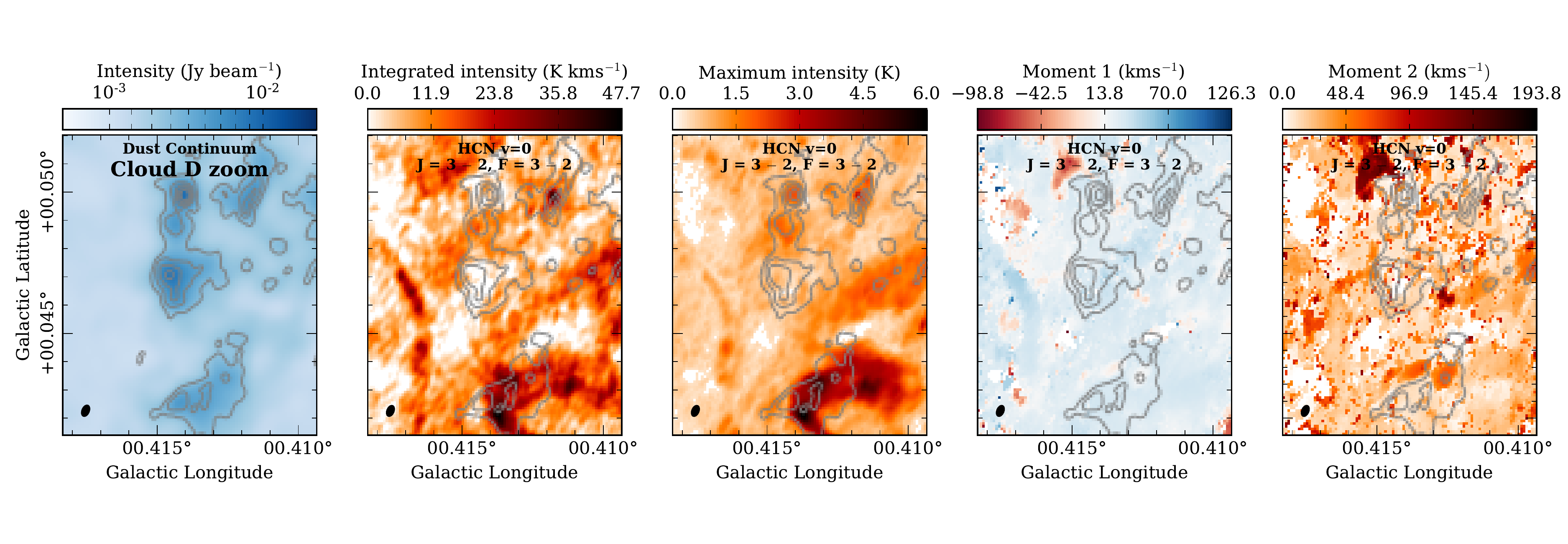}
\includegraphics[trim = 0cm 21mm 0cm 20mm, clip,angle=0,width=1\textwidth]{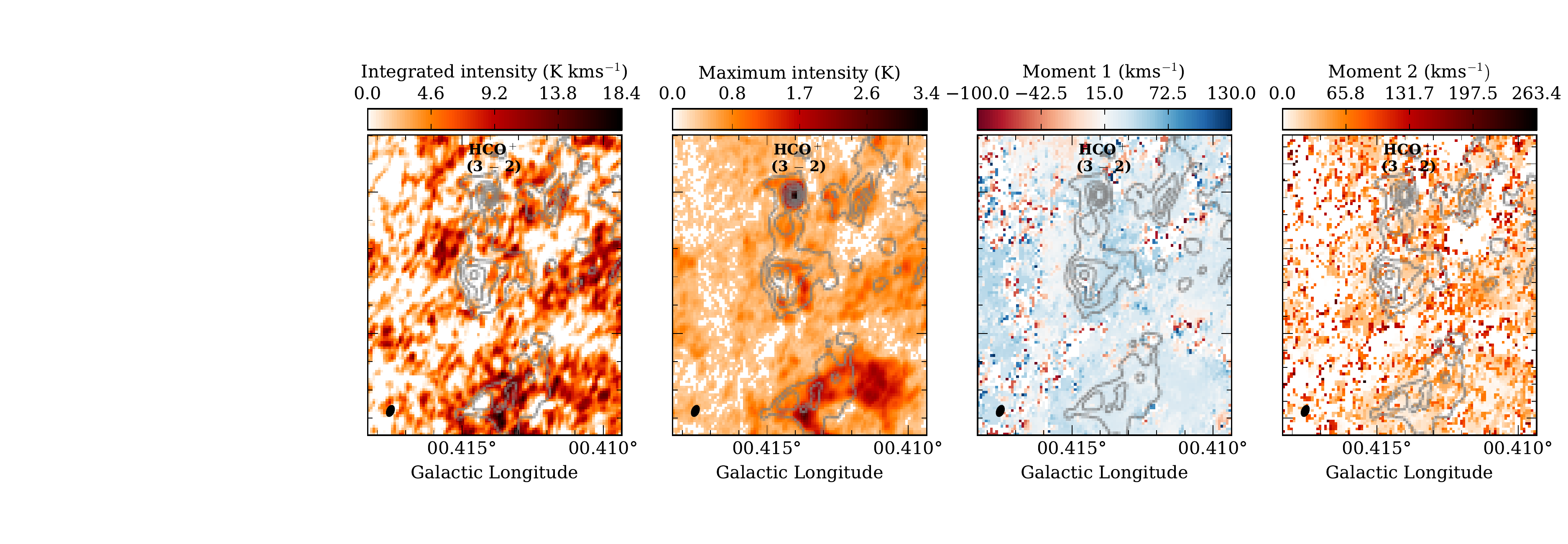}
\includegraphics[trim = 0cm 21mm 0cm 20mm, clip,angle=0,width=1\textwidth]{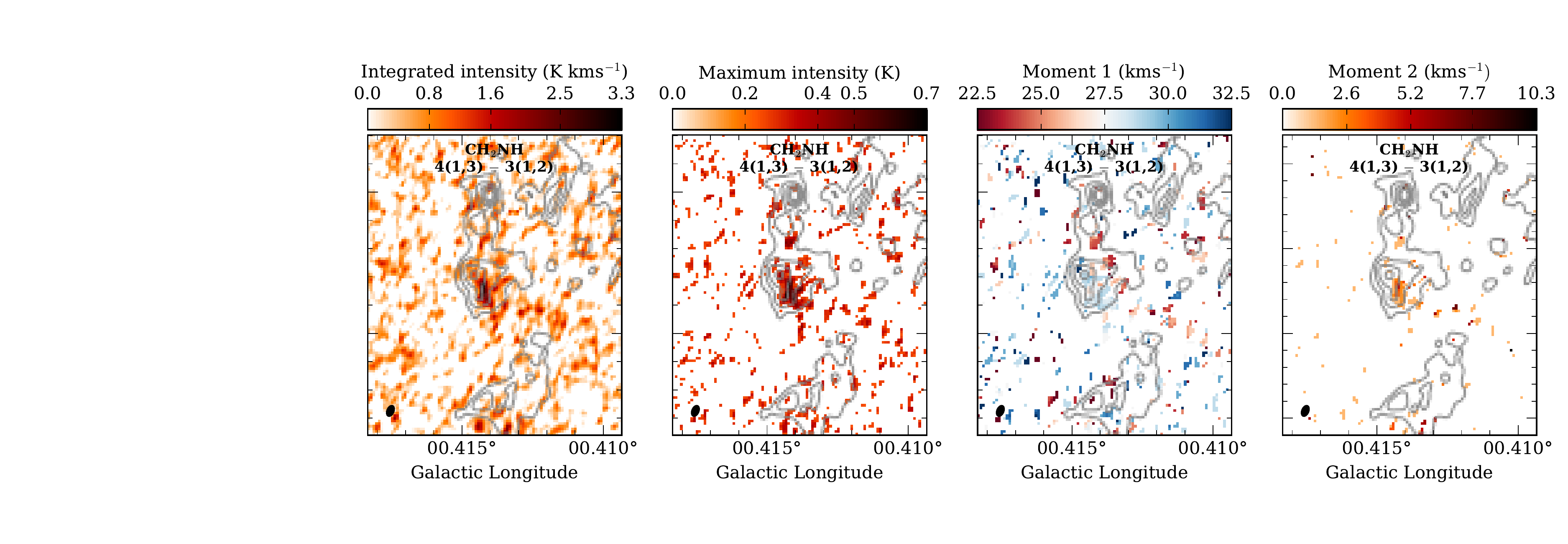}
\includegraphics[trim = 0cm 13mm 0cm 20mm, clip,angle=0,width=1\textwidth]{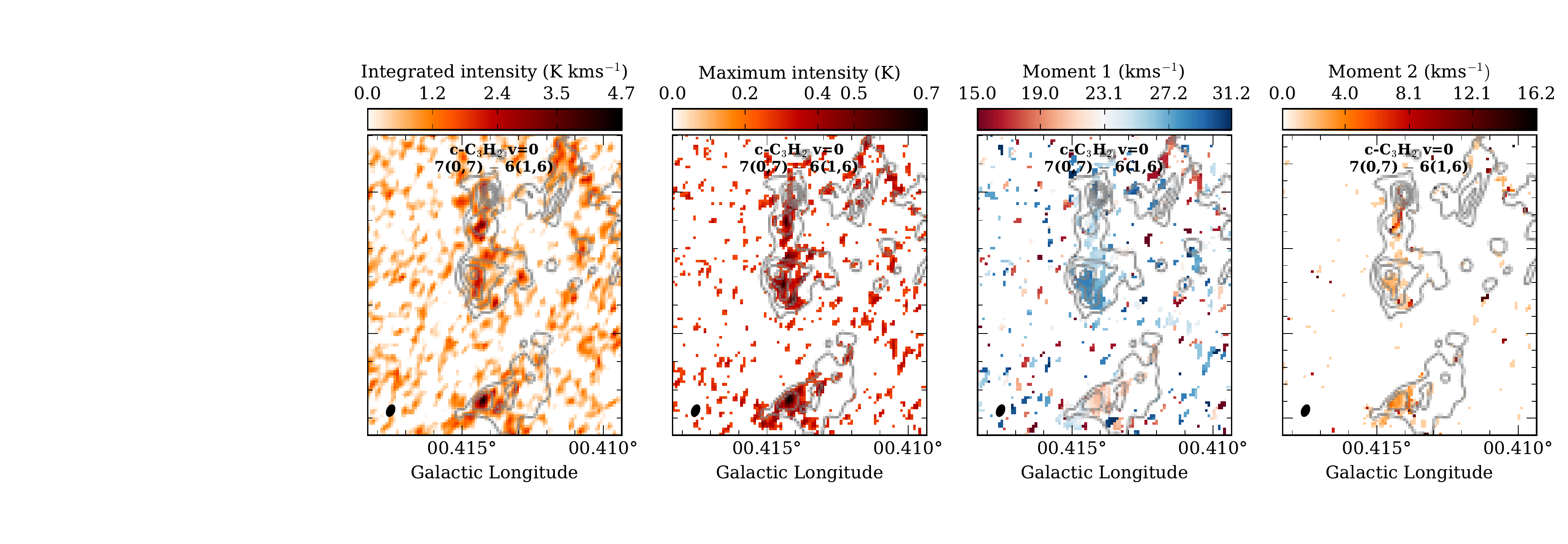}

\caption{Moment maps of the molecular transitions identified towards the zoom region Cloud D (see table\,\ref{line_idents1}). Show is a zoom-in of the region containing cores d2$_{\rm a}$, d4$_{\rm a}$, and d4$_{\rm b}$ (see section\,\ref{virial_cores}). The analysis for the different molecular transitions are presented in each row, with the molecule labeled at the top of each map. Shown in the upper left is the combined 12m, 7m and single dish continuum map, and then from left to right are moment maps of the integrated intensity, peak intensity, intensity weighted centroid velocity, and intensity weighted velocity dispersion for each molecule. Contours on each map are of the continuum shown in levels of [8, 15, 20, 30]\,$\sigma_{\rm rms}$, where $\sigma_{\rm rms}$\,$\sim$\,0.4\,mJy\,beam$^{-1}$.} 
\label{}
\end{figure*}

\begin{figure*}
\centering

\includegraphics[trim = 3mm 23mm 0cm 10mm, clip,angle=0,width=1\textwidth]{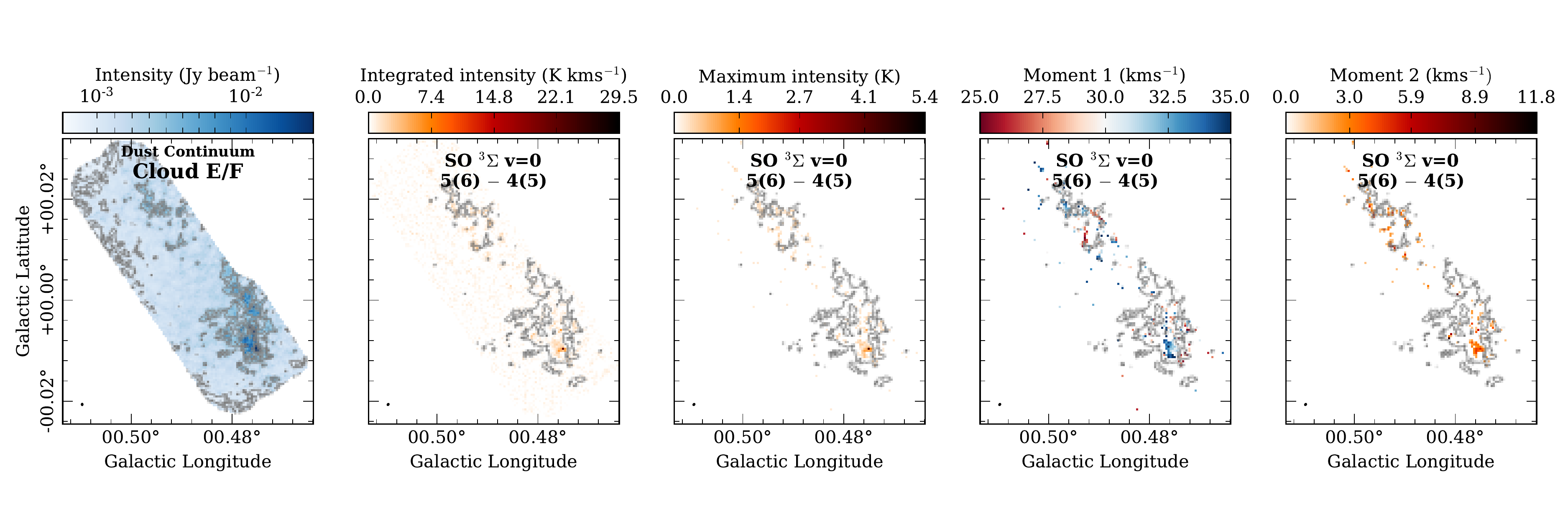} 
\includegraphics[trim = 3mm 23mm 0cm 21mm, clip,angle=0,width=1\textwidth]{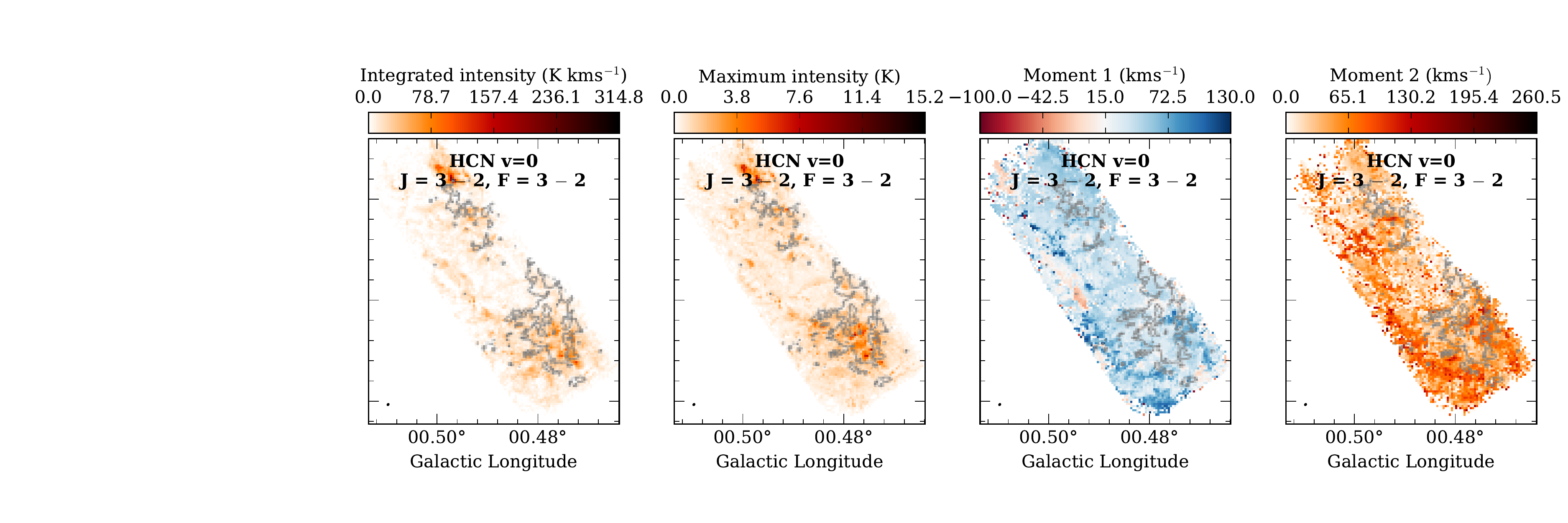} 
\includegraphics[trim = 3mm 23mm 0cm 21mm, clip,angle=0,width=1\textwidth]{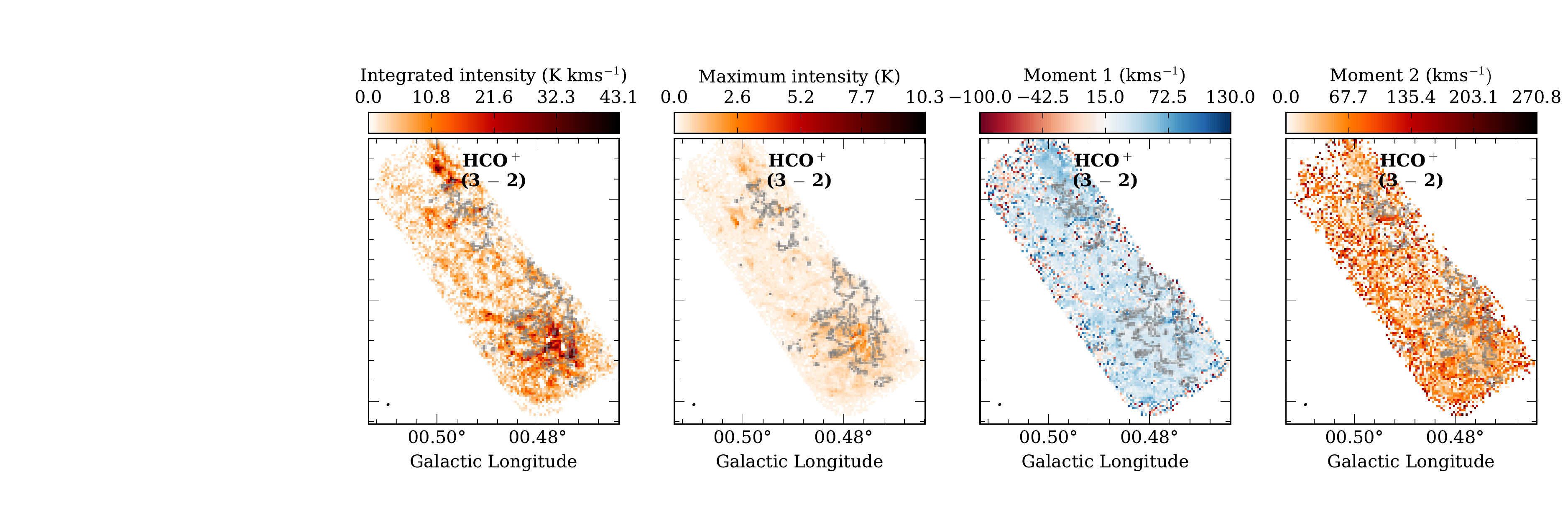} 
\includegraphics[trim = 3mm 23mm 0cm 21mm, clip,angle=0,width=1\textwidth]{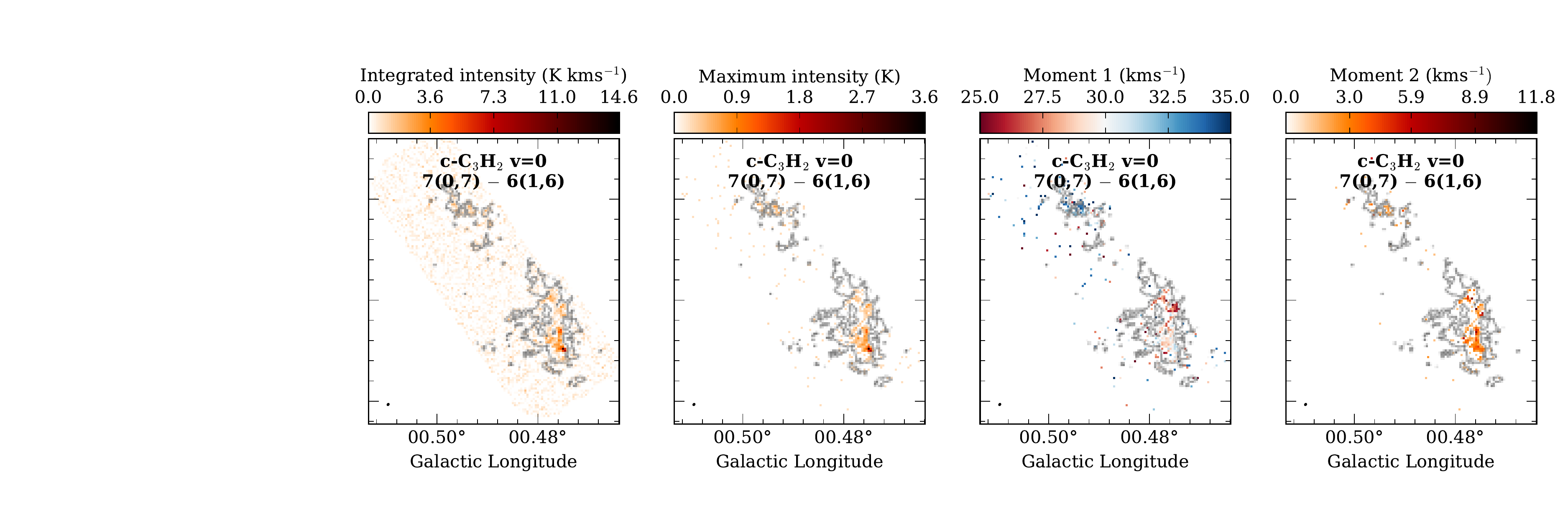} 
\includegraphics[trim = 3mm 13mm 0cm 21mm, clip,angle=0,width=1\textwidth]{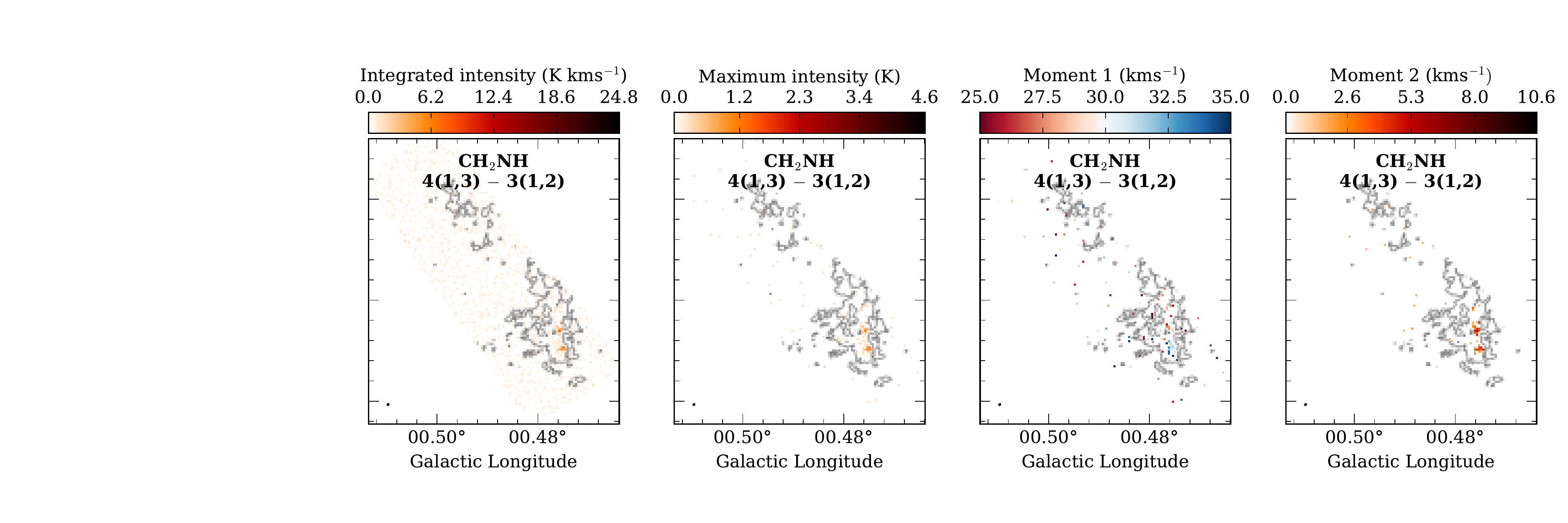} 

\caption{Moment maps of the molecular transitions towards Cloud E/F (see table\,\ref{line_idents1}). Shown here are the molecular transitions that have been classified as either extended or moderately extended (see Figure\,\ref{moments_core_e1}).The analysis for the different molecular transitions are presented in each row, with the molecule labeled at the top of each map. Shown in the upper left is the combined 12m, 7m and single dish continuum map, and then from left to right are moment maps of the integrated intensity, peak intensity, intensity weighted centroid velocity, and intensity weighted velocity dispersion for each molecule. Contours on upper left panel are of the continuum shown in levels of [8, 15, 30, 50]\,$\sigma_{\rm rms}$, where $\sigma_{\rm rms}$\,$\sim$\,0.6\,mJy\,beam$^{-1}$. The lowest of these contours is repeated on each of the moment maps.}

\label{ }
\end{figure*}

\begin{figure*}
\centering

\includegraphics[trim = 3mm 23mm 0cm 10mm, clip,angle=0,width=1\textwidth]{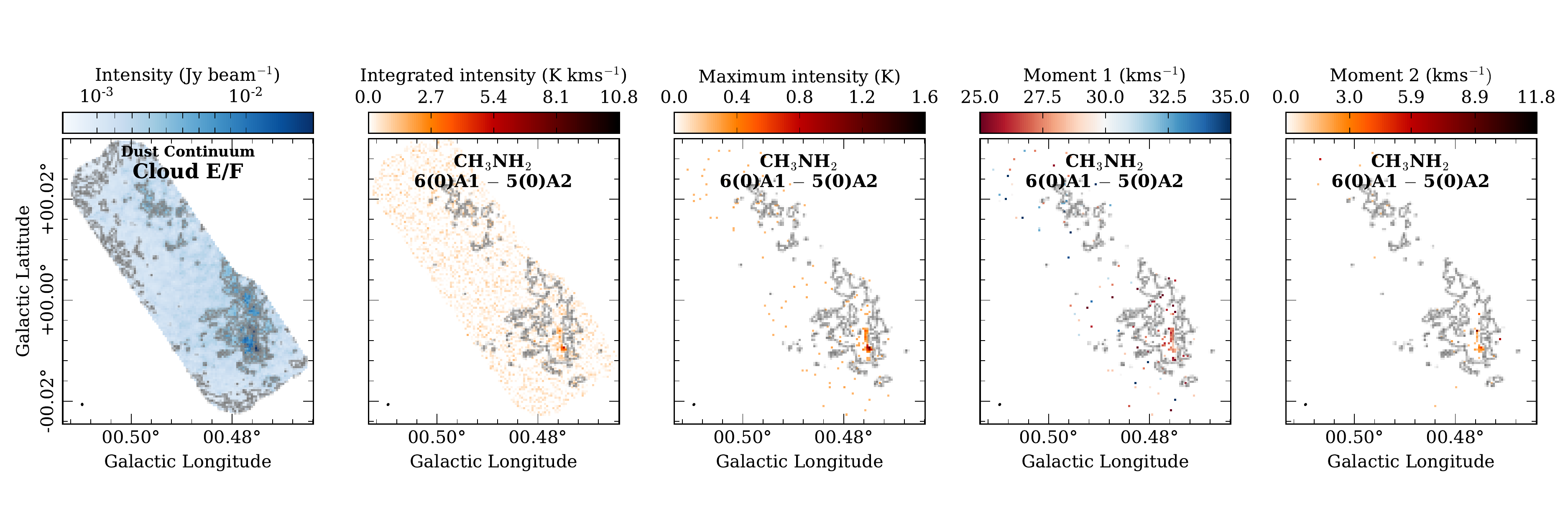} 
\includegraphics[trim = 3mm 23mm 0cm 21mm, clip,angle=0,width=1\textwidth]{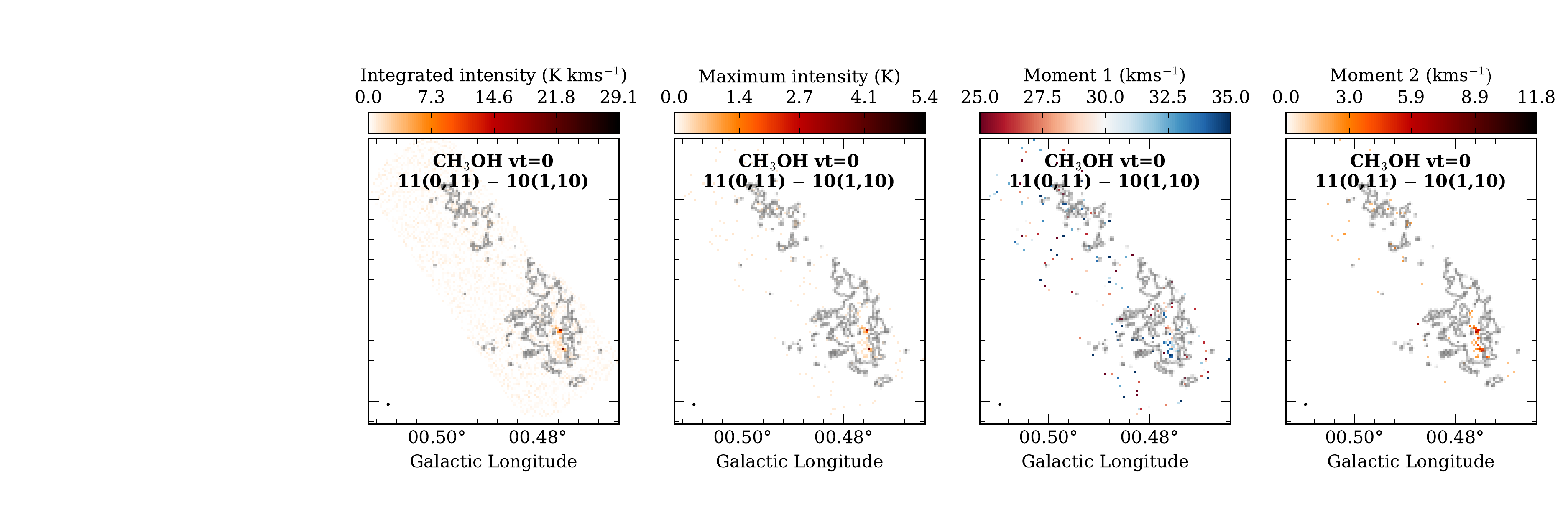} 
\includegraphics[trim = 3mm 23mm 0cm 21mm, clip,angle=0,width=1\textwidth]{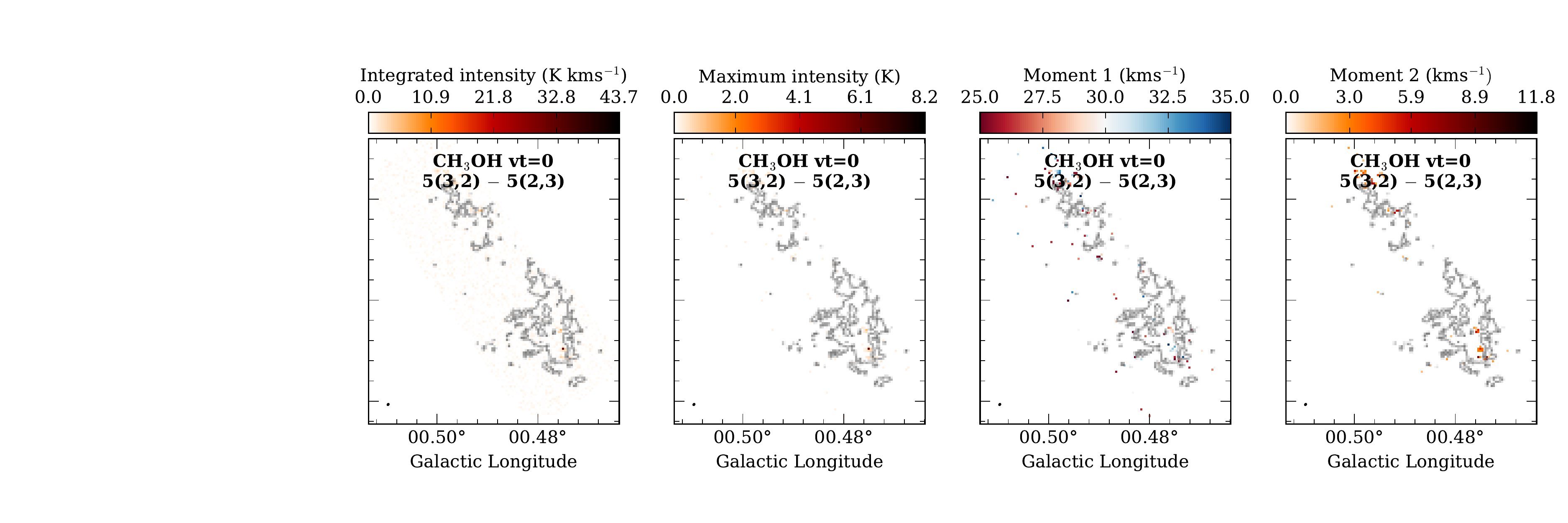} 
\includegraphics[trim = 3mm 23mm 0cm 21mm, clip,angle=0,width=1\textwidth]{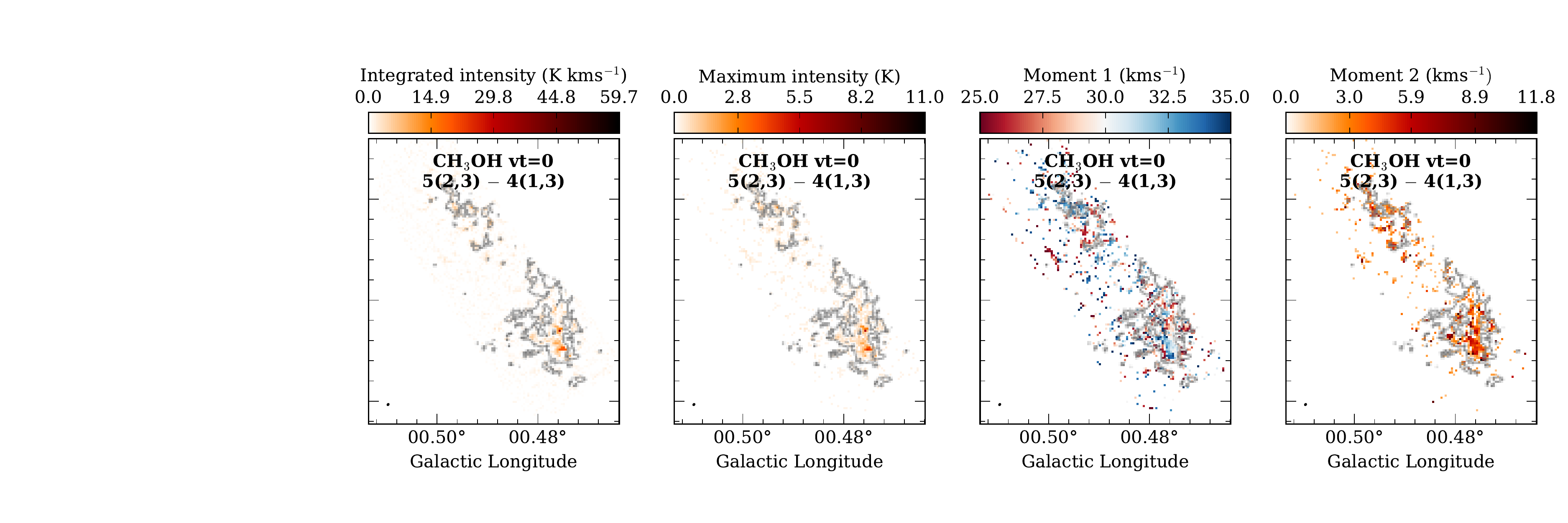} 
\includegraphics[trim = 3mm 13mm 0cm 21mm, clip,angle=0,width=1\textwidth]{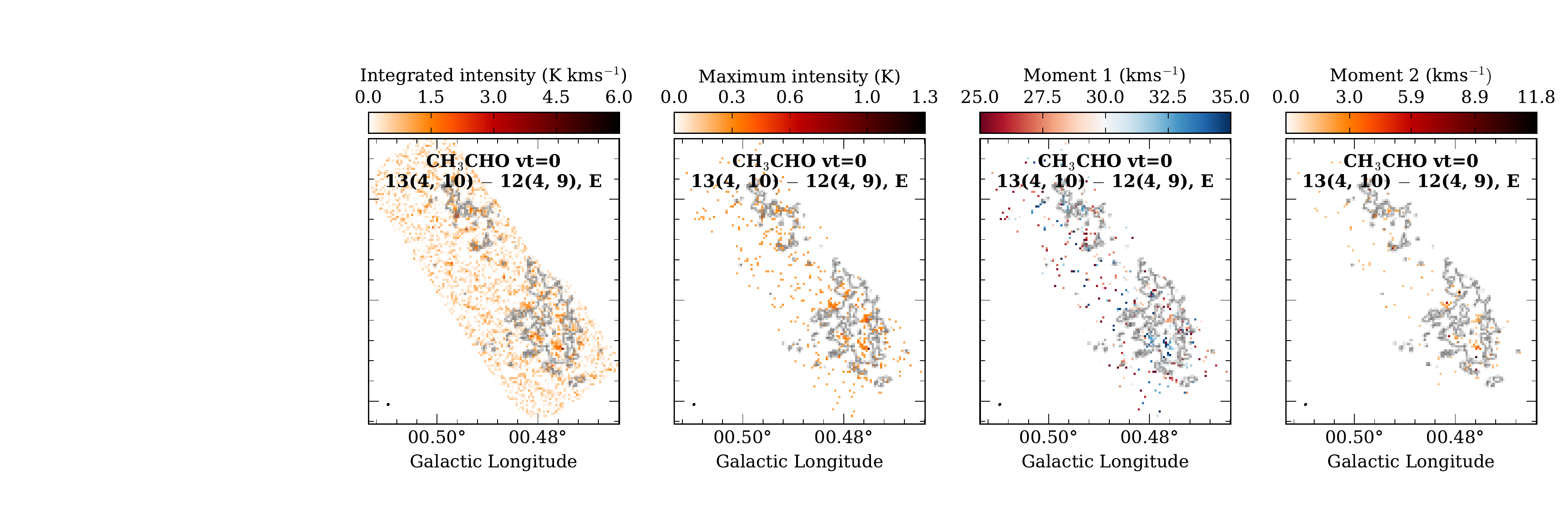}

 \contcaption{}
 
\label{ }
\end{figure*}

\begin{figure*}
\centering

\includegraphics[trim = 3mm 23mm 0cm 10mm, clip,angle=0,width=1\textwidth]{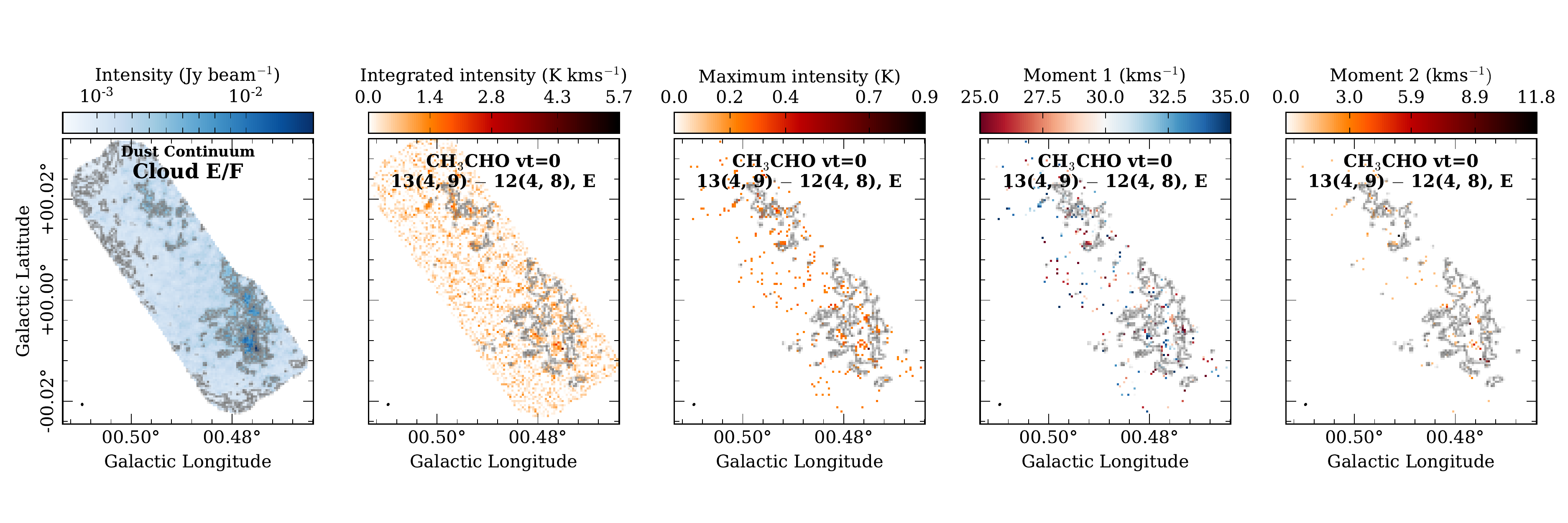} 
\includegraphics[trim = 3mm 23mm 0cm 21mm, clip,angle=0,width=1\textwidth]{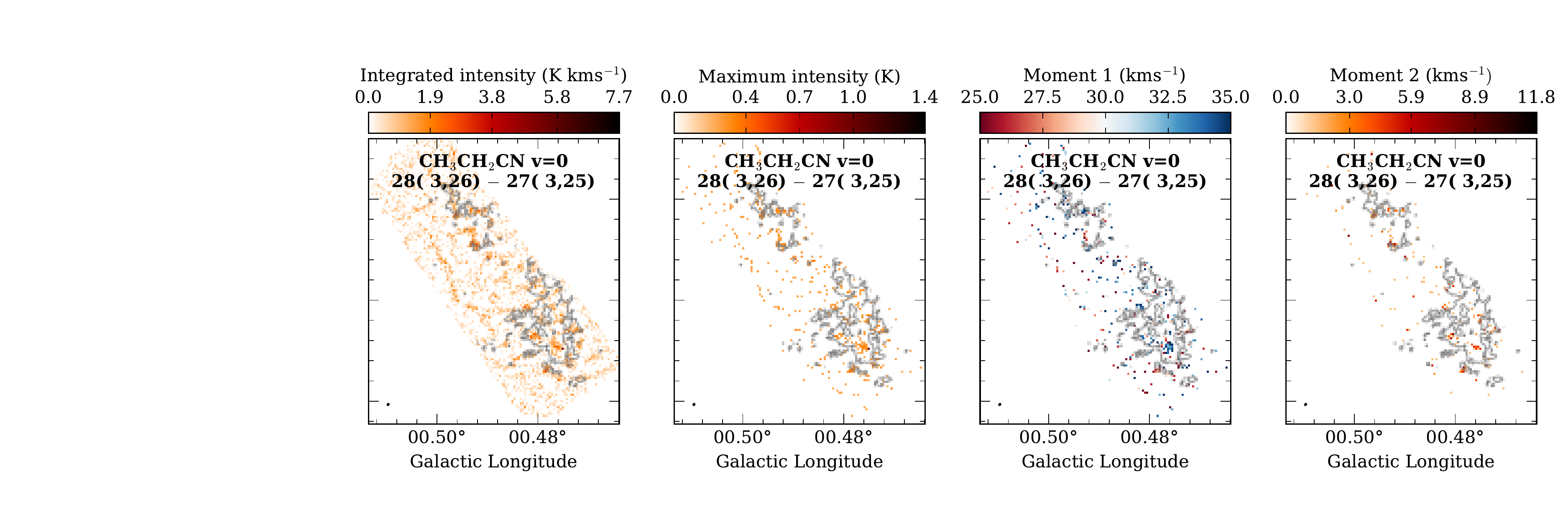} 

 \contcaption{}

\label{ }
\end{figure*}

\begin{figure*}
\centering

\includegraphics[trim = 3mm 26.5mm 3mm 15mm, clip,angle=0,width=1\textwidth]{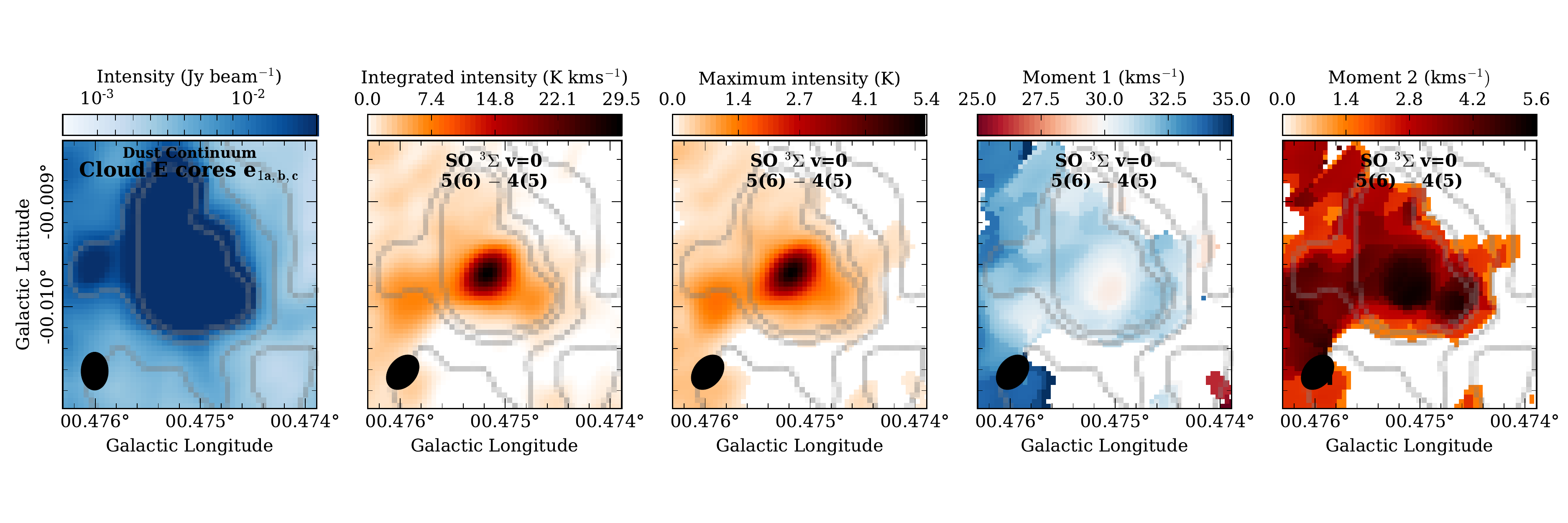} 
\includegraphics[trim = 3mm 14mm 3mm 21.5mm, clip,angle=0,width=1\textwidth]{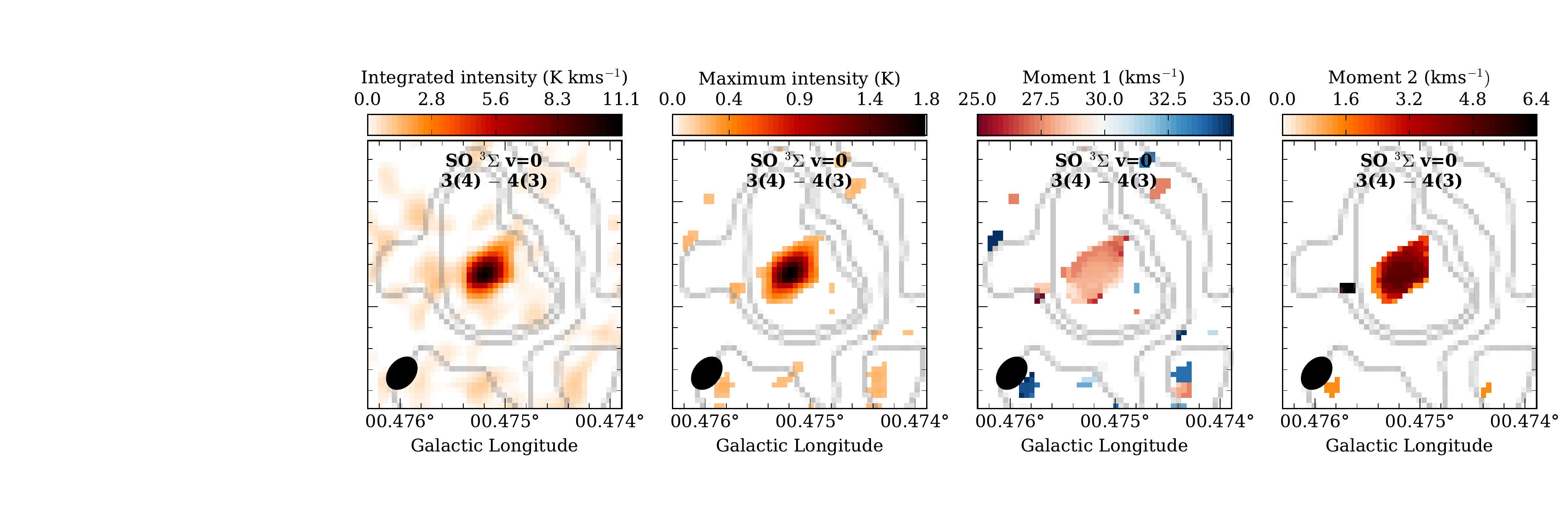}  

\caption{Moment maps of the molecular transitions towards Cloud E/F, which were not noted as being uncertain (see table\,\ref{line_idents1}). Show is a zoom-in of the region containing cores e1$_{\rm a}$, e1$_{\rm b}$, and e1$_{\rm c}$ (see section\,\ref{virial_cores}). The analysis for the different molecular transitions are presented in each row, with the molecule labeled at the top of each map. Shown in the upper left is the combined 12m, 7m and single dish continuum map, and then from left to right are moment maps of the integrated intensity, peak intensity, intensity weighted centroid velocity, and intensity weighted velocity dispersion for each molecule. Contours on each map are of the continuum shown in levels of [8, 15, 30, 50]\,$\sigma_{\rm rms}$, where $\sigma_{\rm rms}$\,$\sim$\,0.6\,mJy\,beam$^{-1}$.} 

\label{moments_core_e1}
\end{figure*}

\begin{figure*} 
\centering 
\includegraphics[trim = 3mm 26.5mm 3mm 15mm, clip,angle=0,width=1\textwidth]{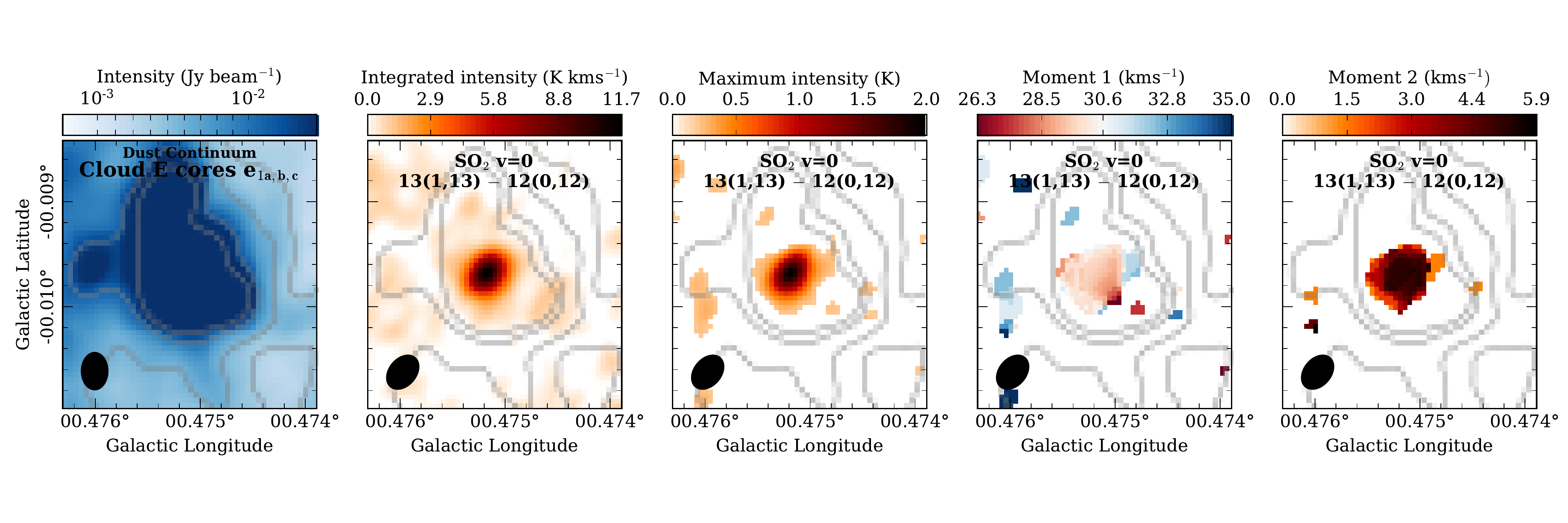} 
\includegraphics[trim = 3mm 26.5mm 3mm 21.5mm, clip,angle=0,width=1\textwidth]{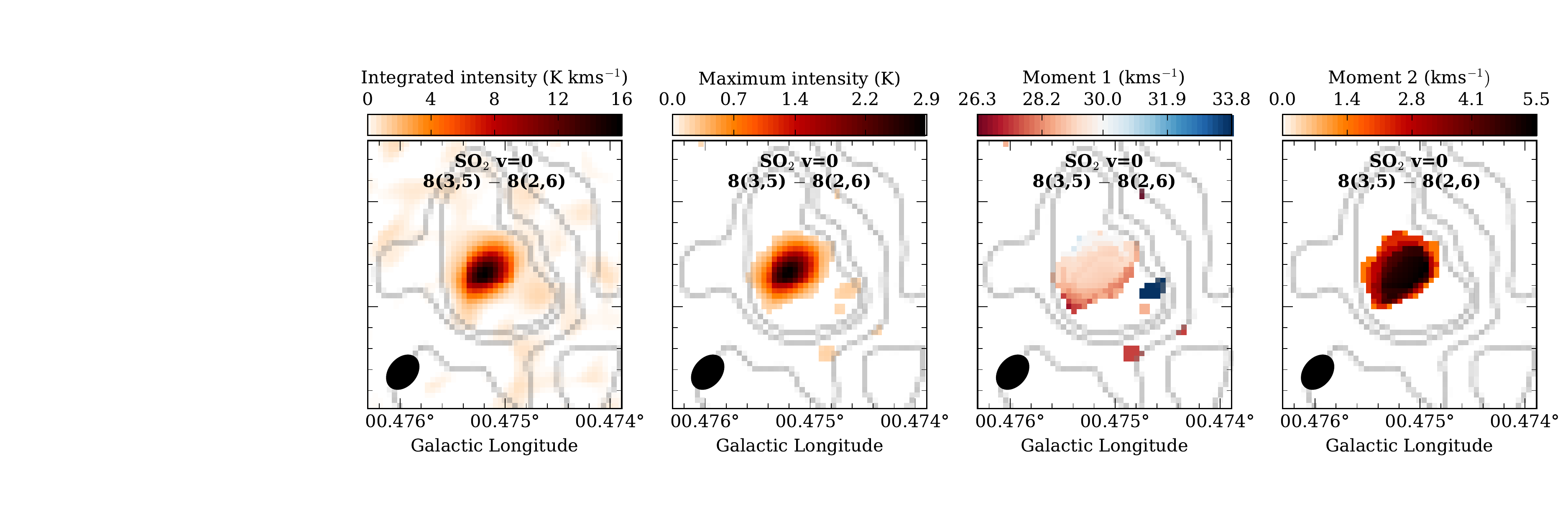} 
\includegraphics[trim = 3mm 26.5mm 3mm 21.5mm, clip,angle=0,width=1\textwidth]{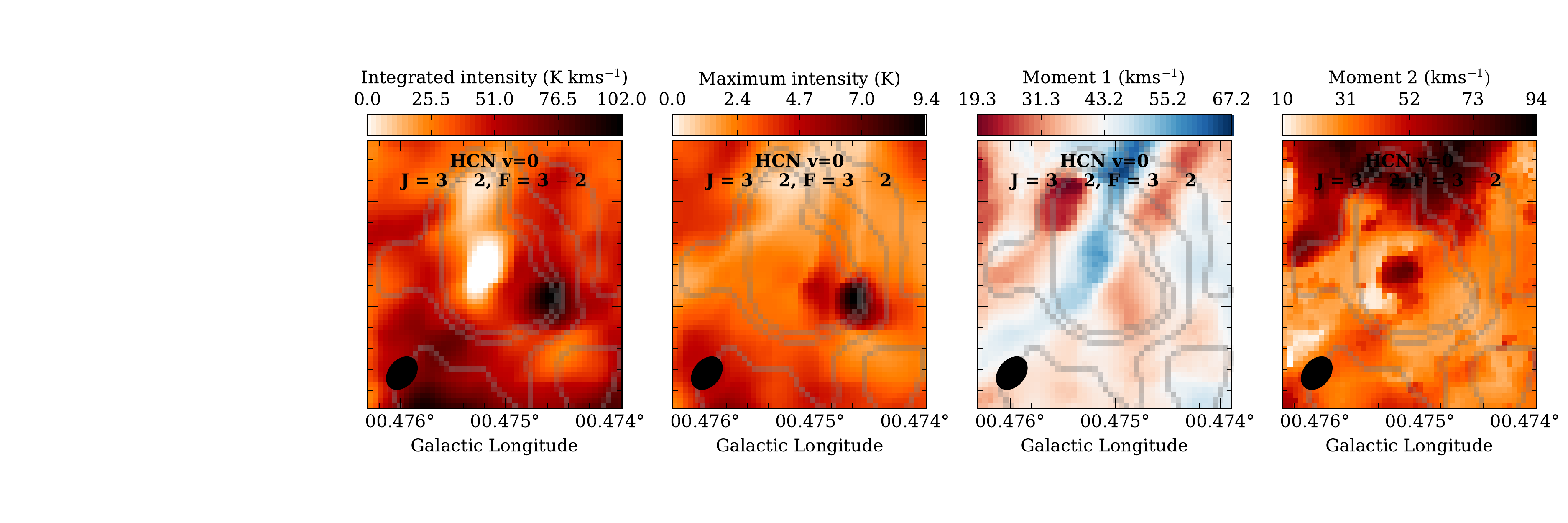} 
\includegraphics[trim = 3mm 26.5mm 3mm 21.5mm, clip,angle=0,width=1\textwidth]{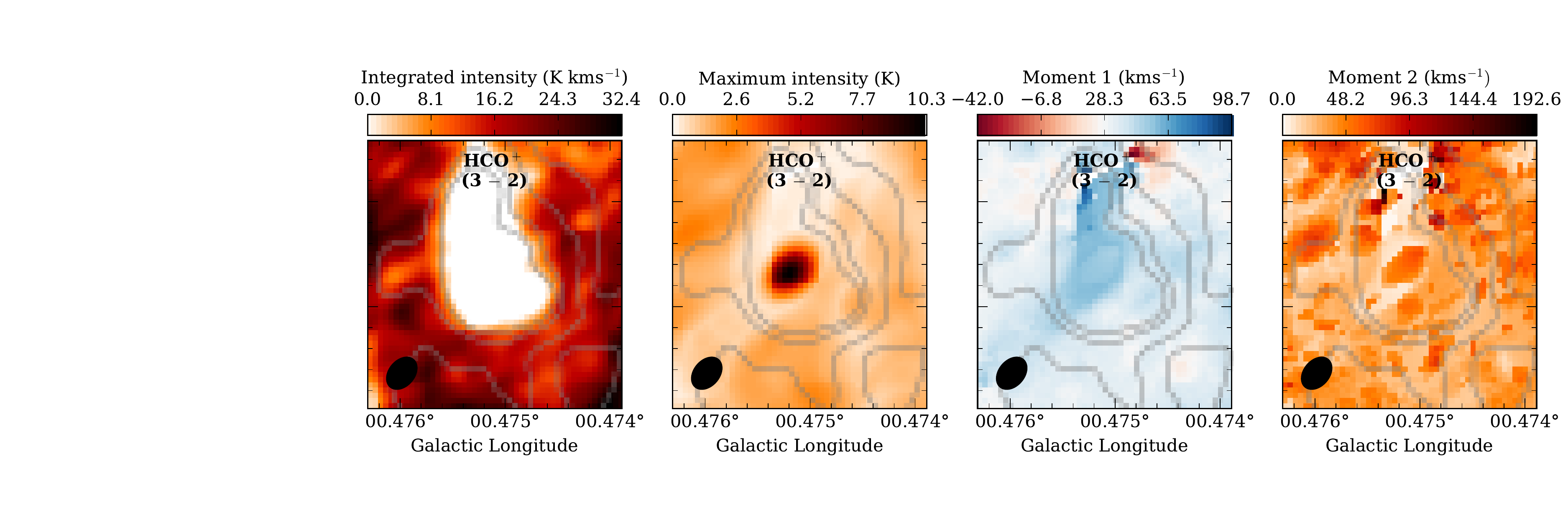} 
\includegraphics[trim = 3mm 26.5mm 3mm 21.5mm, clip,angle=0,width=1\textwidth]{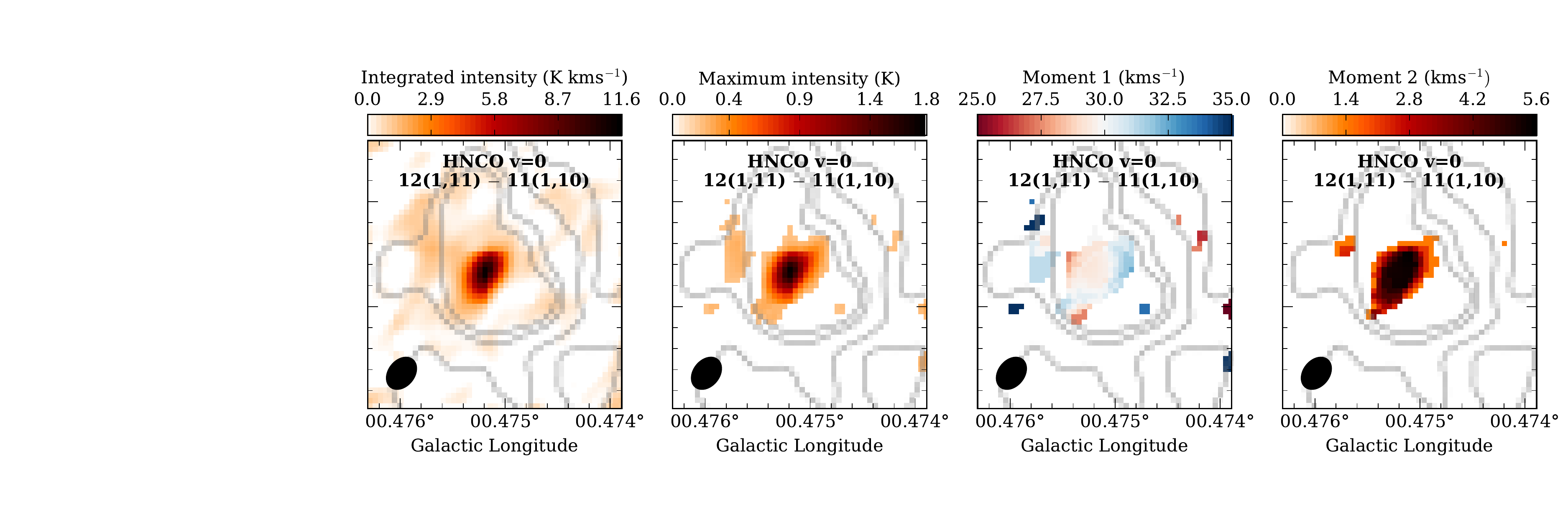} 
\includegraphics[trim = 3mm 14mm 3mm 21.5mm, clip,angle=0,width=1\textwidth]{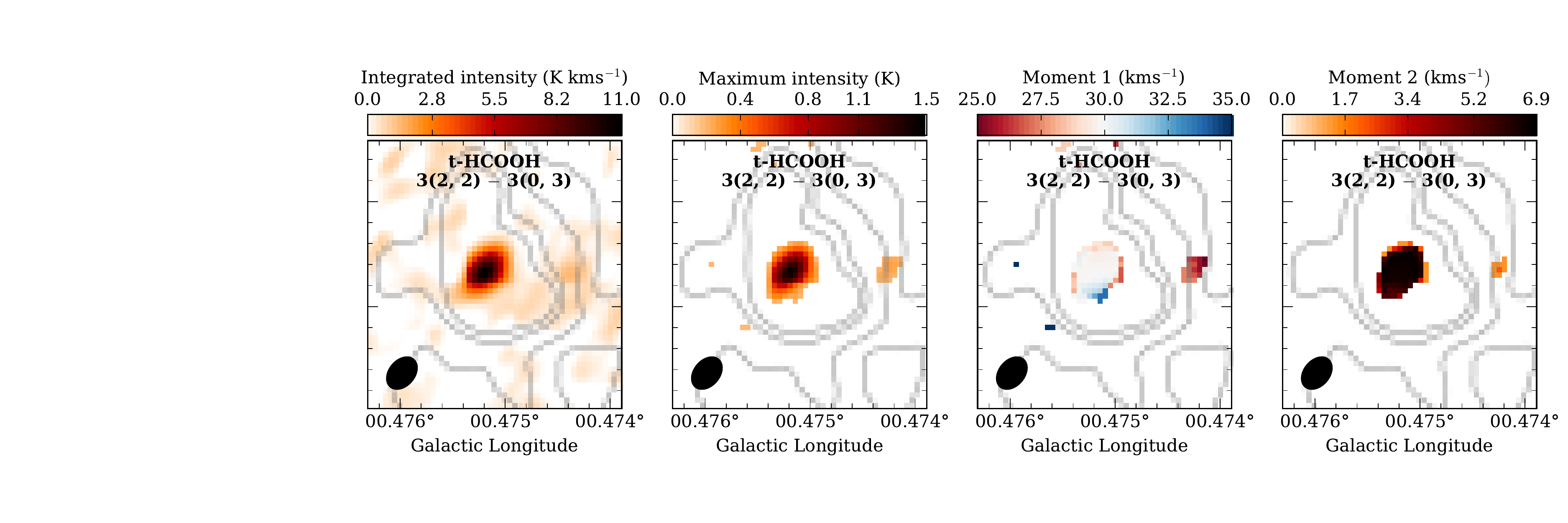} 
\contcaption{} 
\label{ } 
\end{figure*} 

\begin{figure*} 
\centering 
\includegraphics[trim = 3mm 26.5mm 3mm 15mm, clip,angle=0,width=1\textwidth]{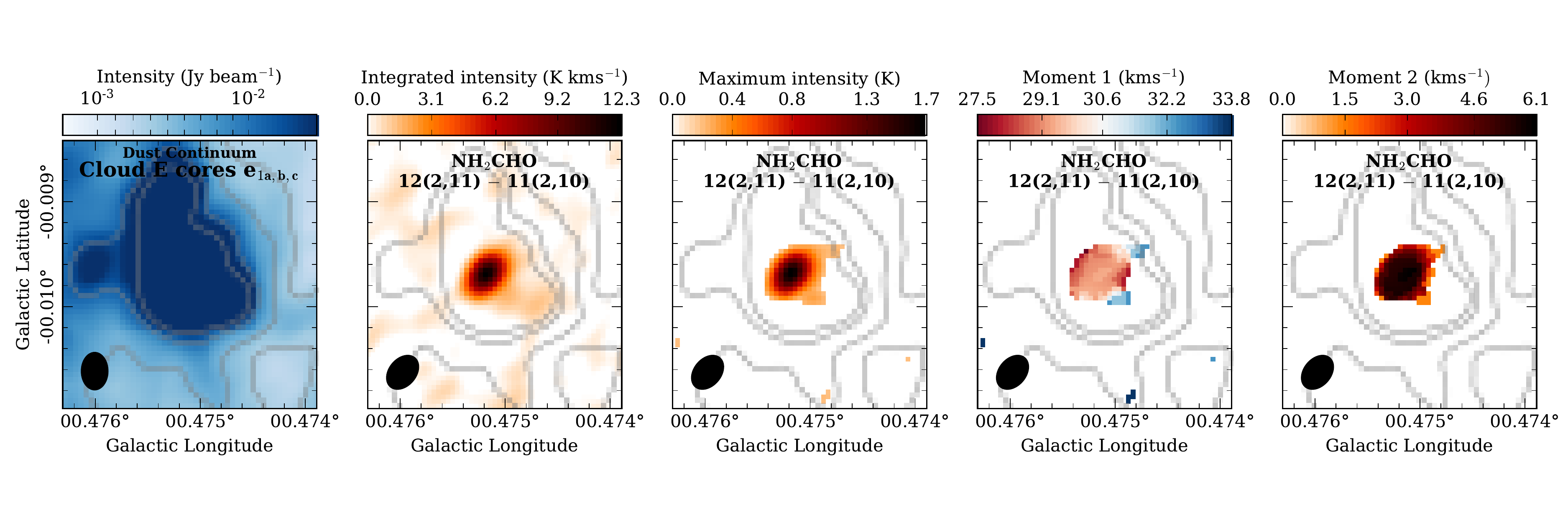} 
\includegraphics[trim = 3mm 26.5mm 3mm 21.5mm, clip,angle=0,width=1\textwidth]{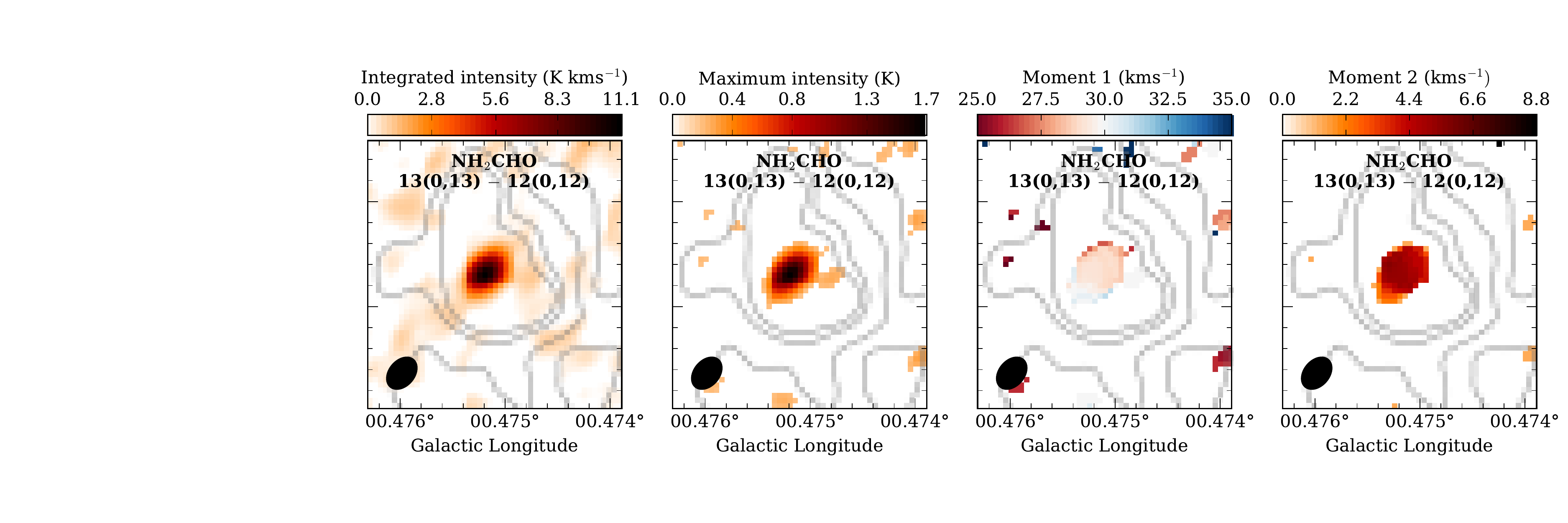} 
\includegraphics[trim = 3mm 26.5mm 3mm 21.5mm, clip,angle=0,width=1\textwidth]{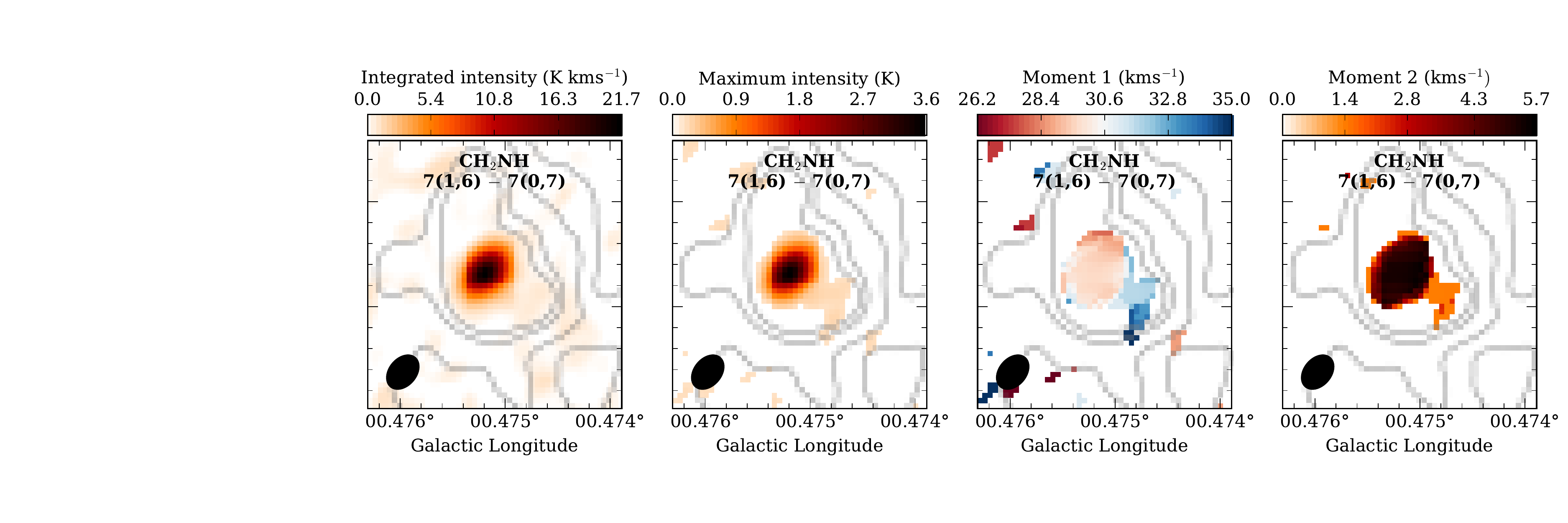} 
\includegraphics[trim = 3mm 26.5mm 3mm 21.5mm, clip,angle=0,width=1\textwidth]{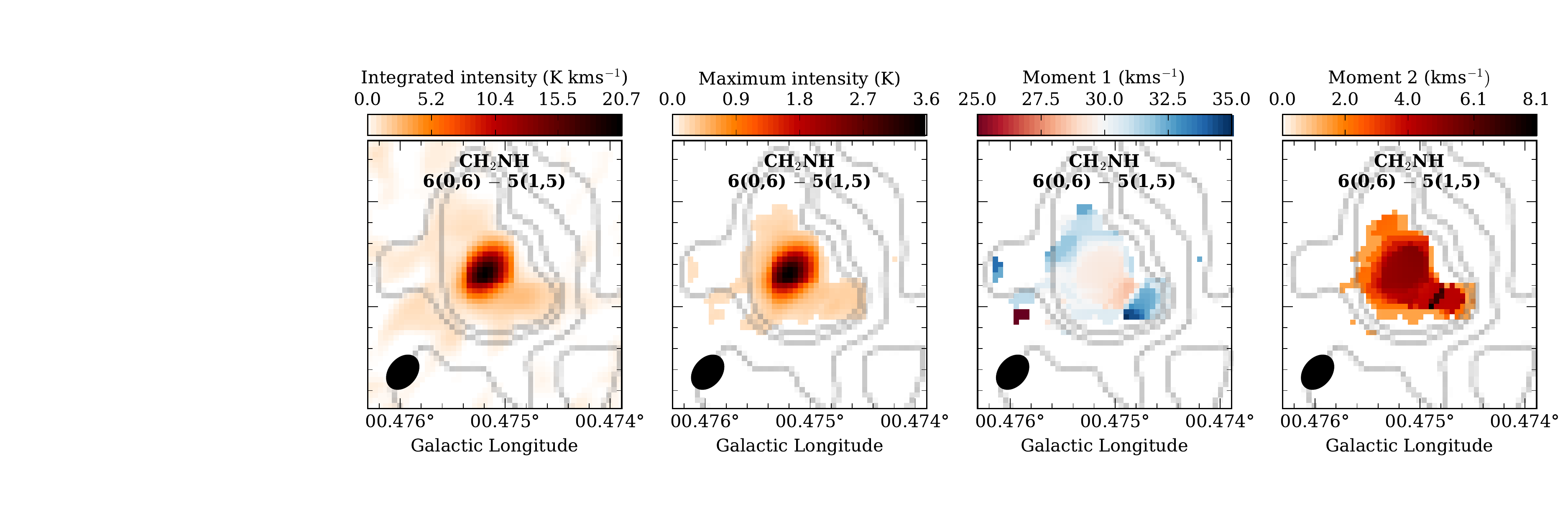} 
\includegraphics[trim = 3mm 26.5mm 3mm 21.5mm, clip,angle=0,width=1\textwidth]{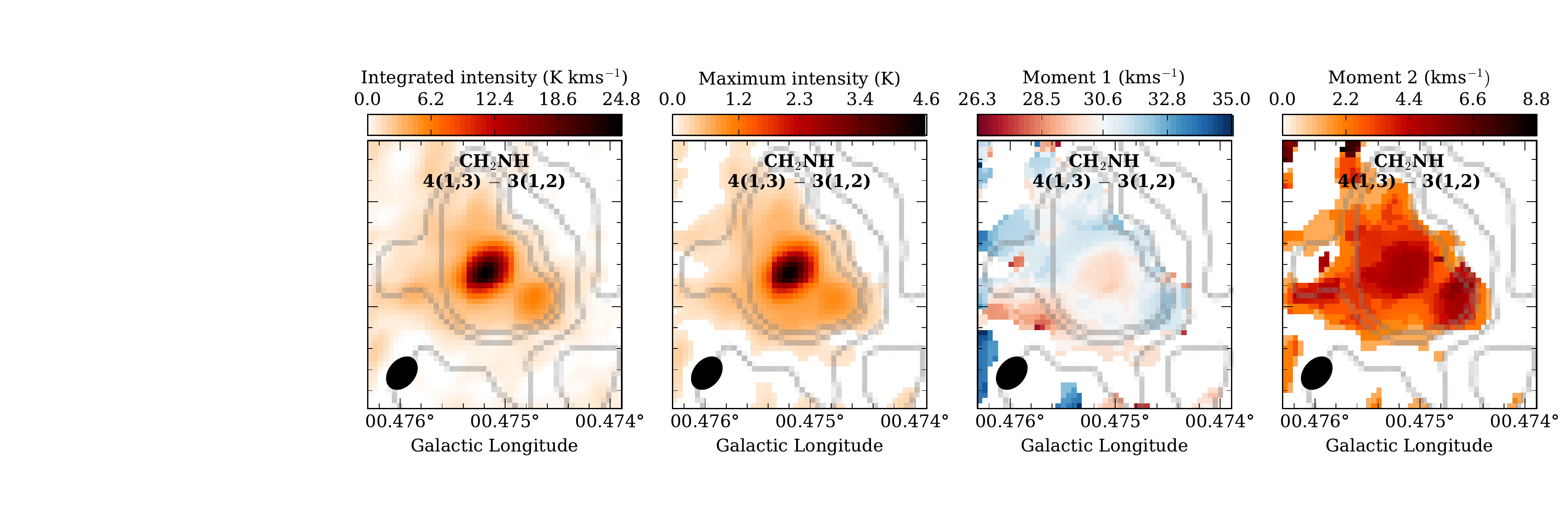} 
\includegraphics[trim = 3mm 14mm 3mm 21.5mm, clip,angle=0,width=1\textwidth]{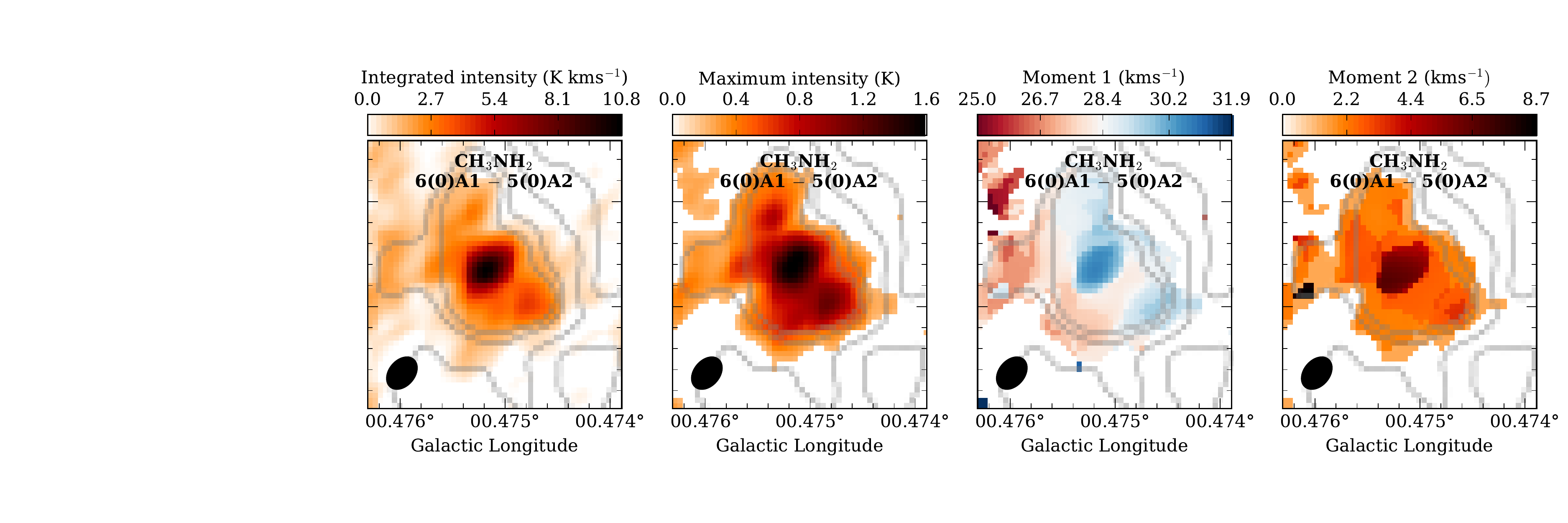} 
\contcaption{} 
\label{ } 
\end{figure*} 

\begin{figure*} 
\centering 
\includegraphics[trim = 3mm 26.5mm 3mm 15mm, clip,angle=0,width=1\textwidth]{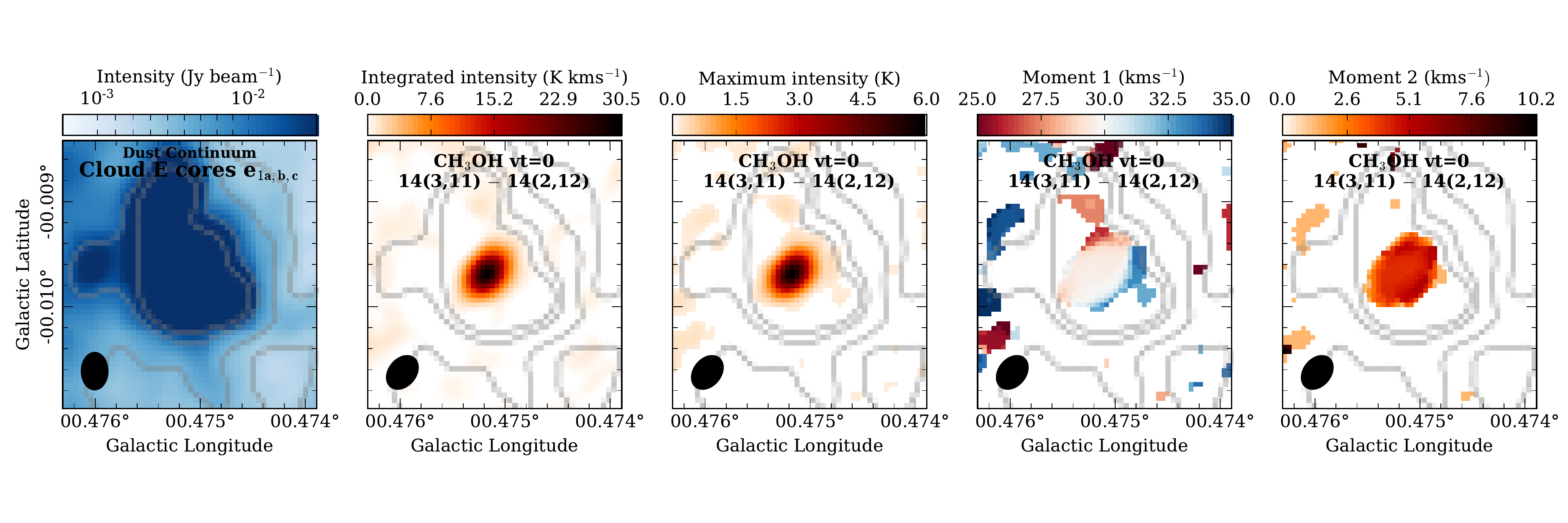} 
\includegraphics[trim = 3mm 26.5mm 3mm 21.5mm, clip,angle=0,width=1\textwidth]{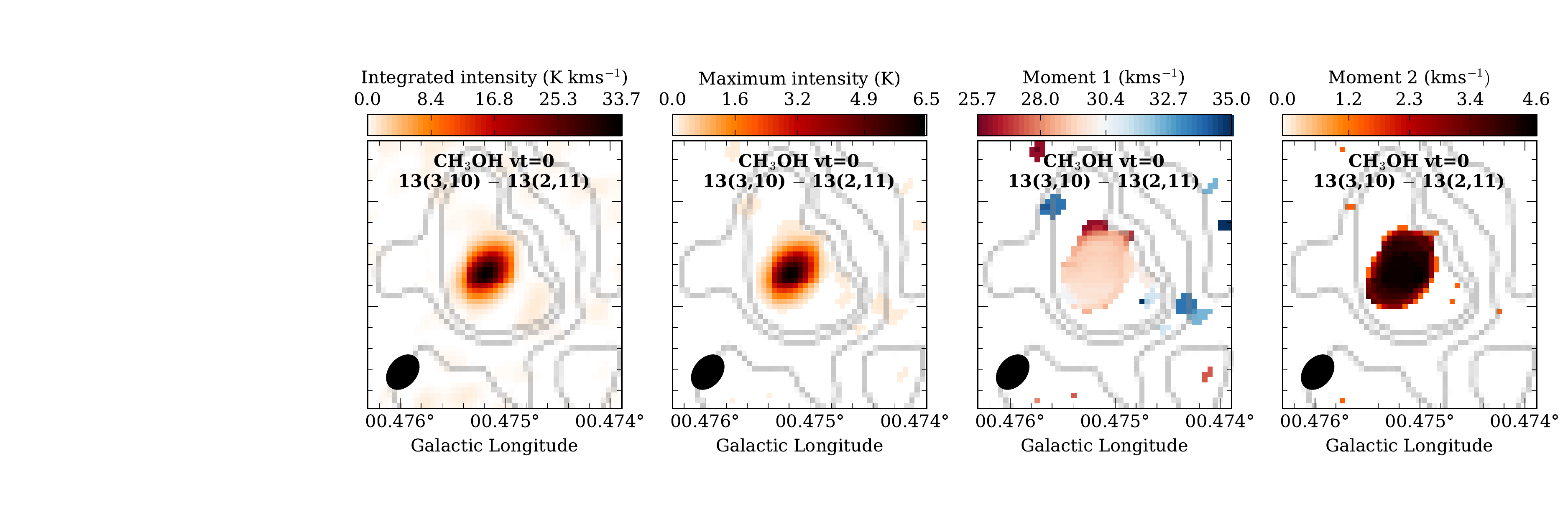} 
\includegraphics[trim = 3mm 26.5mm 3mm 21.5mm, clip,angle=0,width=1\textwidth]{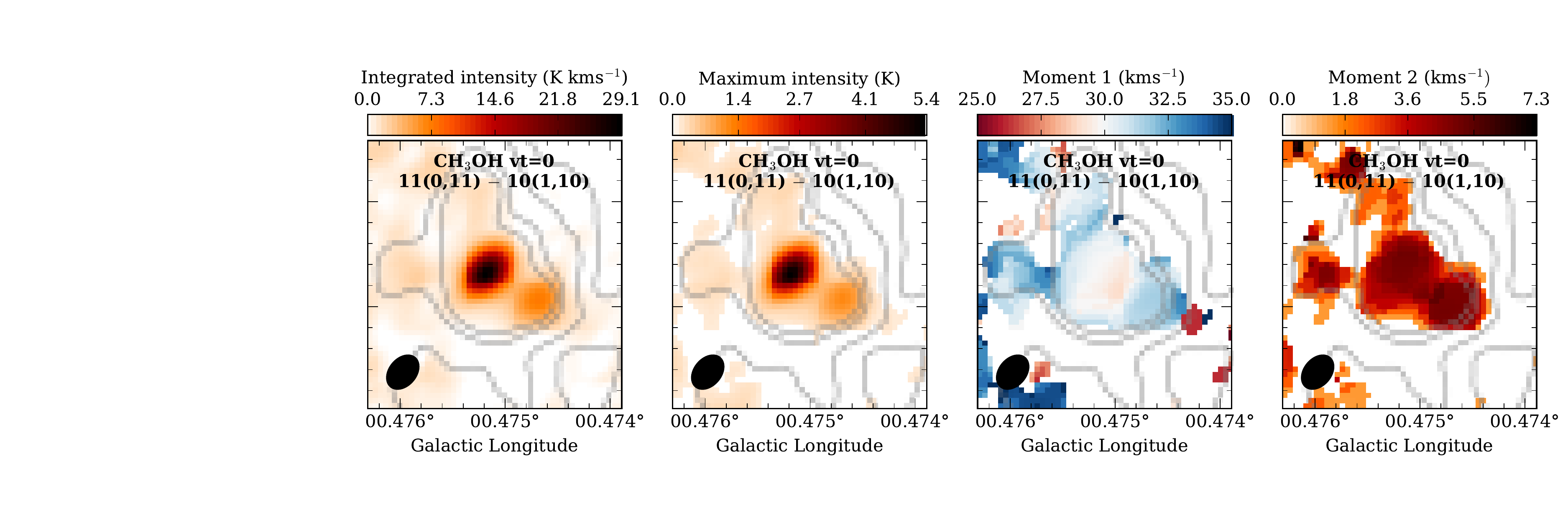} 
\includegraphics[trim = 3mm 26.5mm 3mm 21.5mm, clip,angle=0,width=1\textwidth]{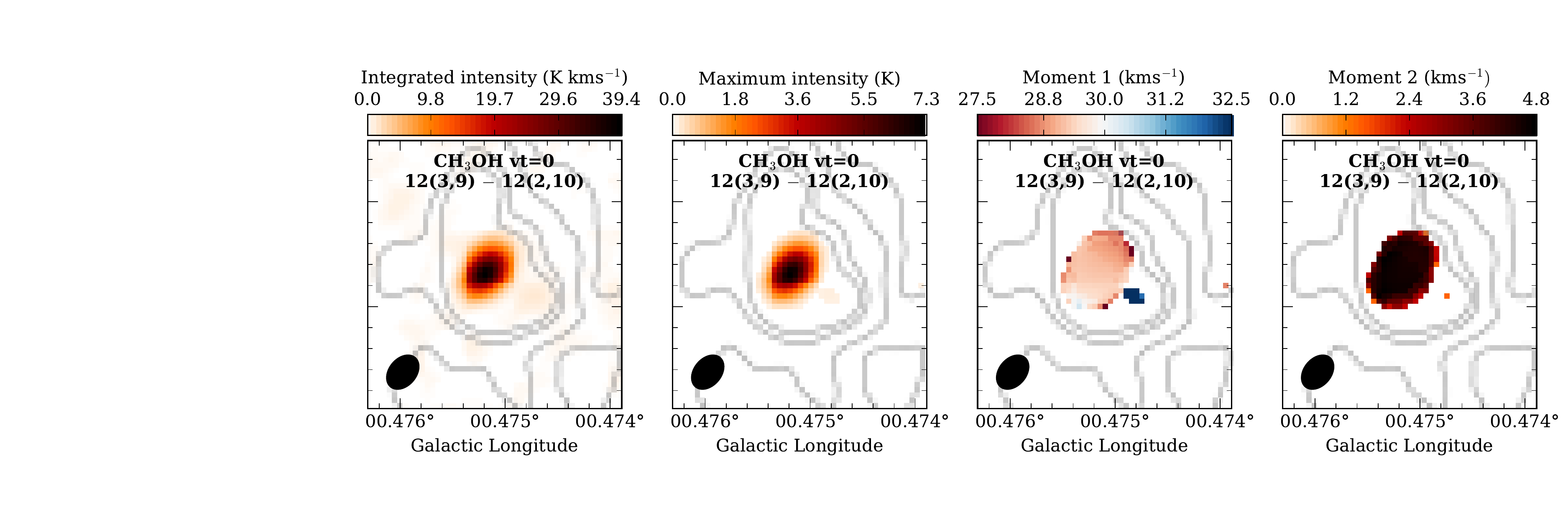} 
\includegraphics[trim = 3mm 26.5mm 3mm 21.5mm, clip,angle=0,width=1\textwidth]{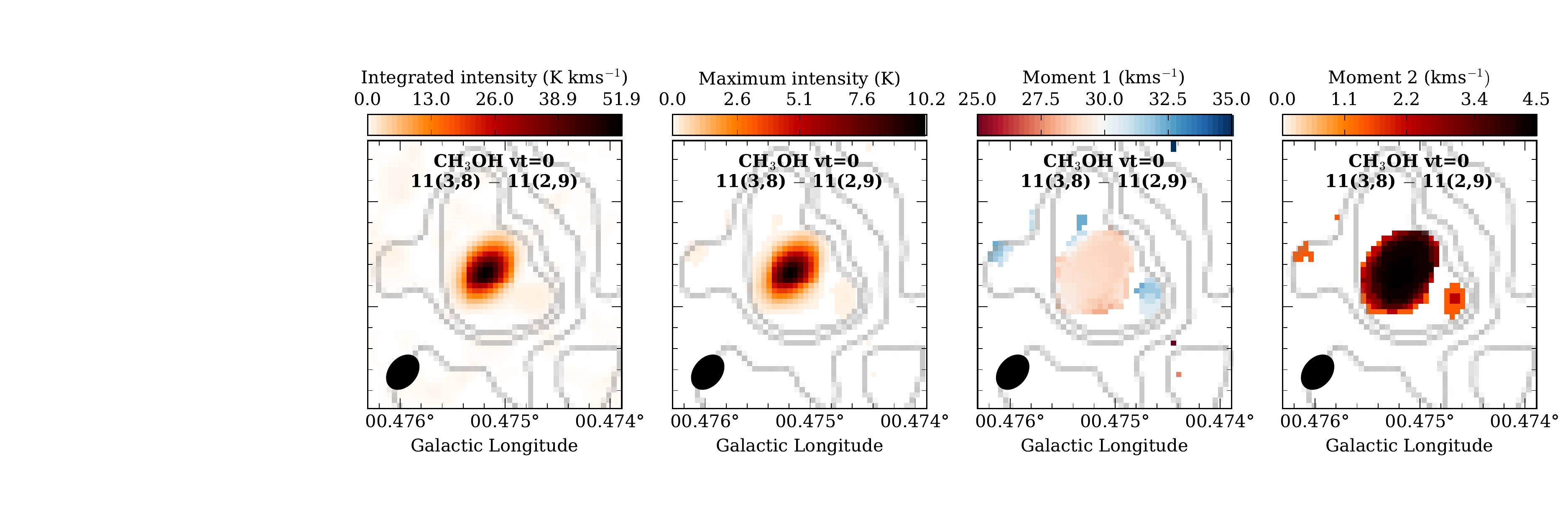} 
\includegraphics[trim = 3mm 14mm 3mm 21.5mm, clip,angle=0,width=1\textwidth]{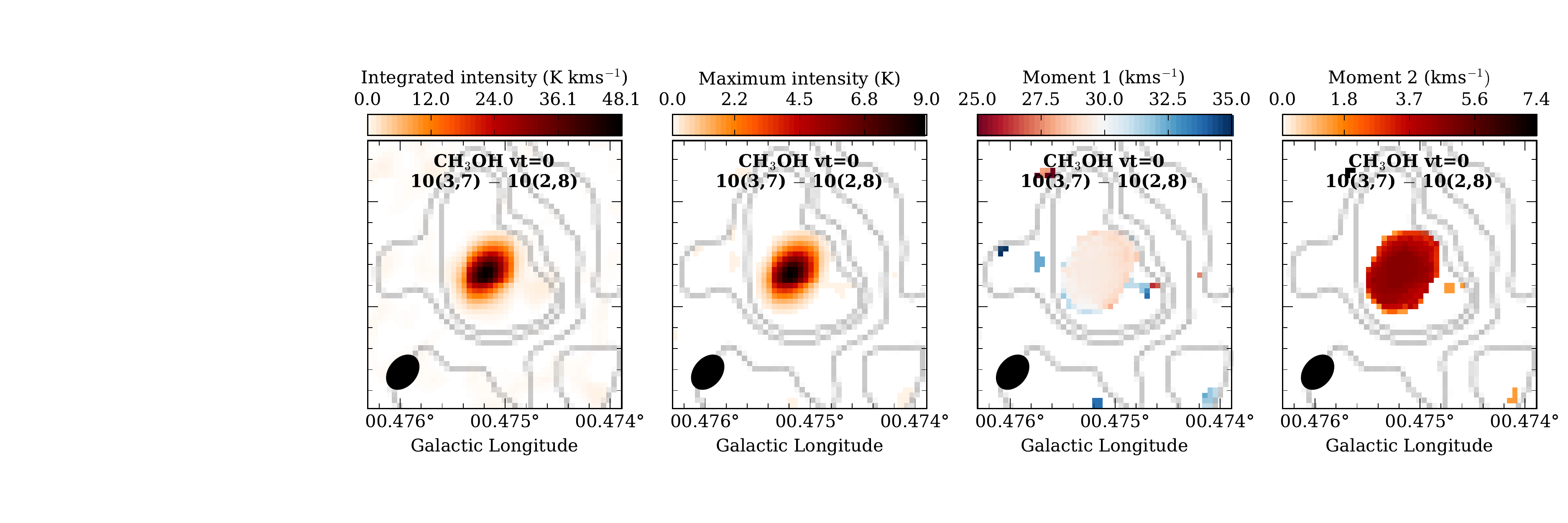} 
\contcaption{} 
\label{ } 
\end{figure*} 

\begin{figure*} 
\centering 
\includegraphics[trim = 3mm 26.5mm 3mm 15mm, clip,angle=0,width=1\textwidth]{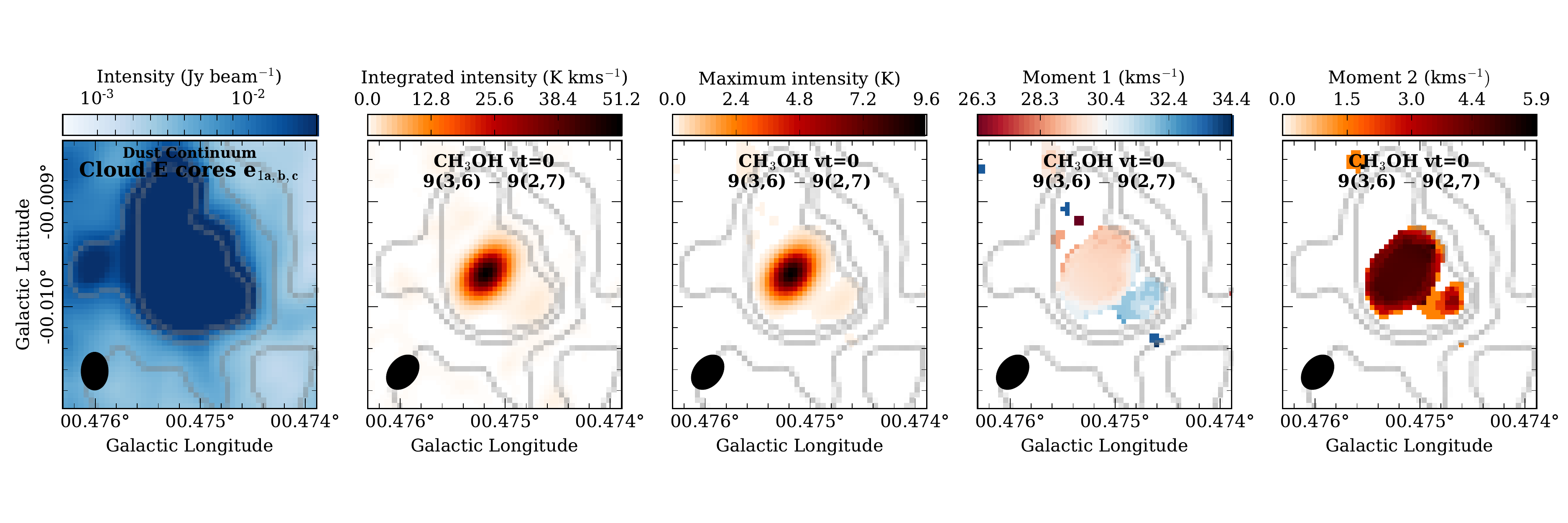} 
\includegraphics[trim = 3mm 26.5mm 3mm 21.5mm, clip,angle=0,width=1\textwidth]{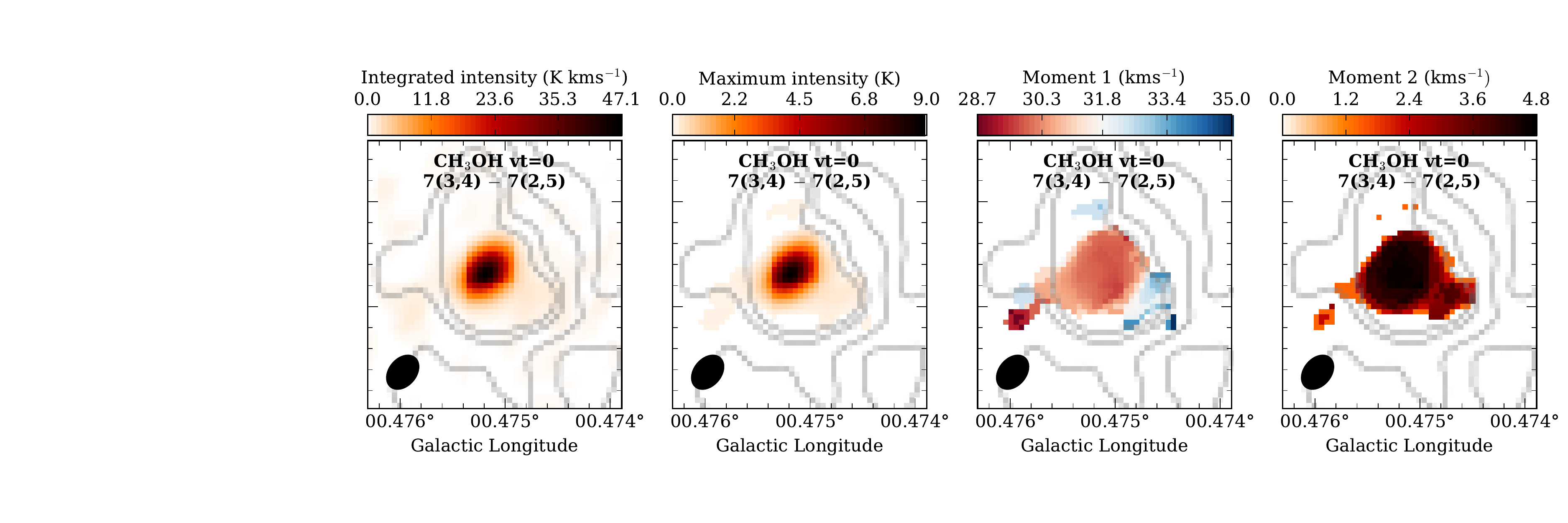} 
\includegraphics[trim = 3mm 26.5mm 3mm 21.5mm, clip,angle=0,width=1\textwidth]{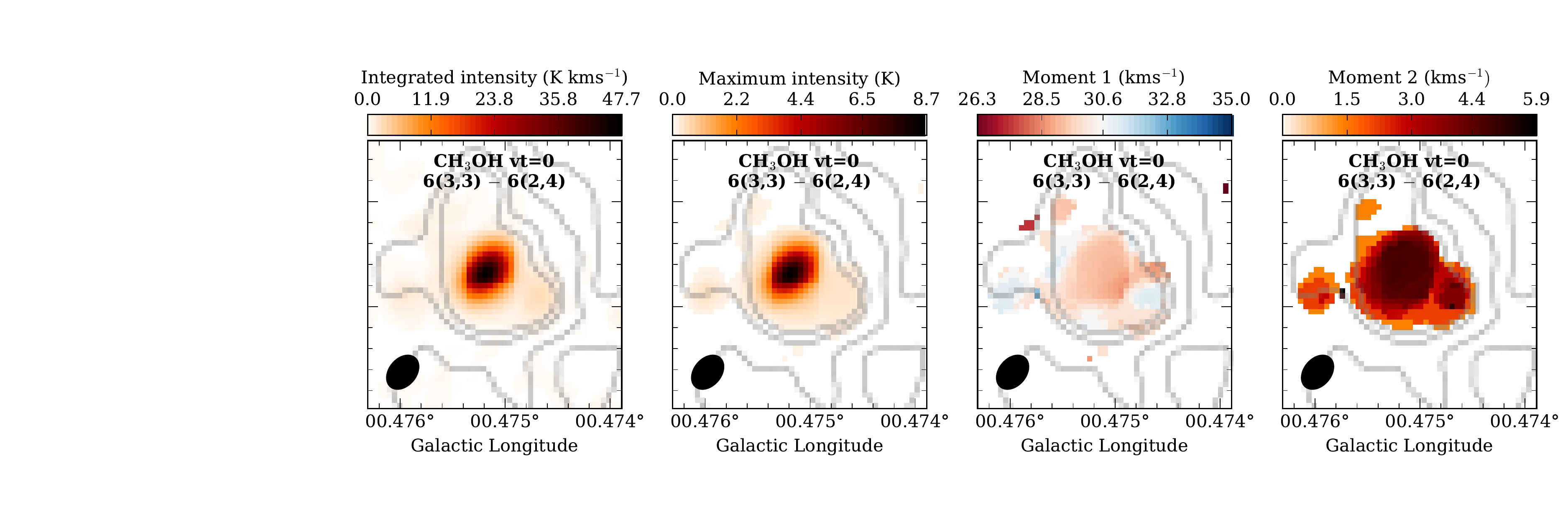} 
\includegraphics[trim = 3mm 26.5mm 3mm 21.5mm, clip,angle=0,width=1\textwidth]{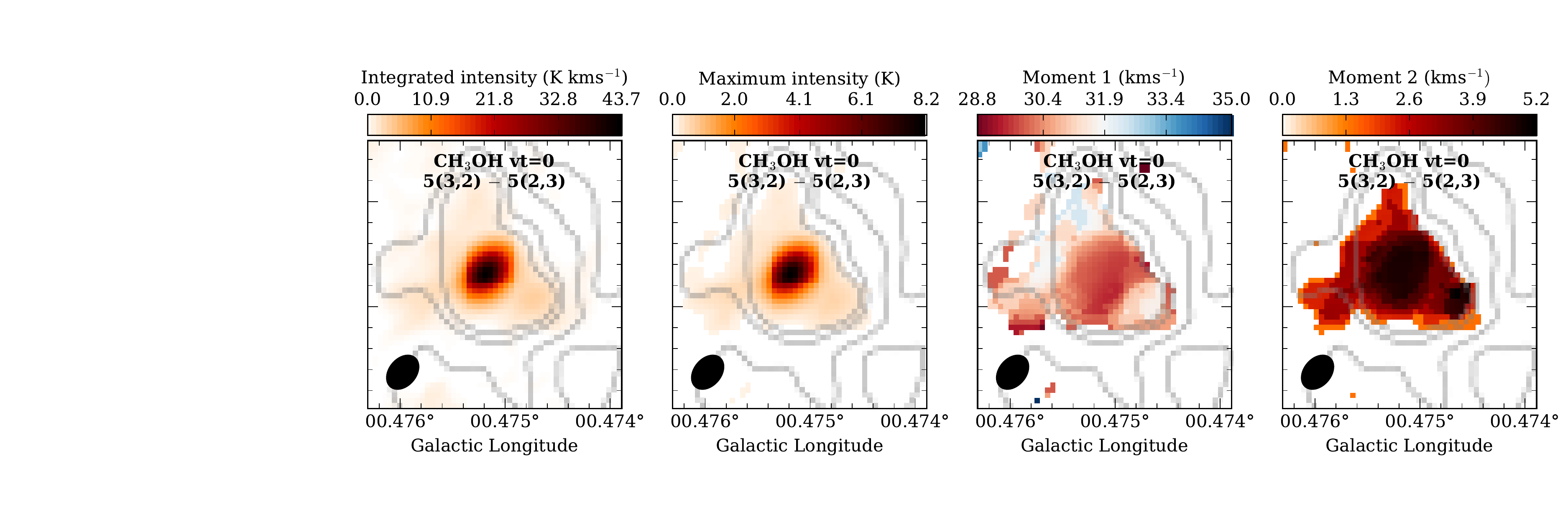} 
\includegraphics[trim = 3mm 26.5mm 3mm 21.5mm, clip,angle=0,width=1\textwidth]{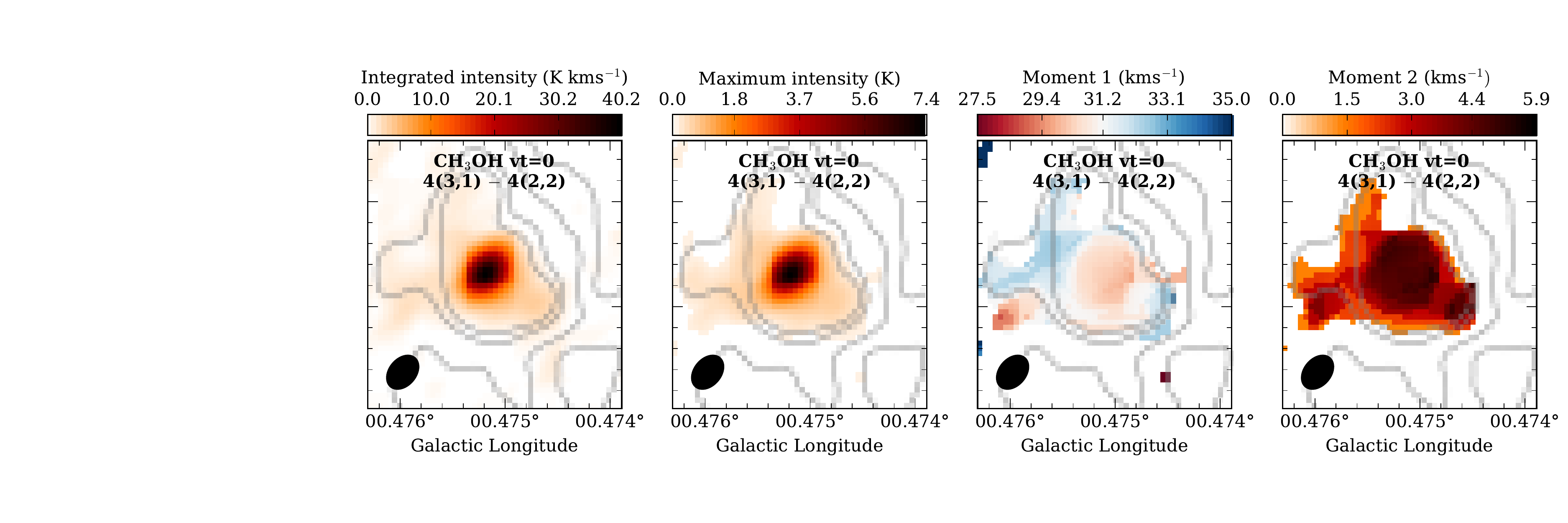} 
\includegraphics[trim = 3mm 14mm 3mm 21.5mm, clip,angle=0,width=1\textwidth]{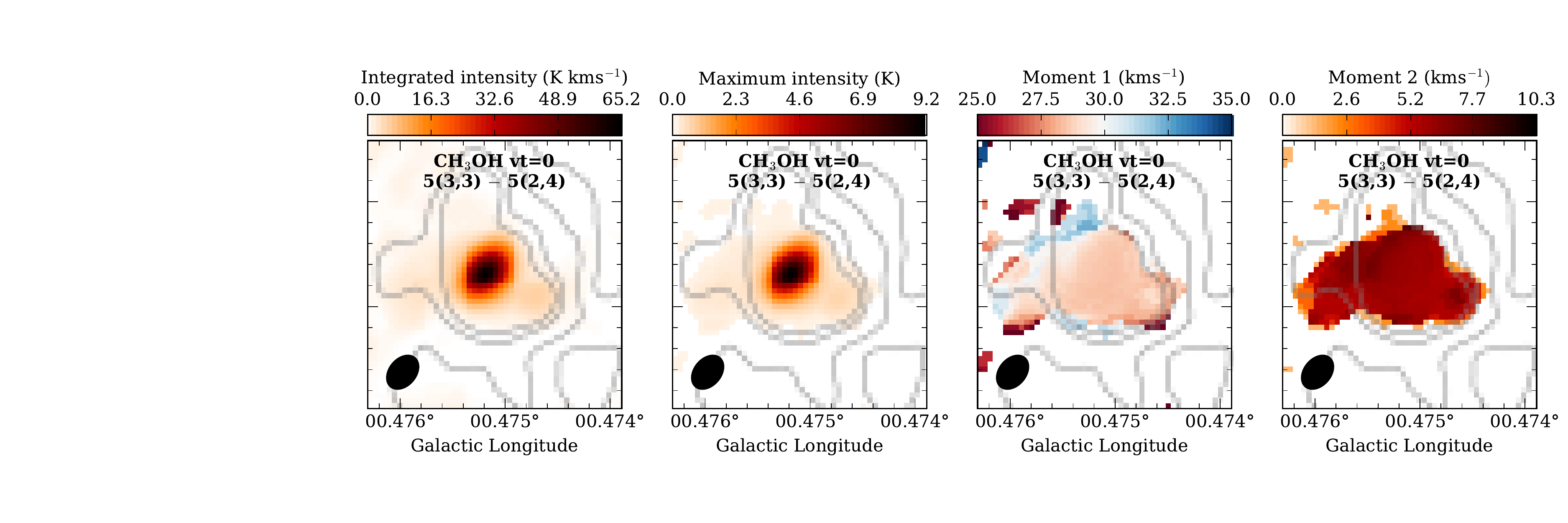} 
\contcaption{} 
\label{ } 
\end{figure*} 

\begin{figure*} 
\centering 
\includegraphics[trim = 3mm 26.5mm 3mm 15mm, clip,angle=0,width=1\textwidth]{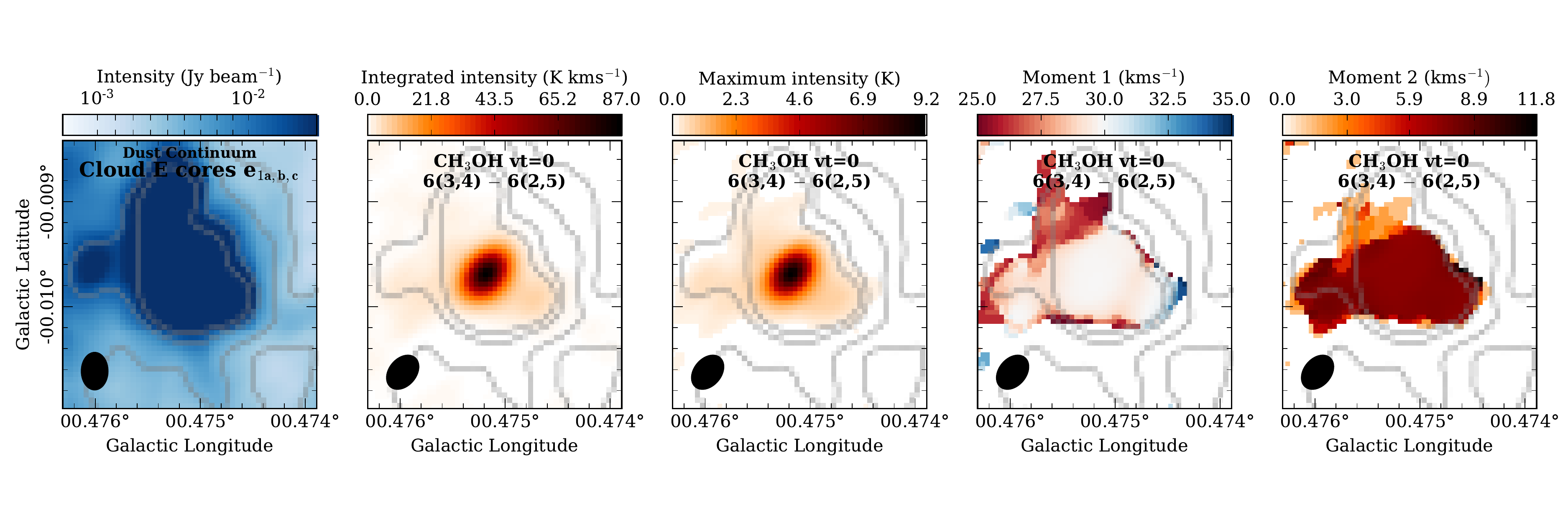} 
\includegraphics[trim = 3mm 26.5mm 3mm 21.5mm, clip,angle=0,width=1\textwidth]{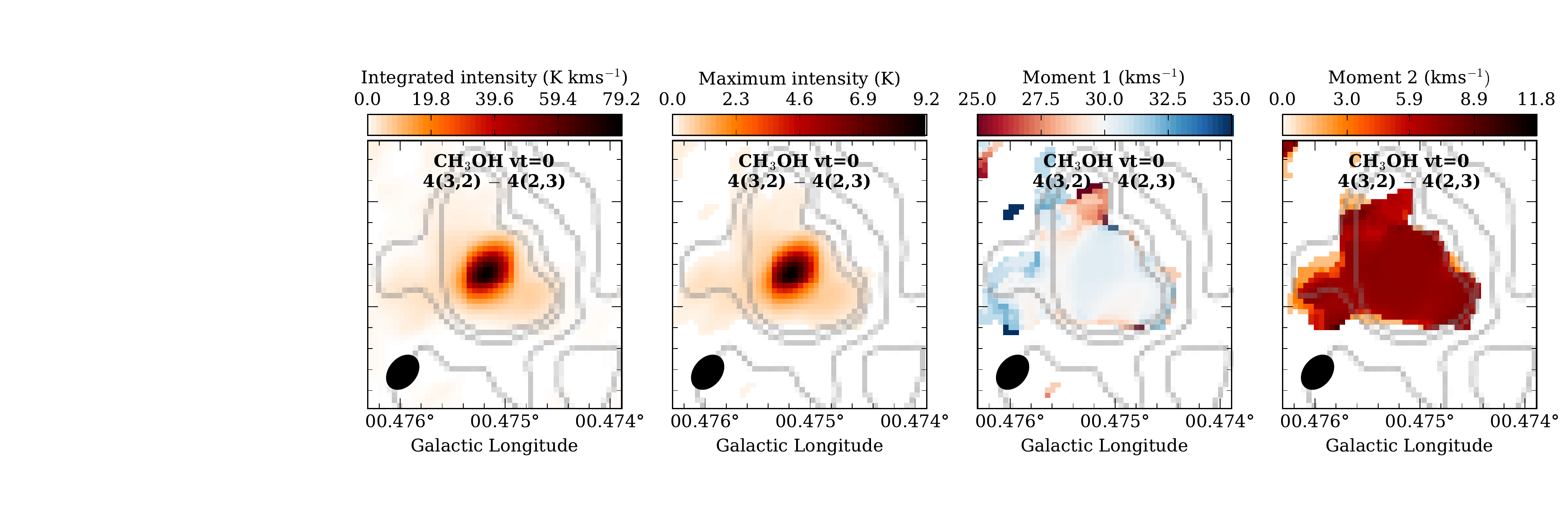} 
\includegraphics[trim = 3mm 26.5mm 3mm 21.5mm, clip,angle=0,width=1\textwidth]{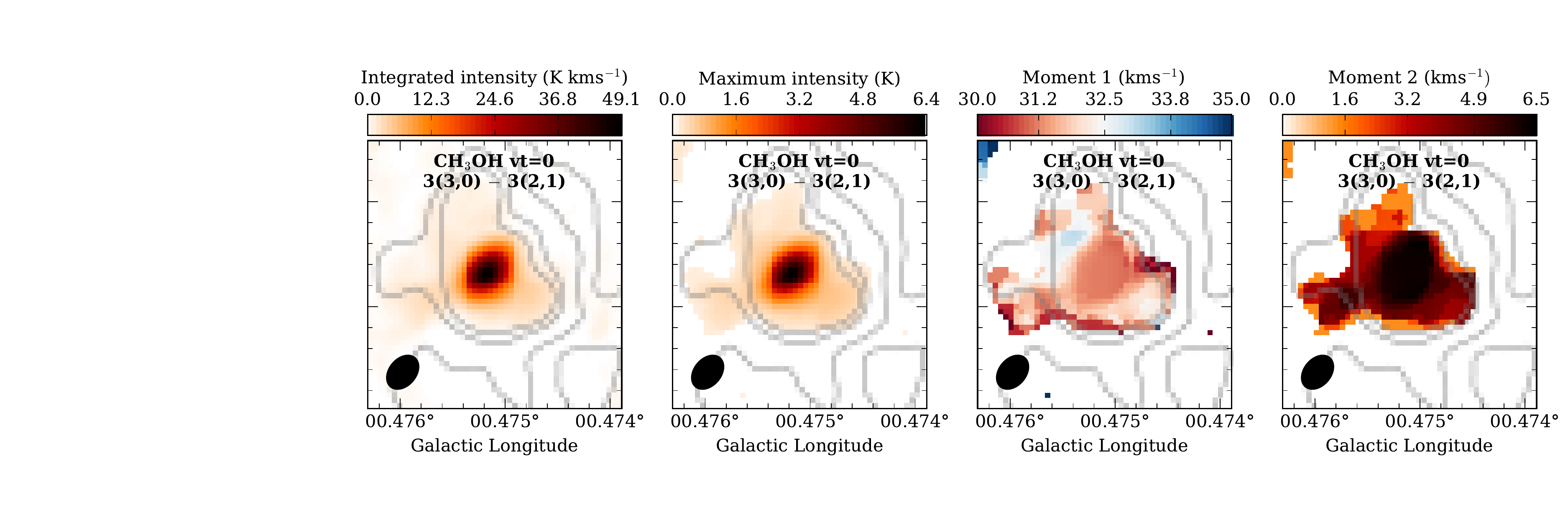} 
\includegraphics[trim = 3mm 26.5mm 3mm 21.5mm, clip,angle=0,width=1\textwidth]{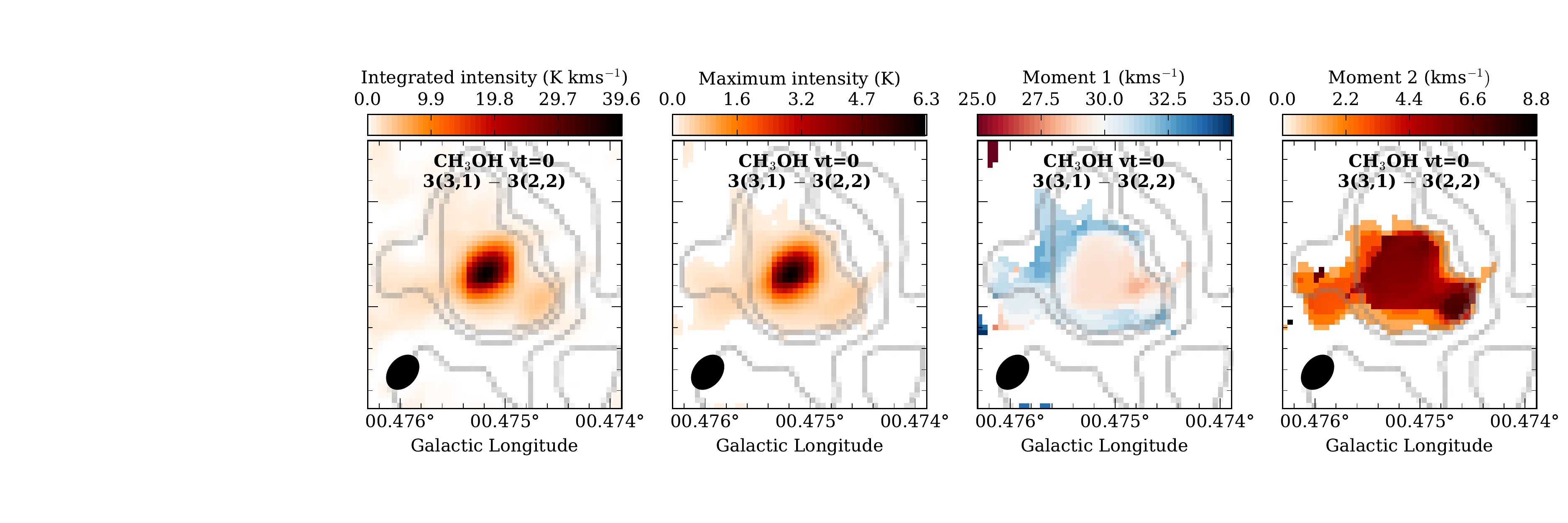} 
\includegraphics[trim = 3mm 26.5mm 3mm 21.5mm, clip,angle=0,width=1\textwidth]{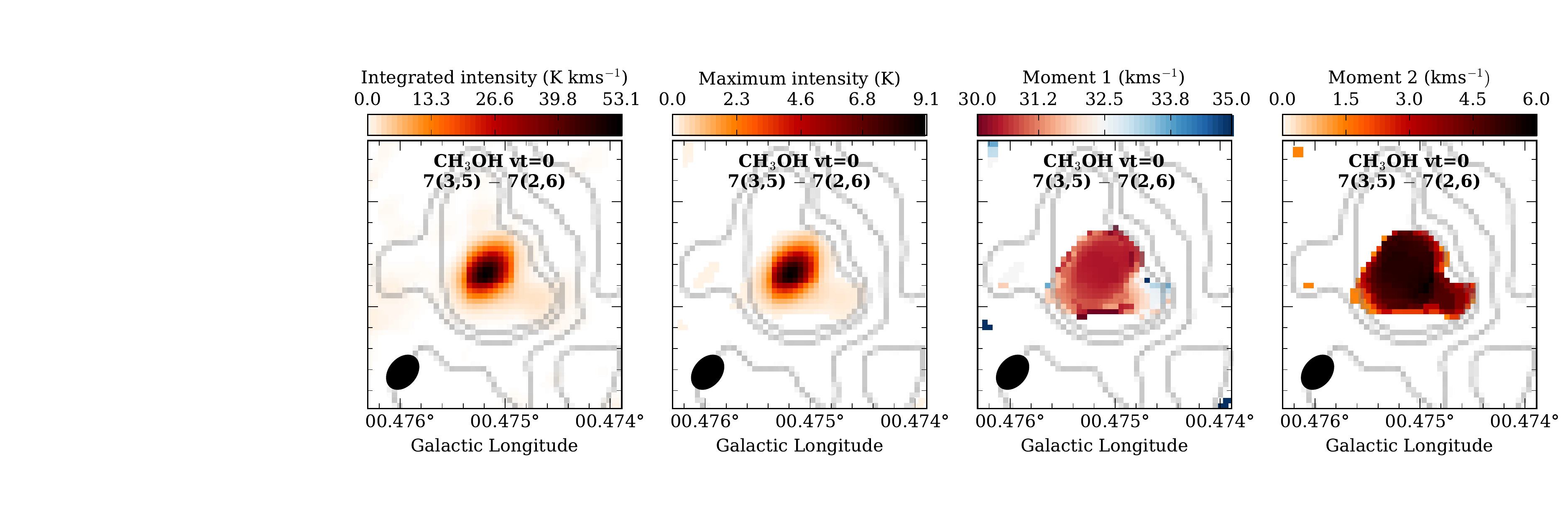} 
\includegraphics[trim = 3mm 14mm 3mm 21.5mm, clip,angle=0,width=1\textwidth]{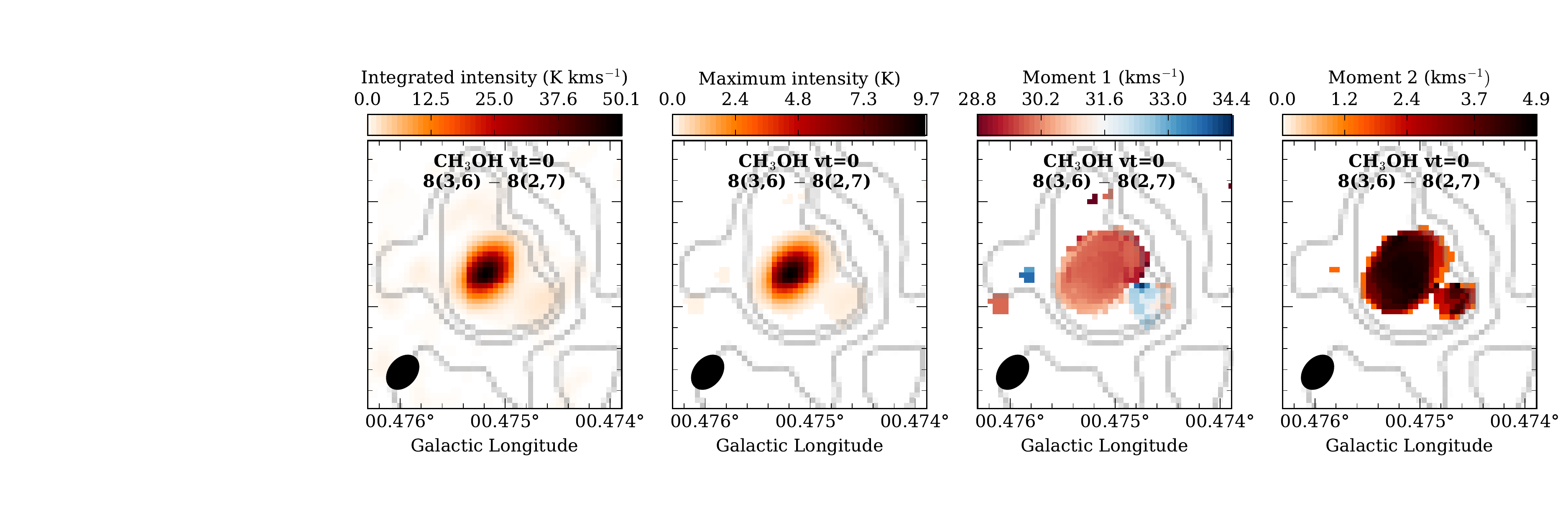} 
\contcaption{} 
\label{ } 
\end{figure*} 

\begin{figure*} 
\centering 
\includegraphics[trim = 3mm 26.5mm 3mm 15mm, clip,angle=0,width=1\textwidth]{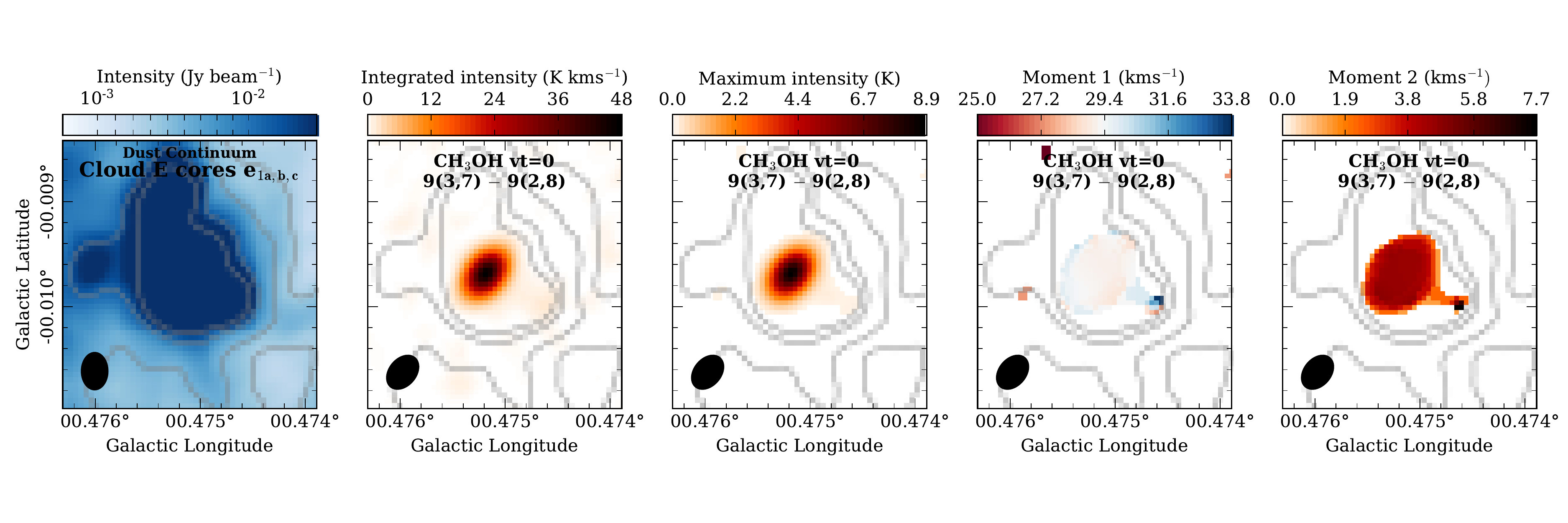} 
\includegraphics[trim = 3mm 26.5mm 3mm 21.5mm, clip,angle=0,width=1\textwidth]{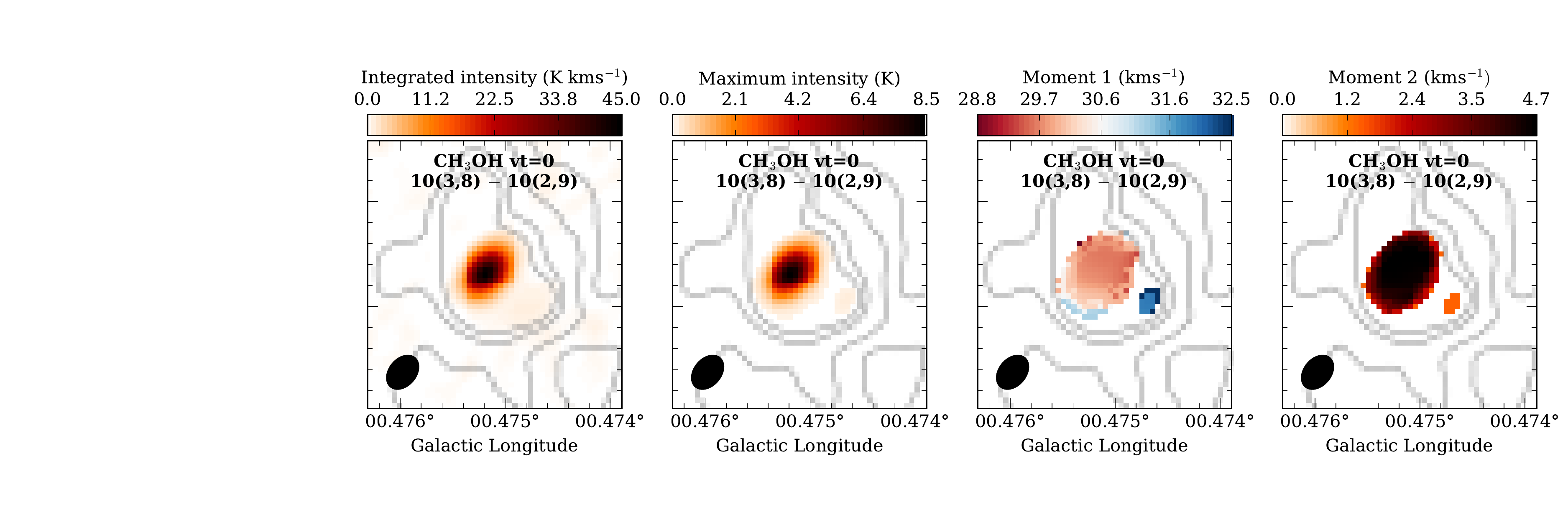} 
\includegraphics[trim = 3mm 26.5mm 3mm 21.5mm, clip,angle=0,width=1\textwidth]{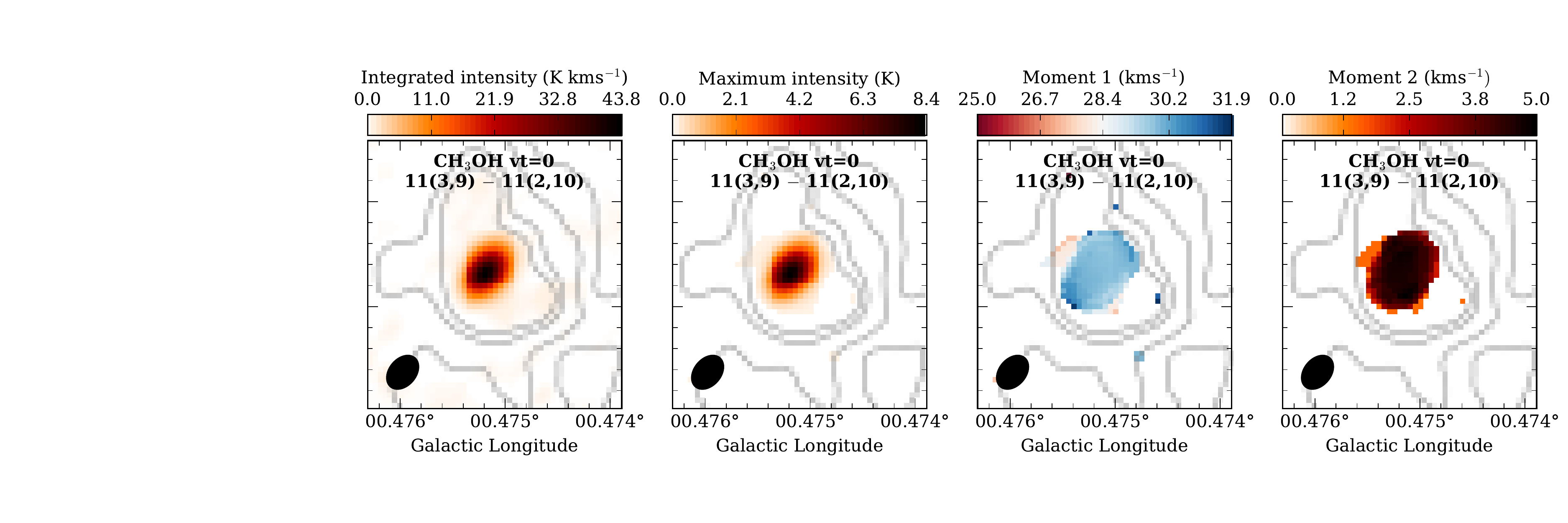} 
\includegraphics[trim = 3mm 26.5mm 3mm 21.5mm, clip,angle=0,width=1\textwidth]{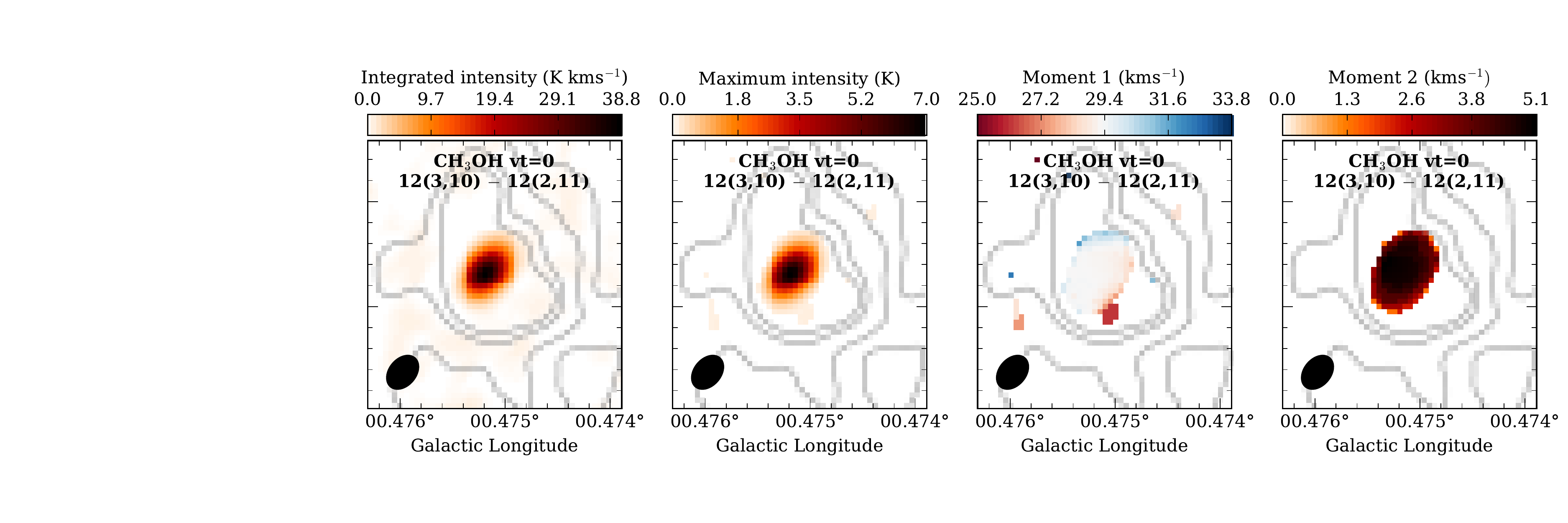} 
\includegraphics[trim = 3mm 26.5mm 3mm 21.5mm, clip,angle=0,width=1\textwidth]{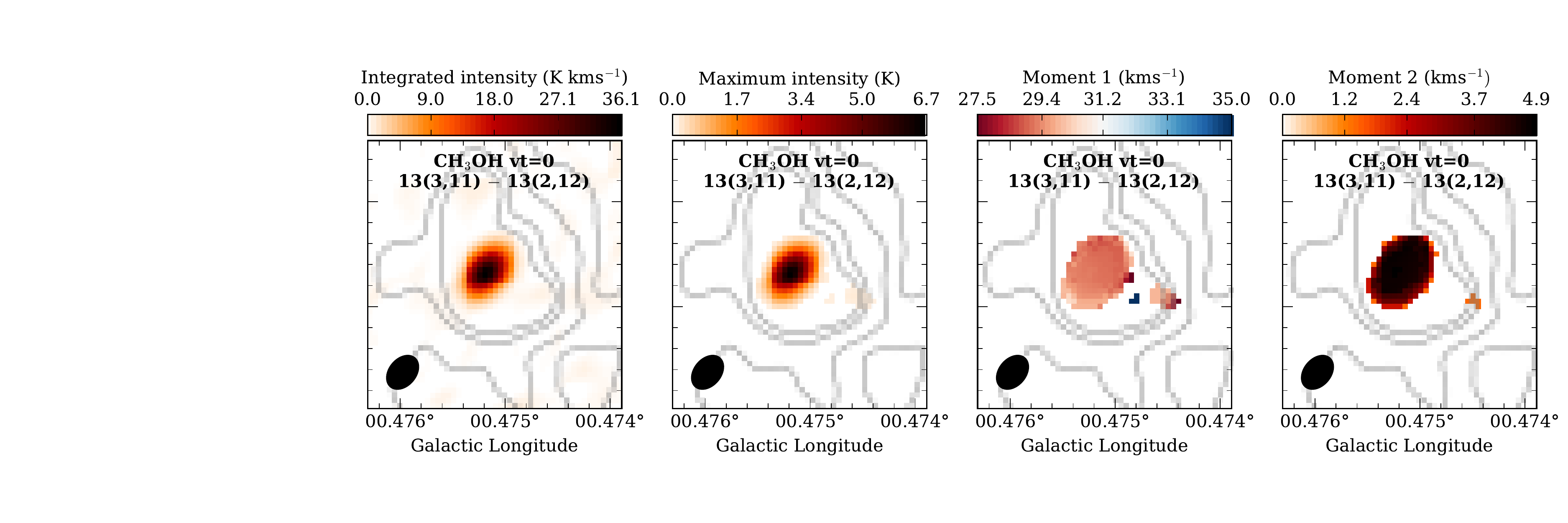} 
\includegraphics[trim = 3mm 14mm 3mm 21.5mm, clip,angle=0,width=1\textwidth]{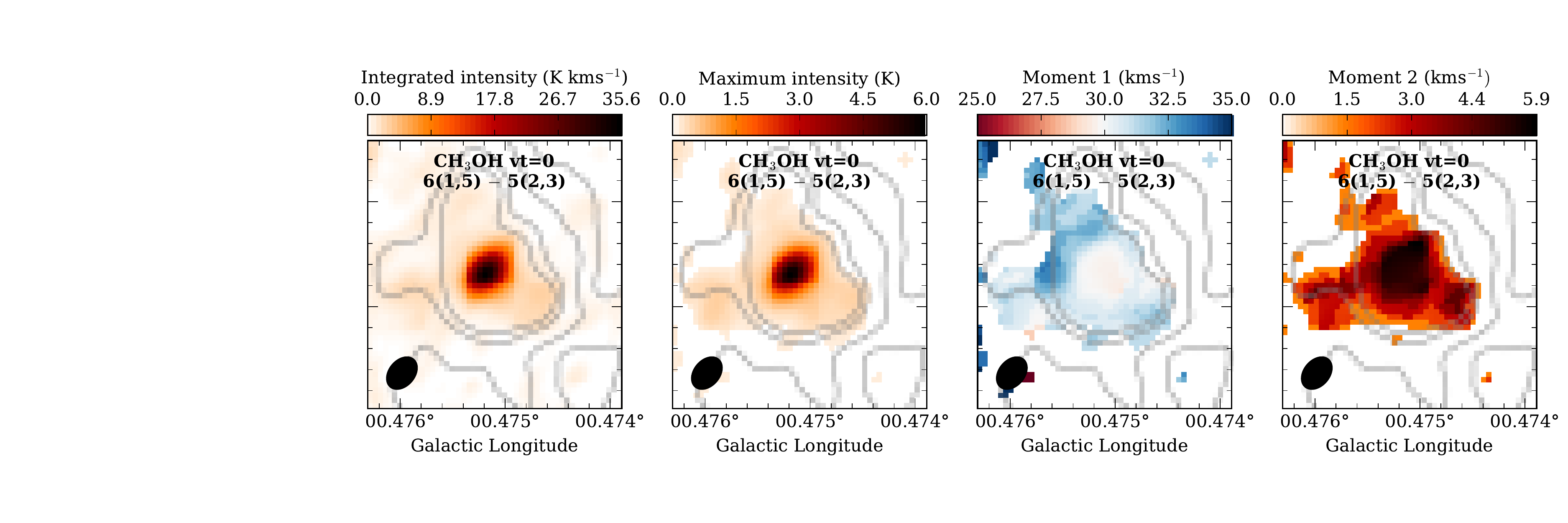} 
\contcaption{} 
\label{ } 
\end{figure*} 

\begin{figure*} 
\centering 
\includegraphics[trim = 3mm 26.5mm 3mm 15mm, clip,angle=0,width=1\textwidth]{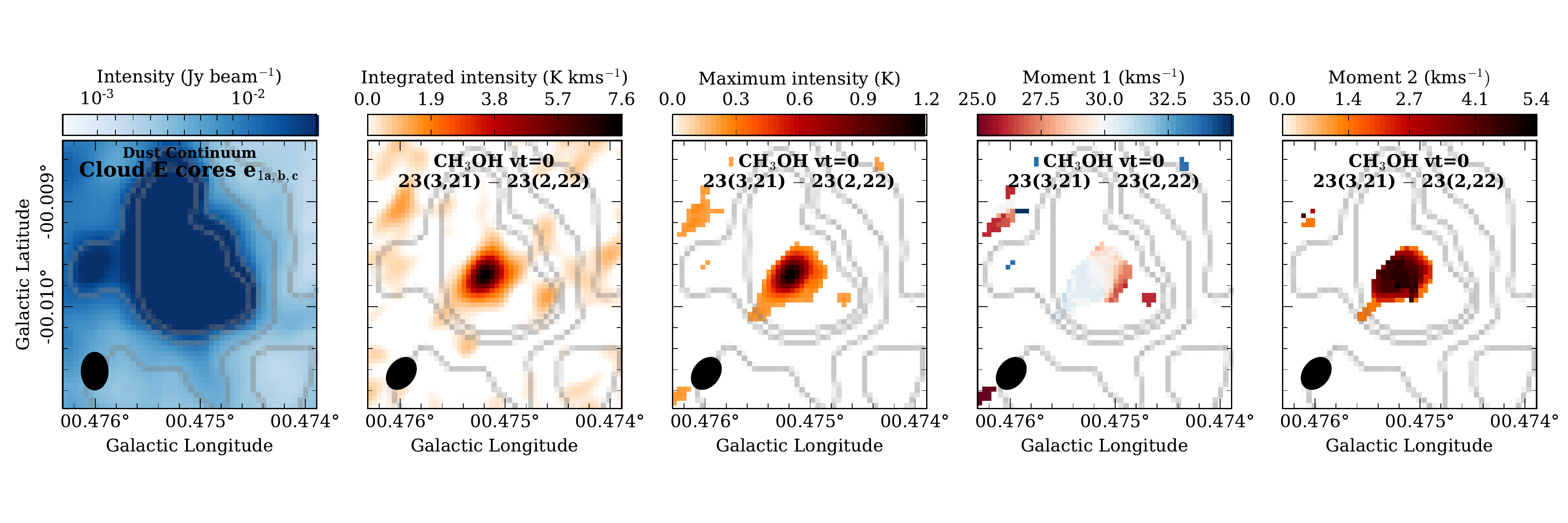} 
\includegraphics[trim = 3mm 26.5mm 3mm 21.5mm, clip,angle=0,width=1\textwidth]{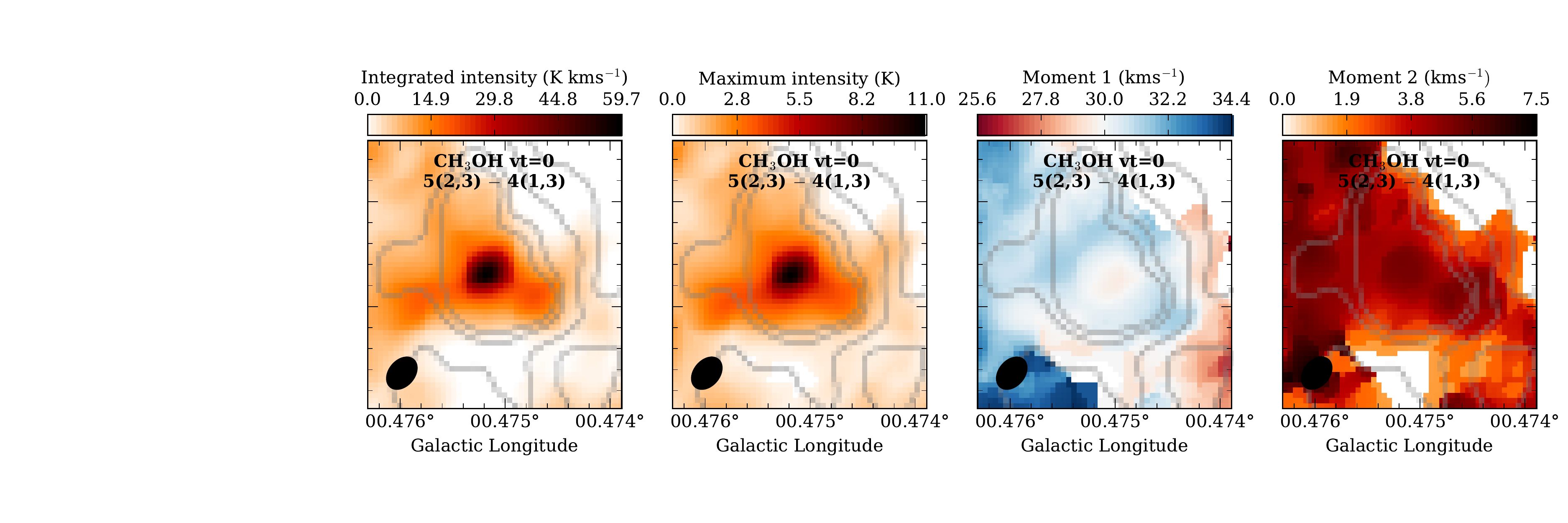} 
\includegraphics[trim = 3mm 26.5mm 3mm 21.5mm, clip,angle=0,width=1\textwidth]{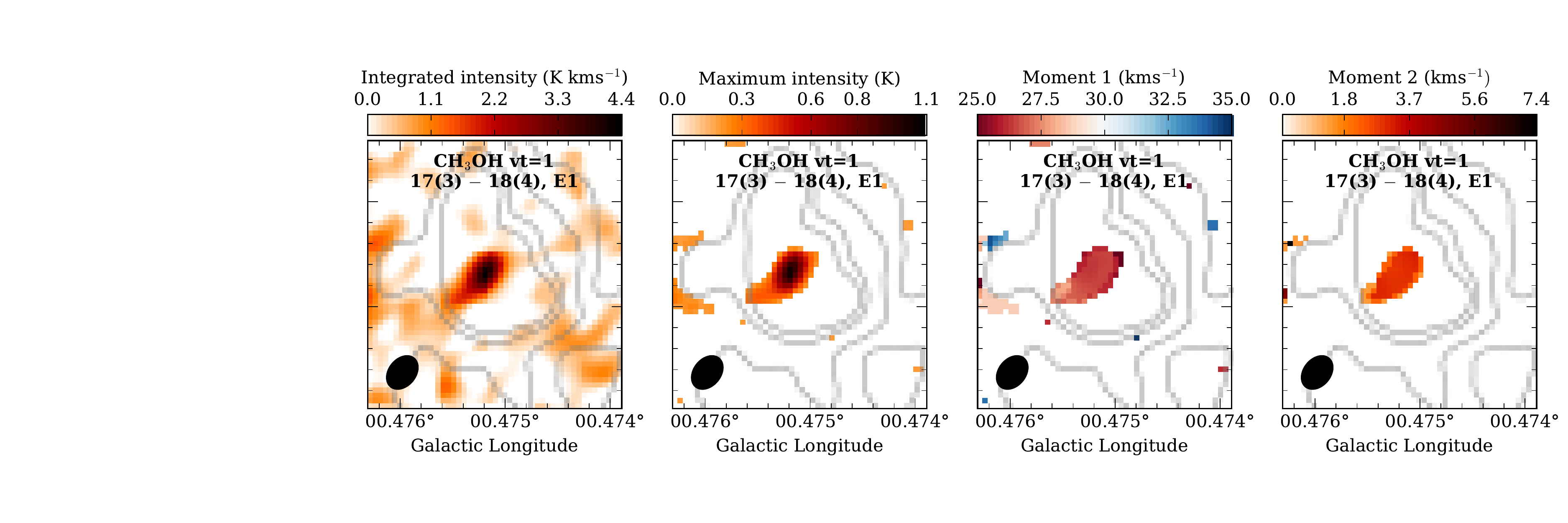} 
\includegraphics[trim = 3mm 26.5mm 3mm 21.5mm, clip,angle=0,width=1\textwidth]{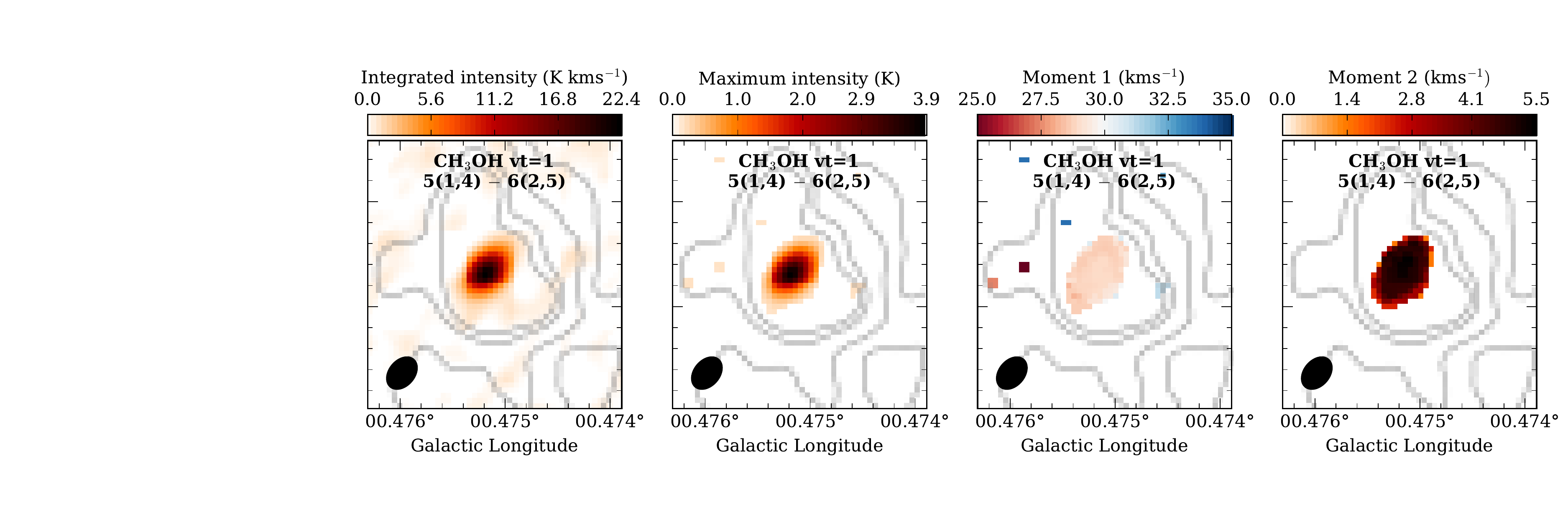} 
\includegraphics[trim = 3mm 26.5mm 3mm 21.5mm, clip,angle=0,width=1\textwidth]{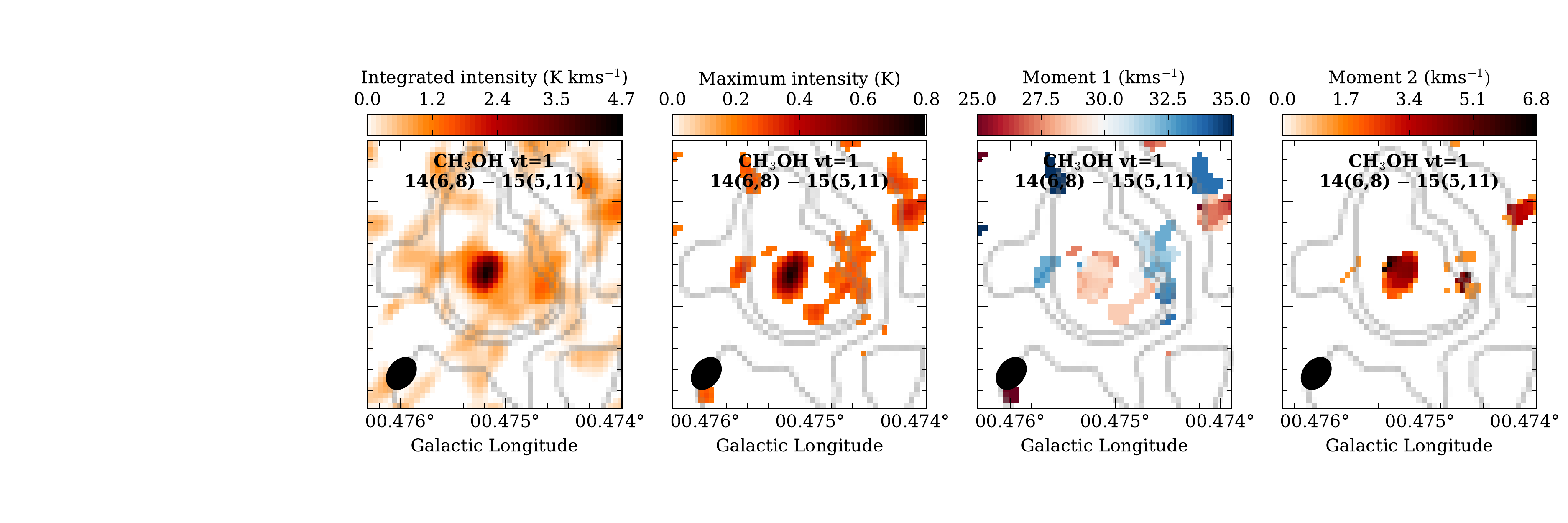} 
\includegraphics[trim = 3mm 14mm 3mm 21.5mm, clip,angle=0,width=1\textwidth]{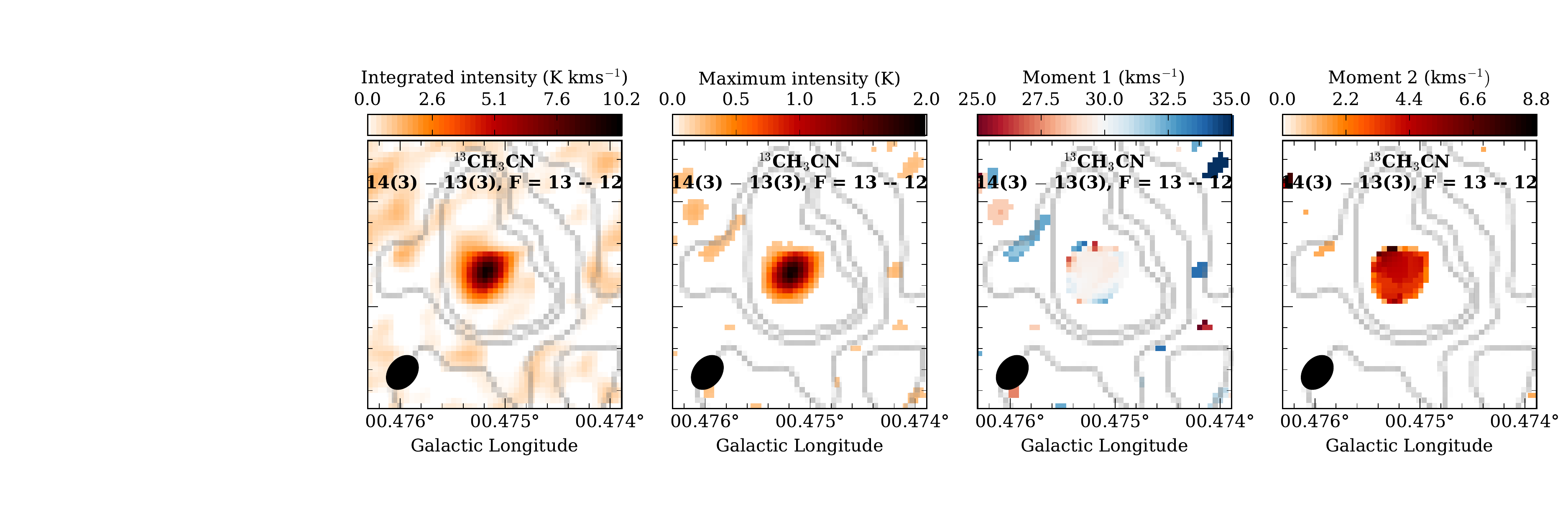} 
\contcaption{} 
\label{ } 
\end{figure*} 

\begin{figure*} 
\centering 
\includegraphics[trim = 3mm 26.5mm 3mm 15mm, clip,angle=0,width=1\textwidth]{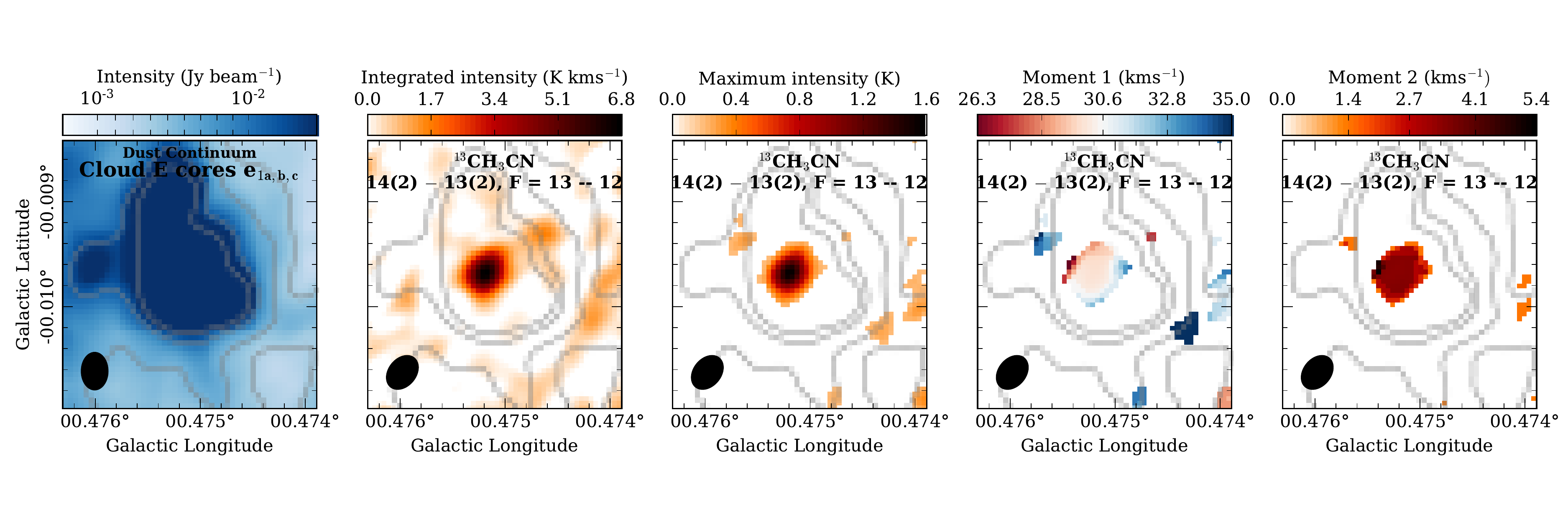} 
\includegraphics[trim = 3mm 26.5mm 3mm 21.5mm, clip,angle=0,width=1\textwidth]{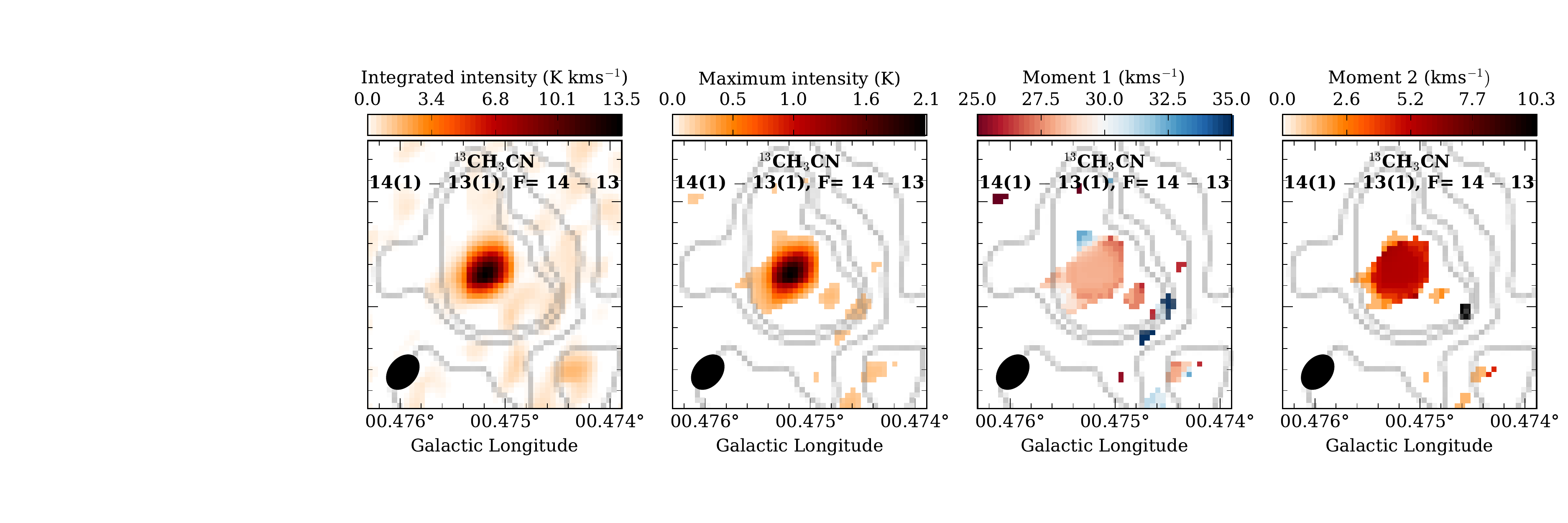} 
\includegraphics[trim = 3mm 26.5mm 3mm 21.5mm, clip,angle=0,width=1\textwidth]{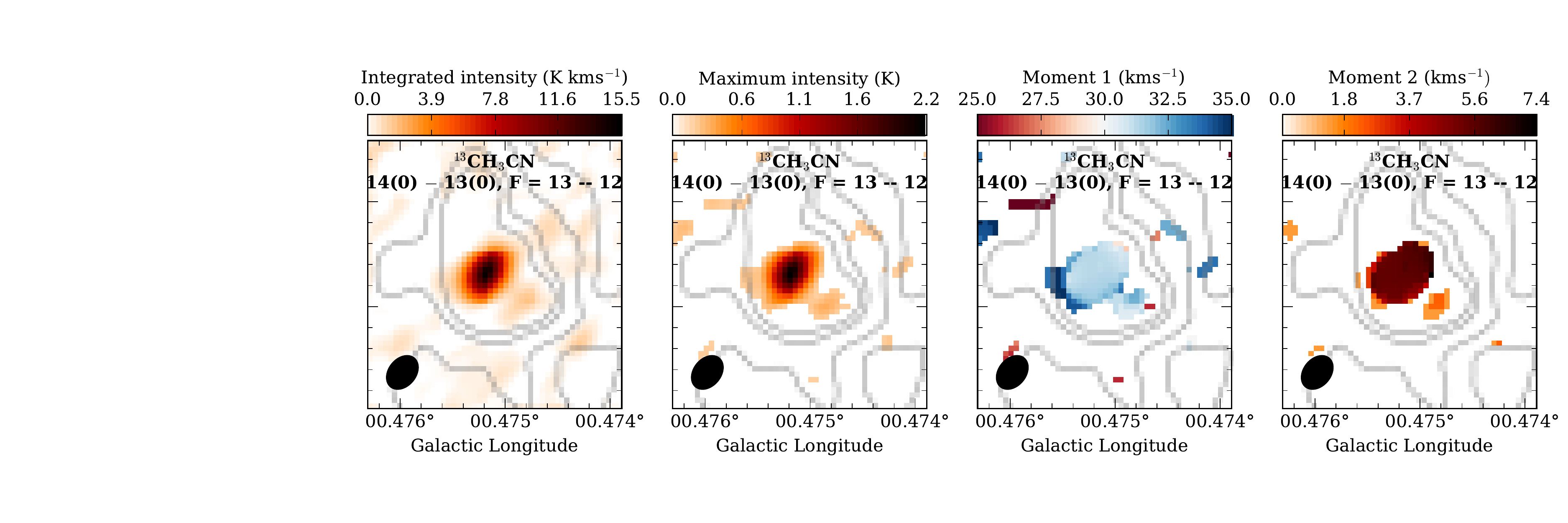} 
\includegraphics[trim = 3mm 26.5mm 3mm 21.5mm, clip,angle=0,width=1\textwidth]{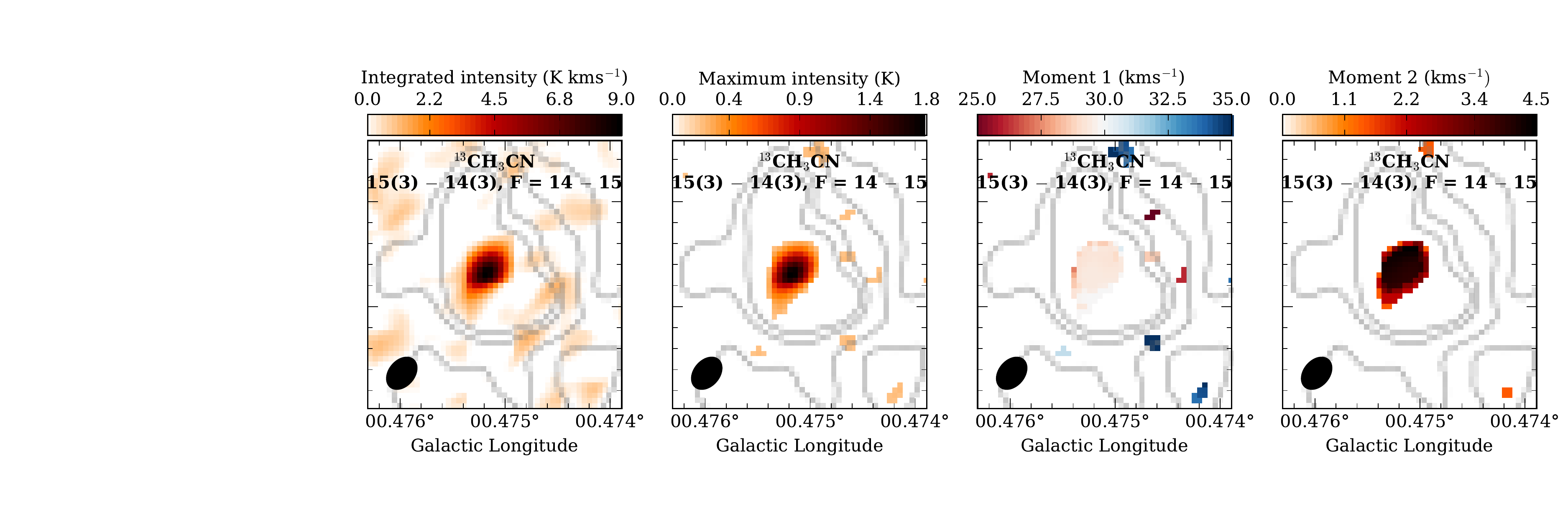} 
\includegraphics[trim = 3mm 26.5mm 3mm 21.5mm, clip,angle=0,width=1\textwidth]{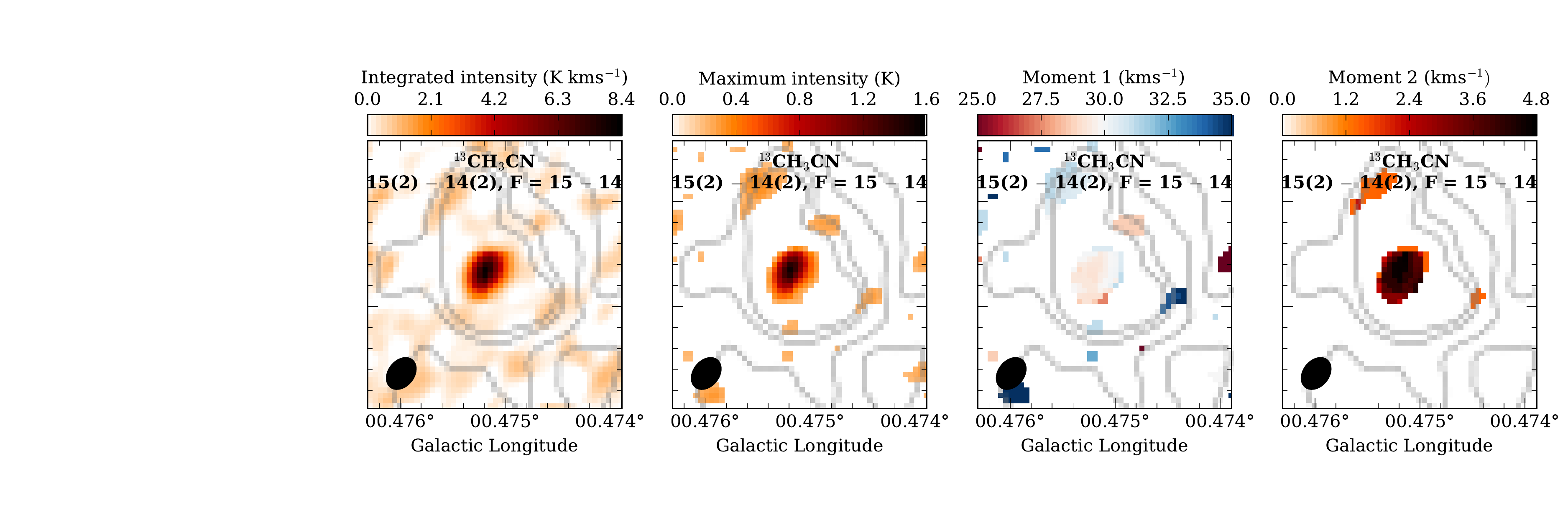} 
\includegraphics[trim = 3mm 14mm 3mm 21.5mm, clip,angle=0,width=1\textwidth]{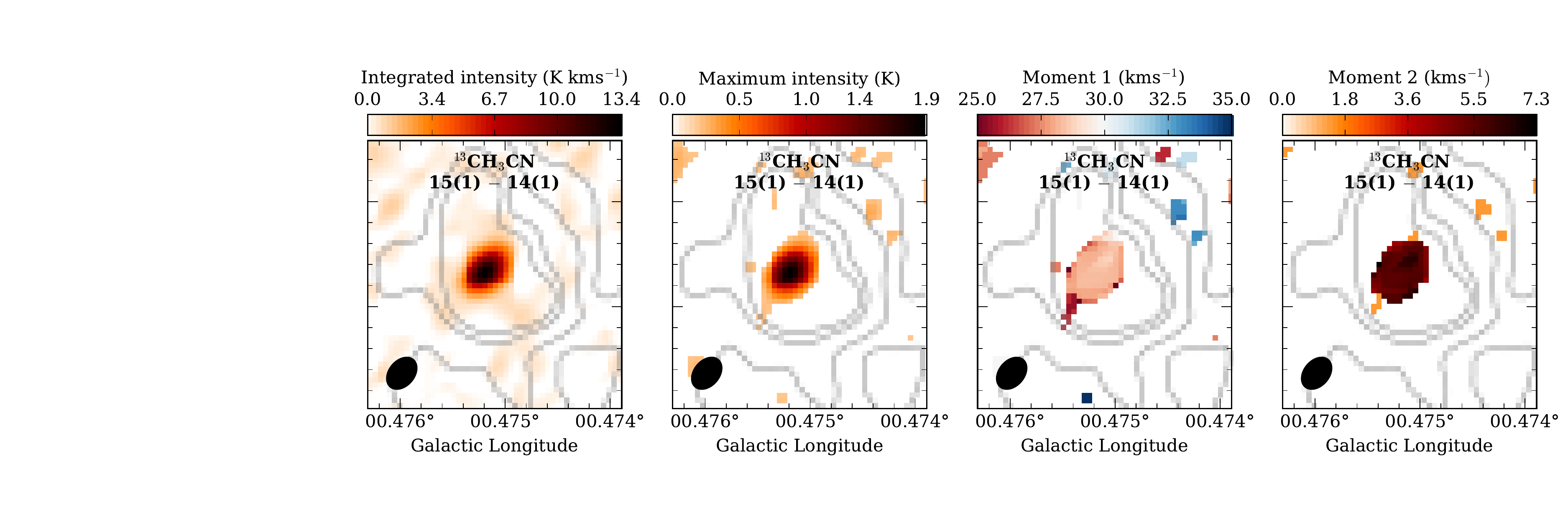} 
\contcaption{} 
\label{ } 
\end{figure*} 

\begin{figure*} 
\centering 
\includegraphics[trim = 3mm 26.5mm 3mm 15mm, clip,angle=0,width=1\textwidth]{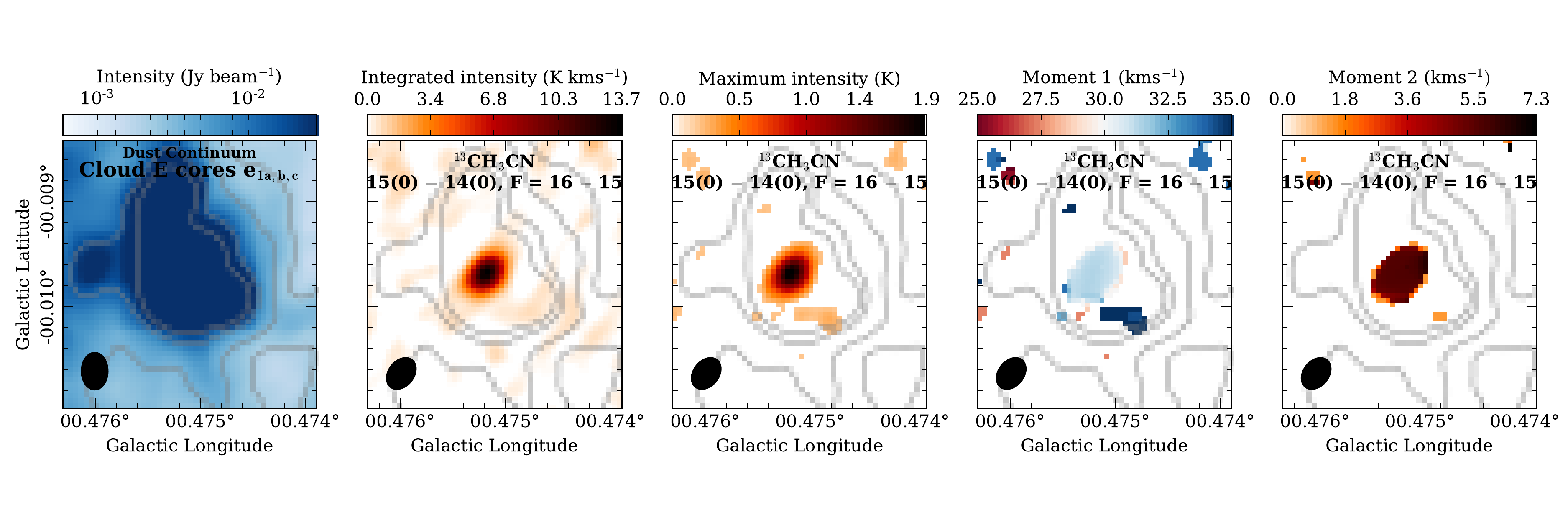} 
\includegraphics[trim = 3mm 26.5mm 3mm 21.5mm, clip,angle=0,width=1\textwidth]{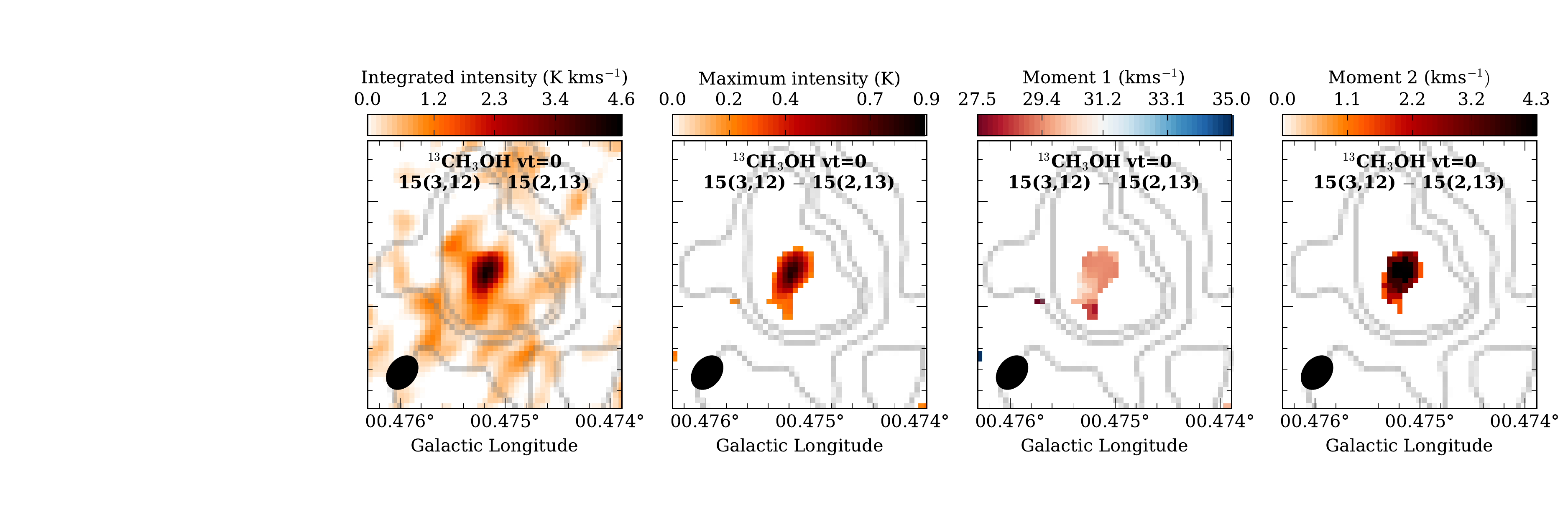} 
\includegraphics[trim = 3mm 26.5mm 3mm 21.5mm, clip,angle=0,width=1\textwidth]{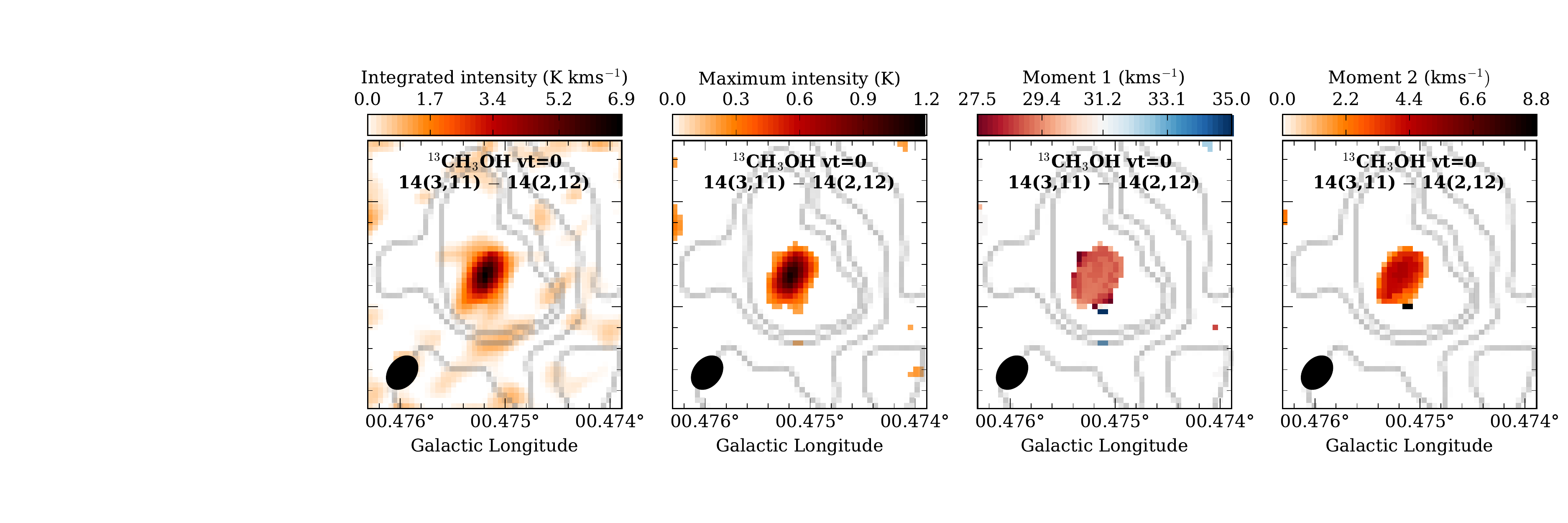} 
\includegraphics[trim = 3mm 26.5mm 3mm 21.5mm, clip,angle=0,width=1\textwidth]{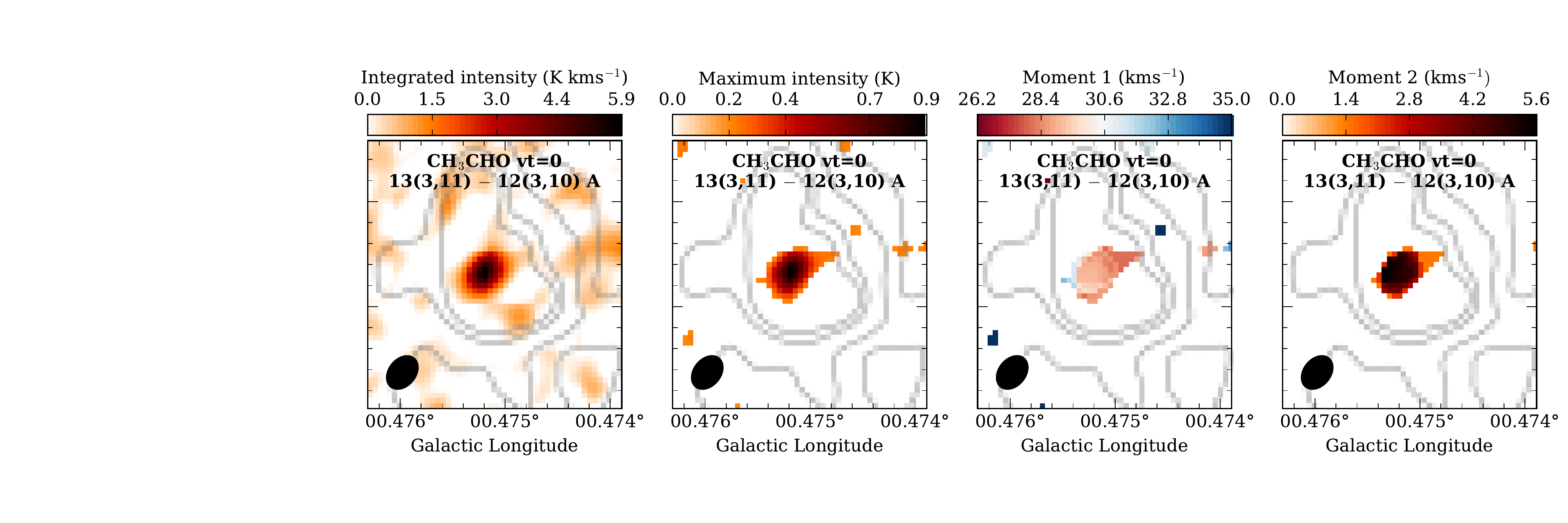} 
\includegraphics[trim = 3mm 26.5mm 3mm 21.5mm, clip,angle=0,width=1\textwidth]{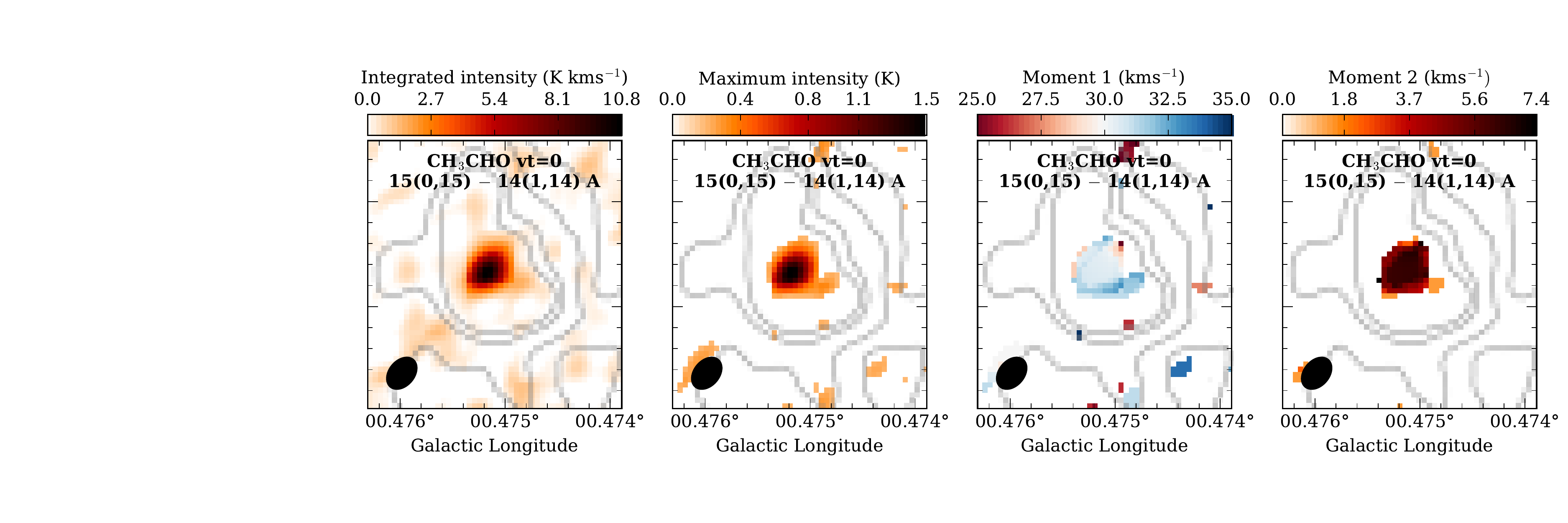} 
\includegraphics[trim = 3mm 14mm 3mm 21.5mm, clip,angle=0,width=1\textwidth]{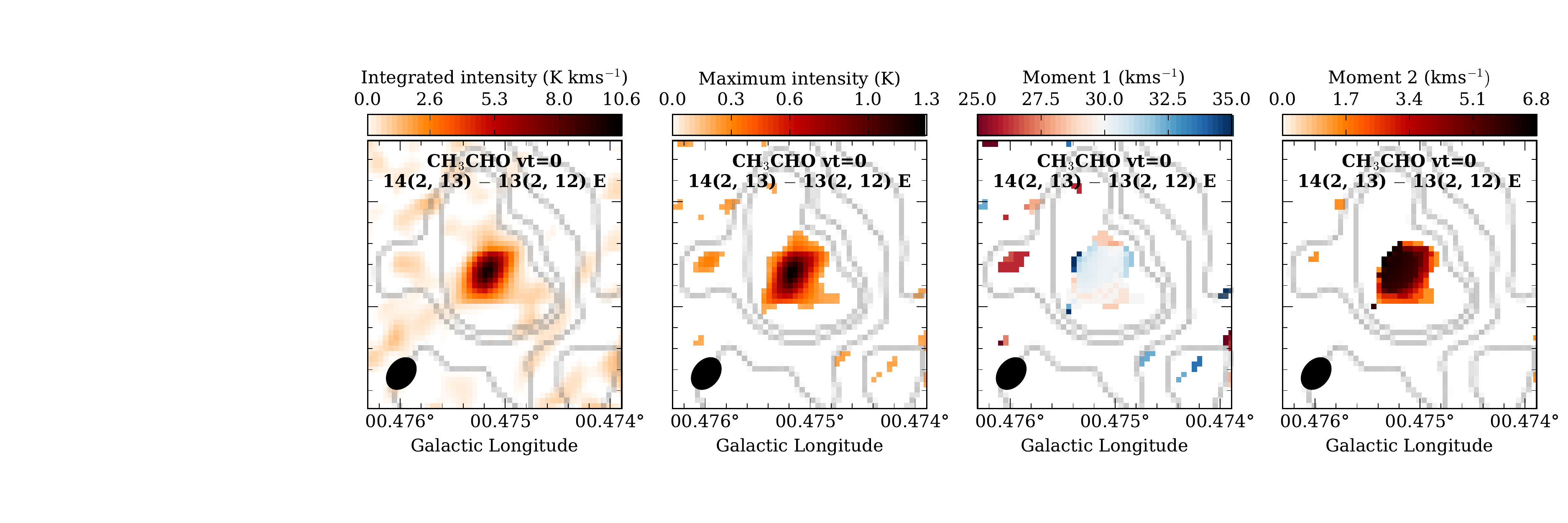} 
\contcaption{} 
\label{ } 
\end{figure*} 

\begin{figure*} 
\centering 
\includegraphics[trim = 3mm 26.5mm 3mm 15mm, clip,angle=0,width=1\textwidth]{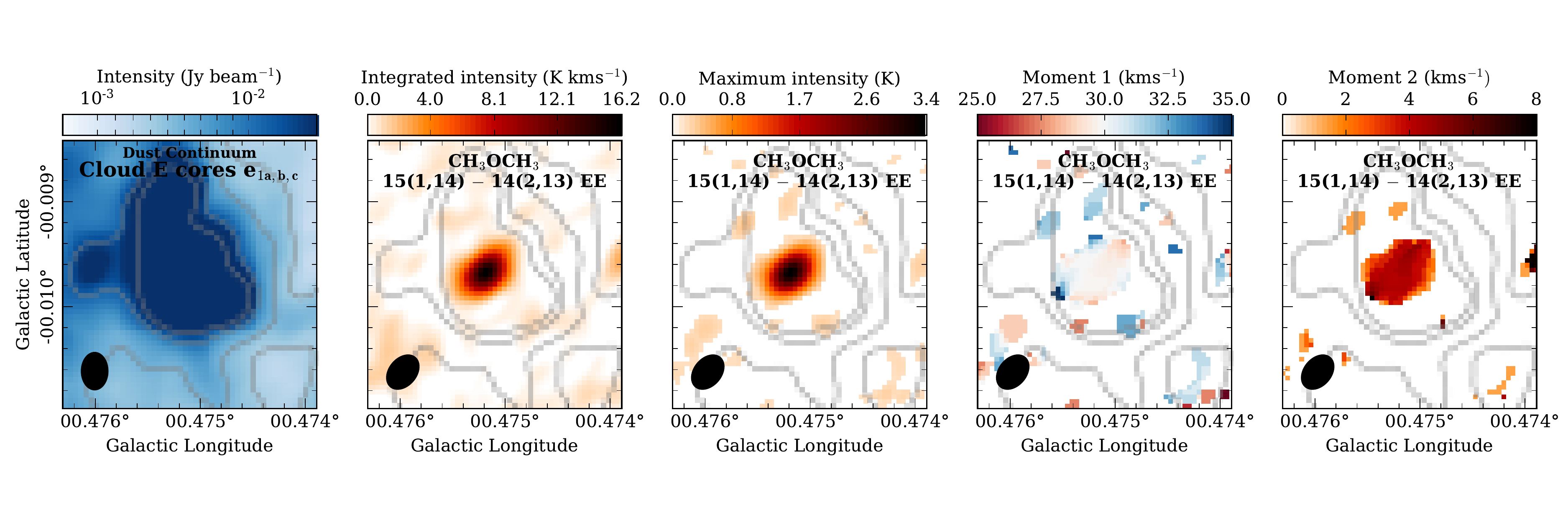} 
\includegraphics[trim = 3mm 26.5mm 3mm 21.5mm, clip,angle=0,width=1\textwidth]{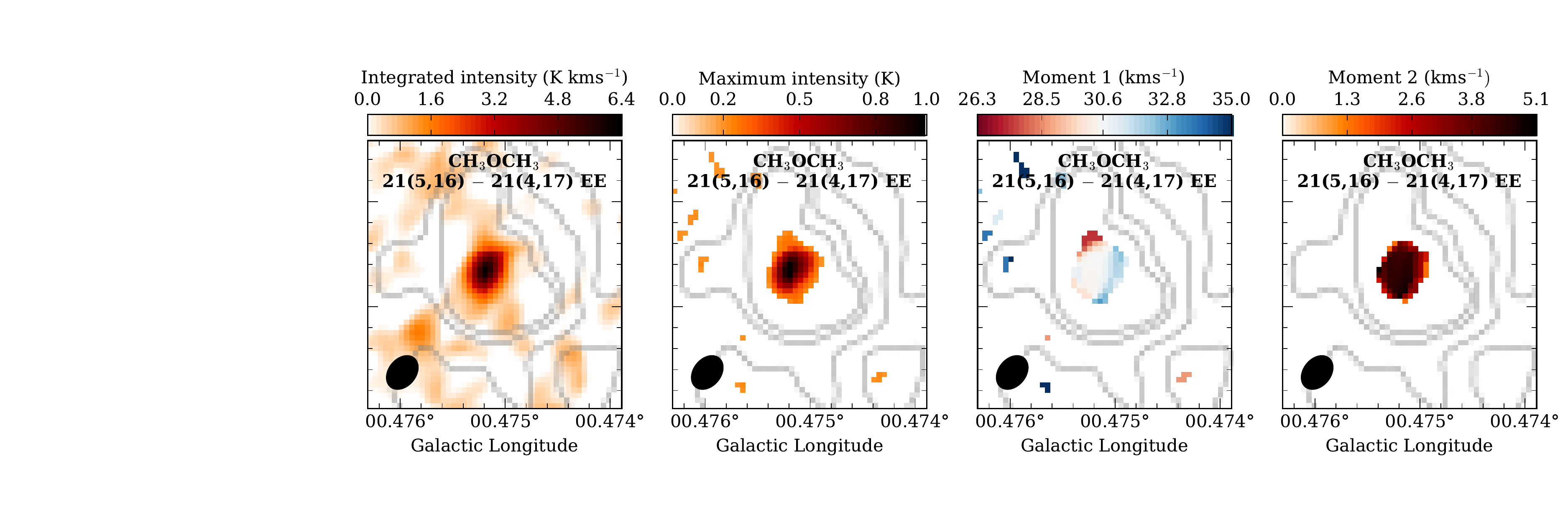} 
\includegraphics[trim = 3mm 26.5mm 3mm 21.5mm, clip,angle=0,width=1\textwidth]{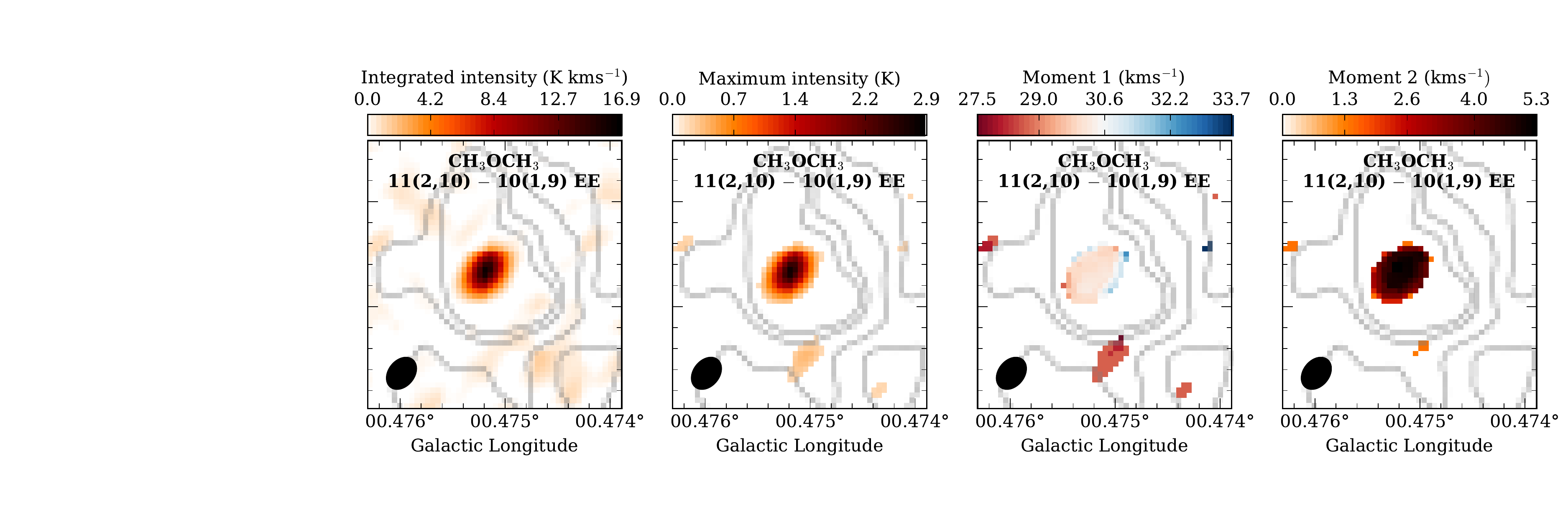} 
\includegraphics[trim = 3mm 26.5mm 3mm 21.5mm, clip,angle=0,width=1\textwidth]{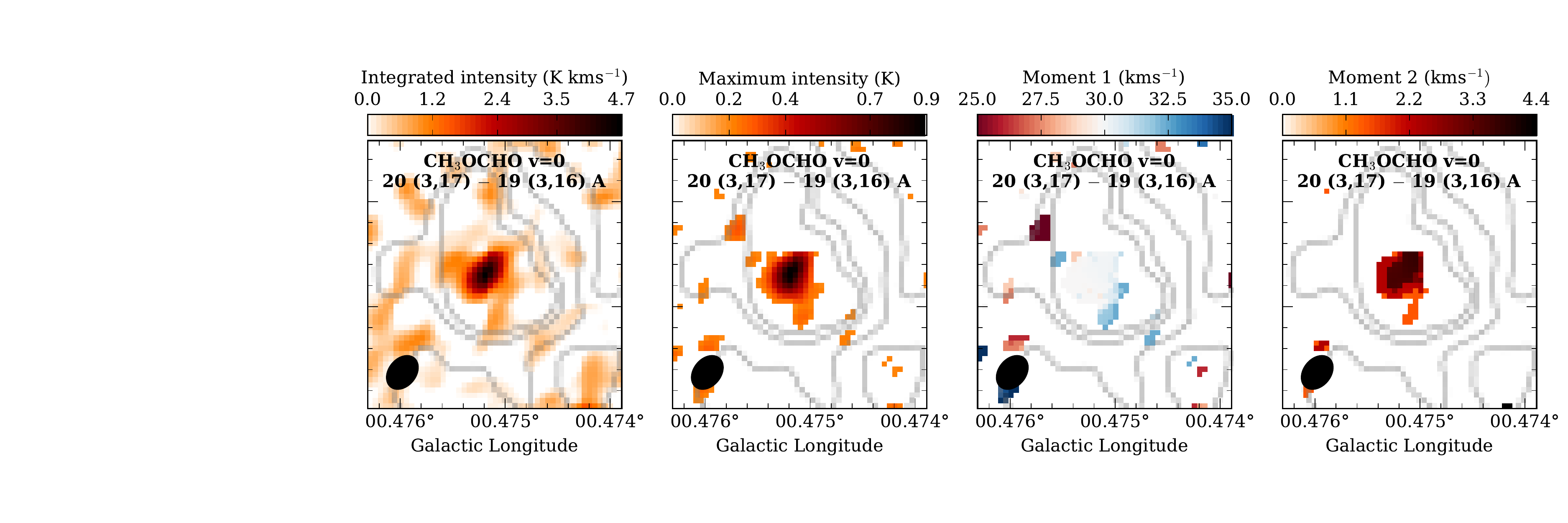} 
\includegraphics[trim = 3mm 26.5mm 3mm 21.5mm, clip,angle=0,width=1\textwidth]{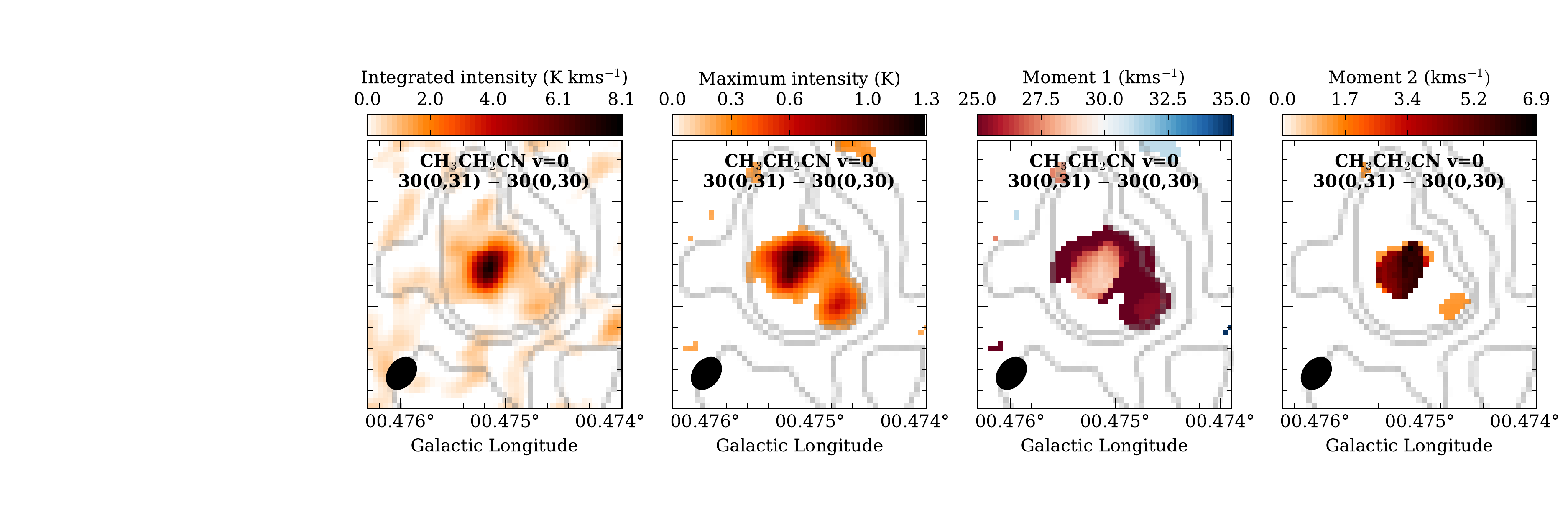} 
\includegraphics[trim = 3mm 14mm 3mm 21.5mm, clip,angle=0,width=1\textwidth]{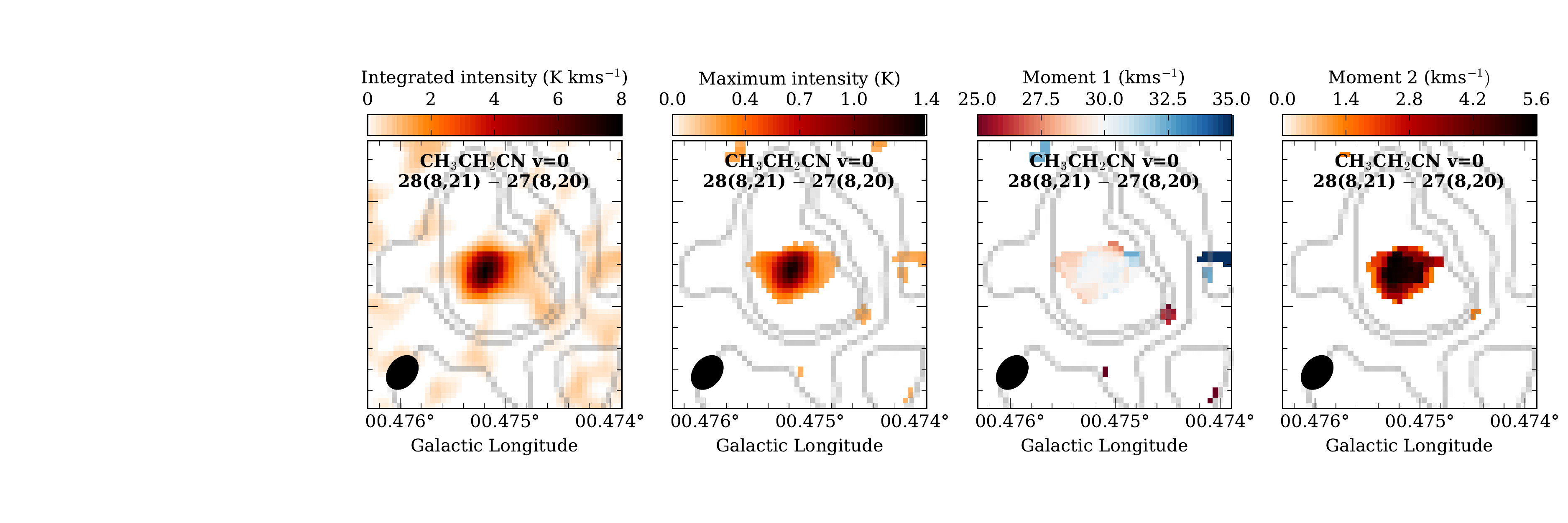} 
\contcaption{} 
\label{ } 
\end{figure*} 

\begin{figure*} 
\centering 
\includegraphics[trim = 3mm 26.5mm 3mm 15mm, clip,angle=0,width=1\textwidth]{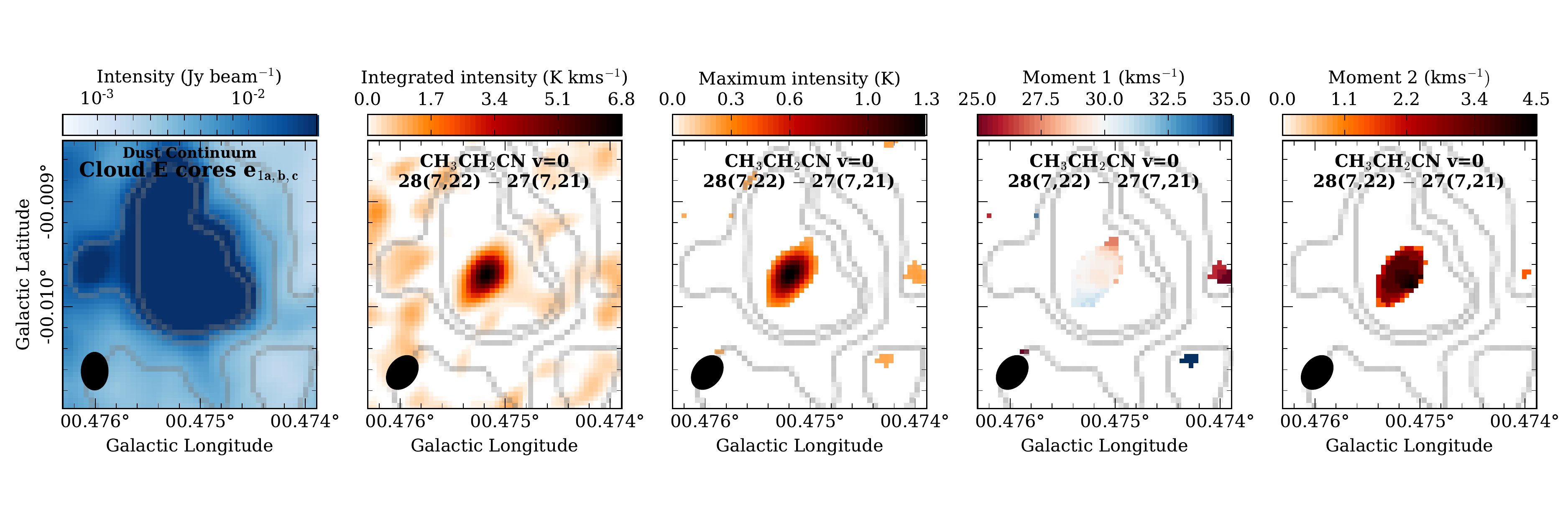} 
\includegraphics[trim = 3mm 26.5mm 3mm 21.5mm, clip,angle=0,width=1\textwidth]{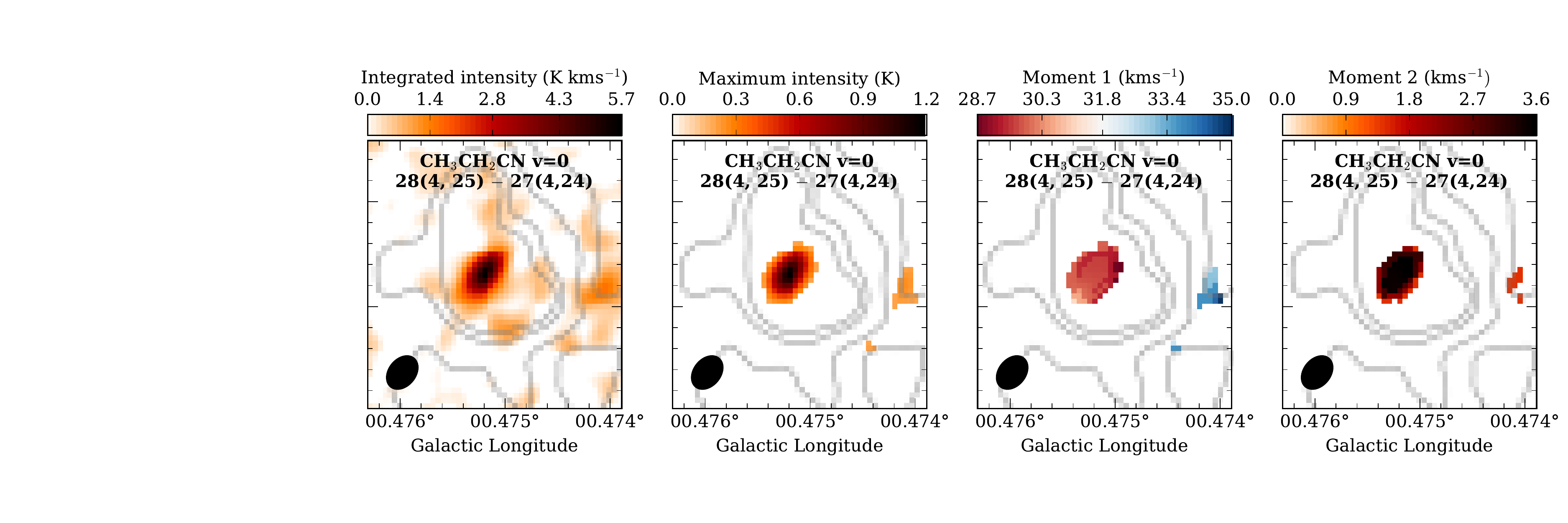} 
\includegraphics[trim = 3mm 26.5mm 3mm 21.5mm, clip,angle=0,width=1\textwidth]{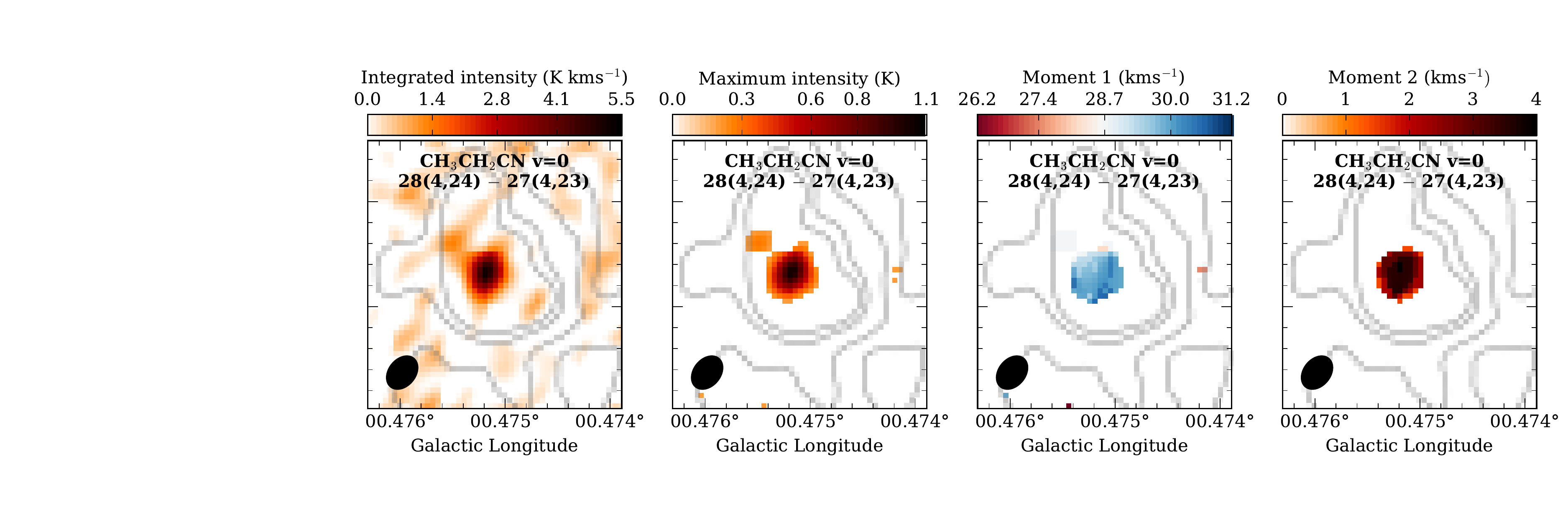} 
\includegraphics[trim = 3mm 26.5mm 3mm 21.5mm, clip,angle=0,width=1\textwidth]{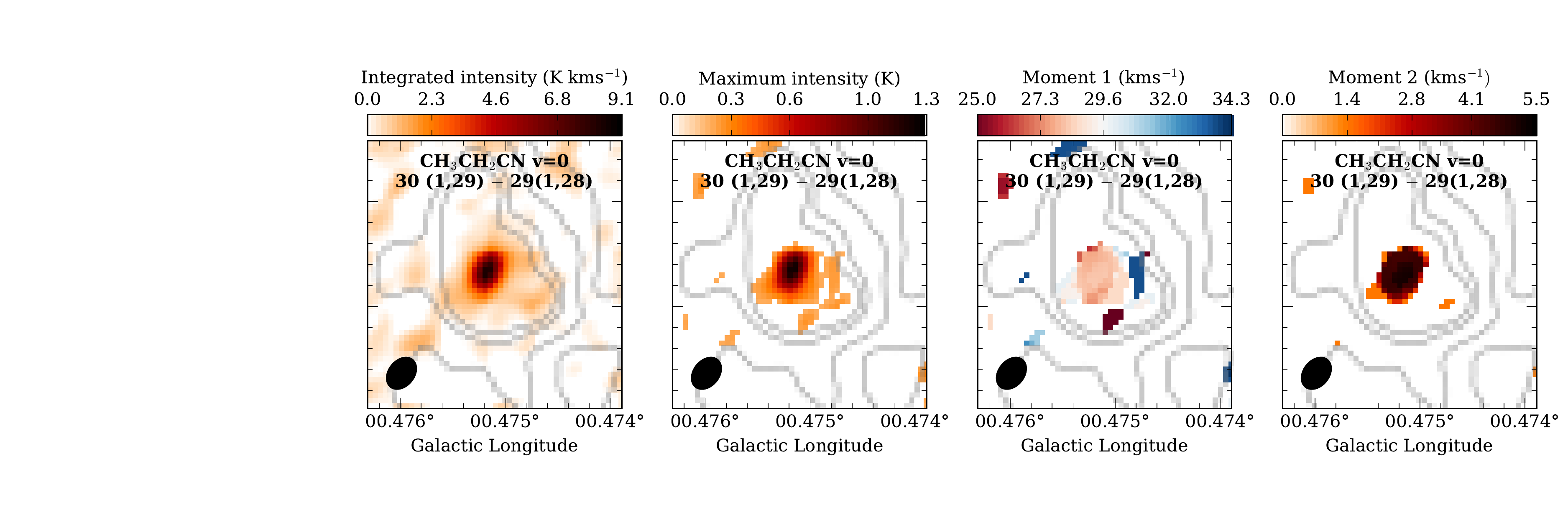} 
\includegraphics[trim = 3mm 14mm 3mm 21.5mm, clip,angle=0,width=1\textwidth]{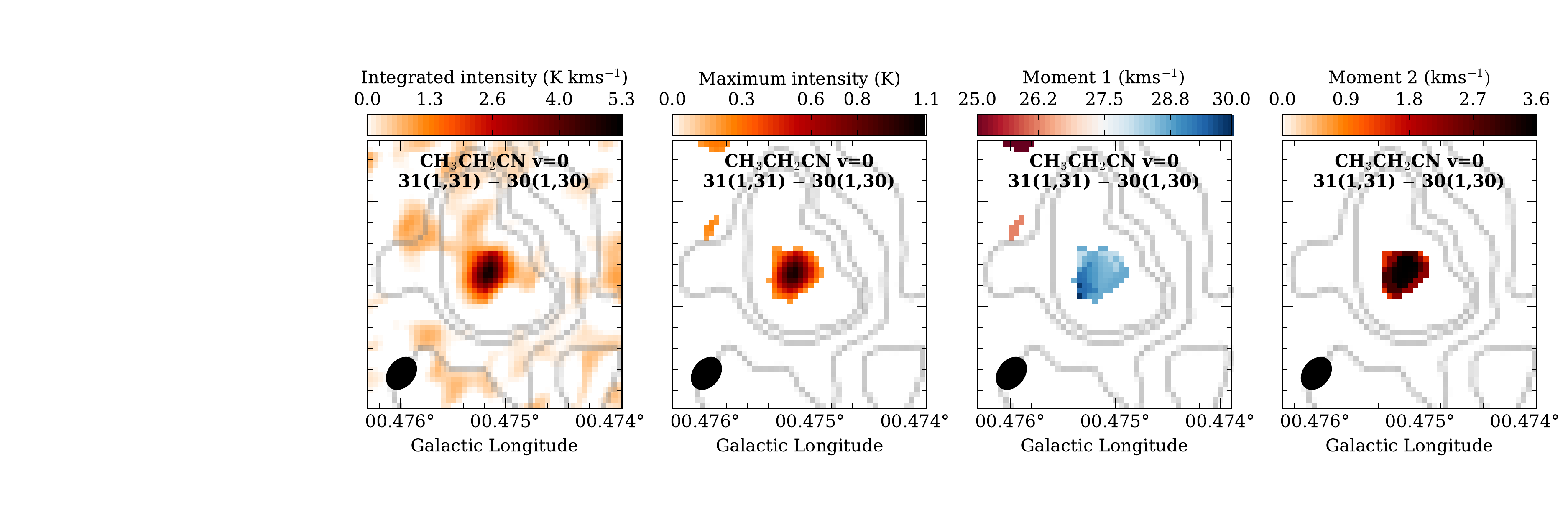} 

\contcaption{}
\label{ }
\end{figure*}

\begin{figure*}
\centering
\includegraphics[trim =  3mm 28mm 2mm 15mm, clip,angle=0,width=1\textwidth]{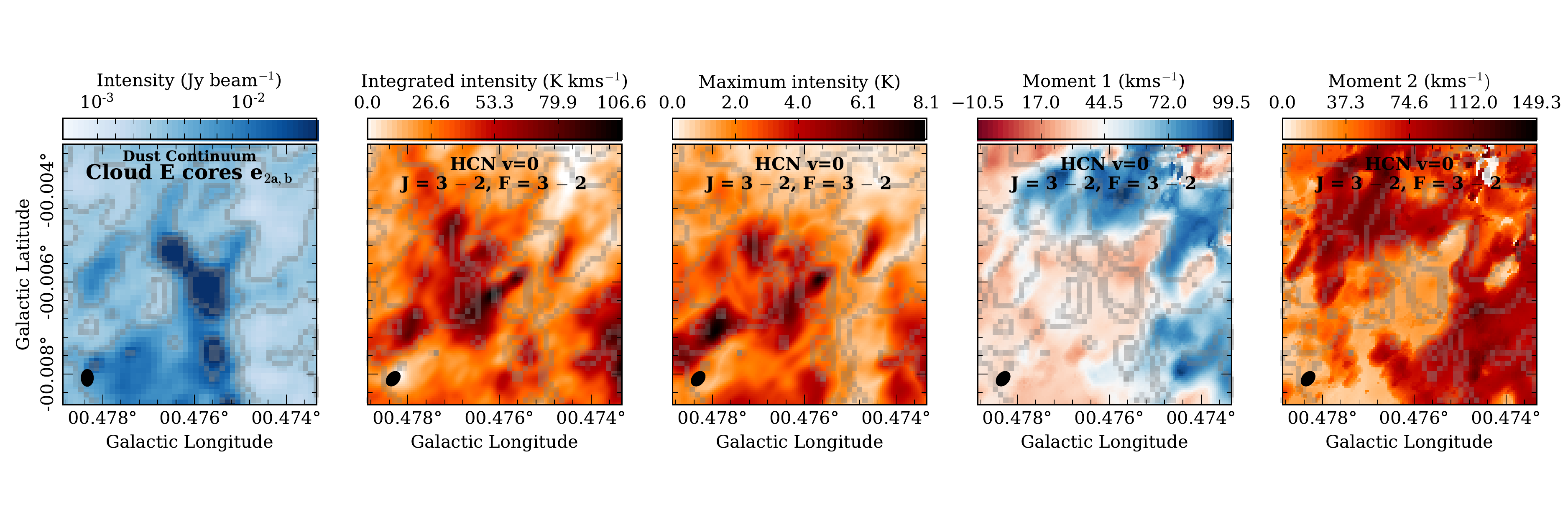}
\includegraphics[trim = 3mm 28mm 2mm 22mm, clip,angle=0,width=1\textwidth]{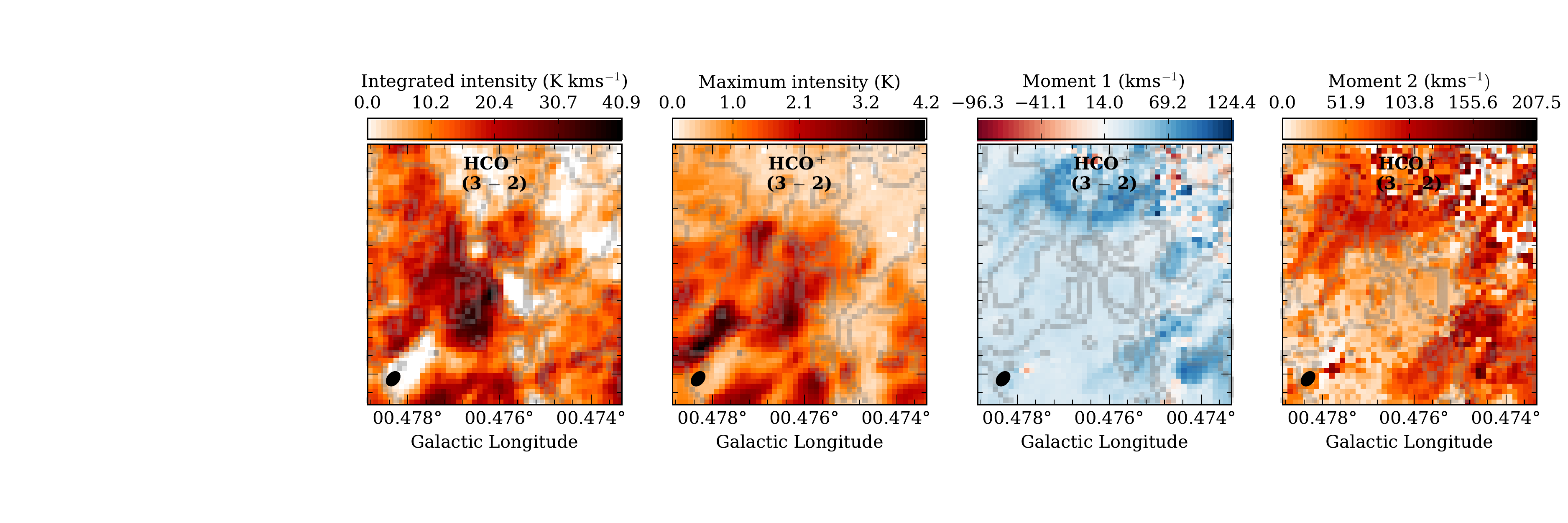}
\includegraphics[trim = 3mm 28mm 3mm 22mm, clip,angle=0,width=1\textwidth]{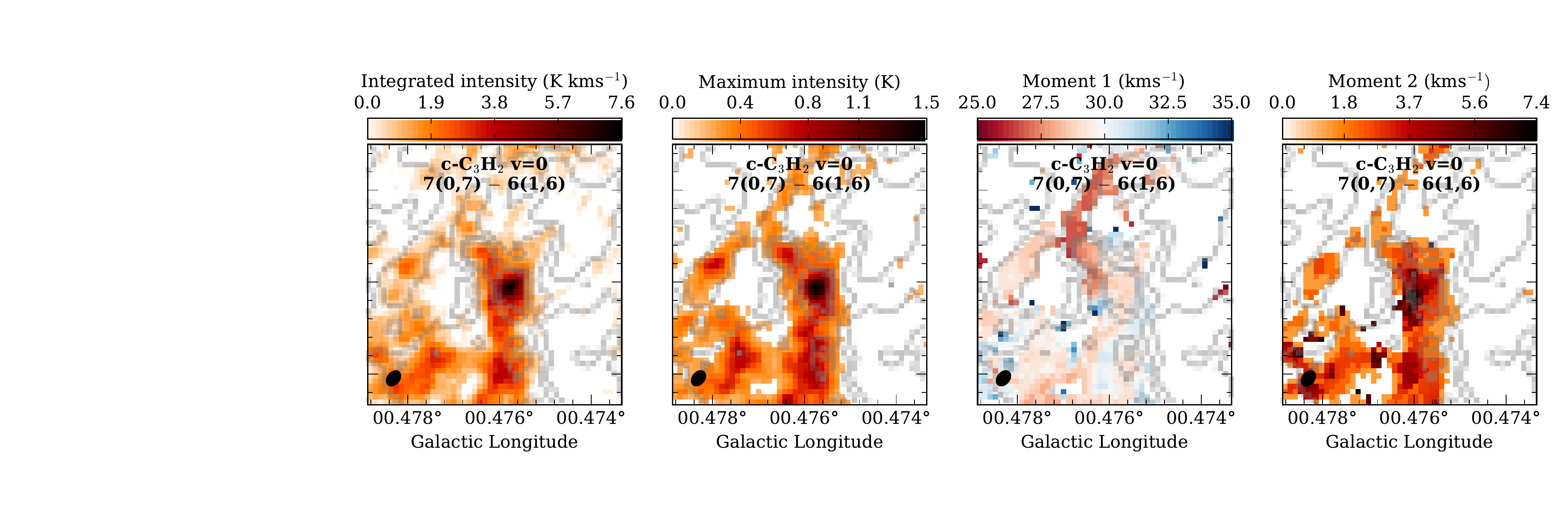}
\includegraphics[trim = 3mm 14mm 3mm 22mm, clip,angle=0,width=1\textwidth]{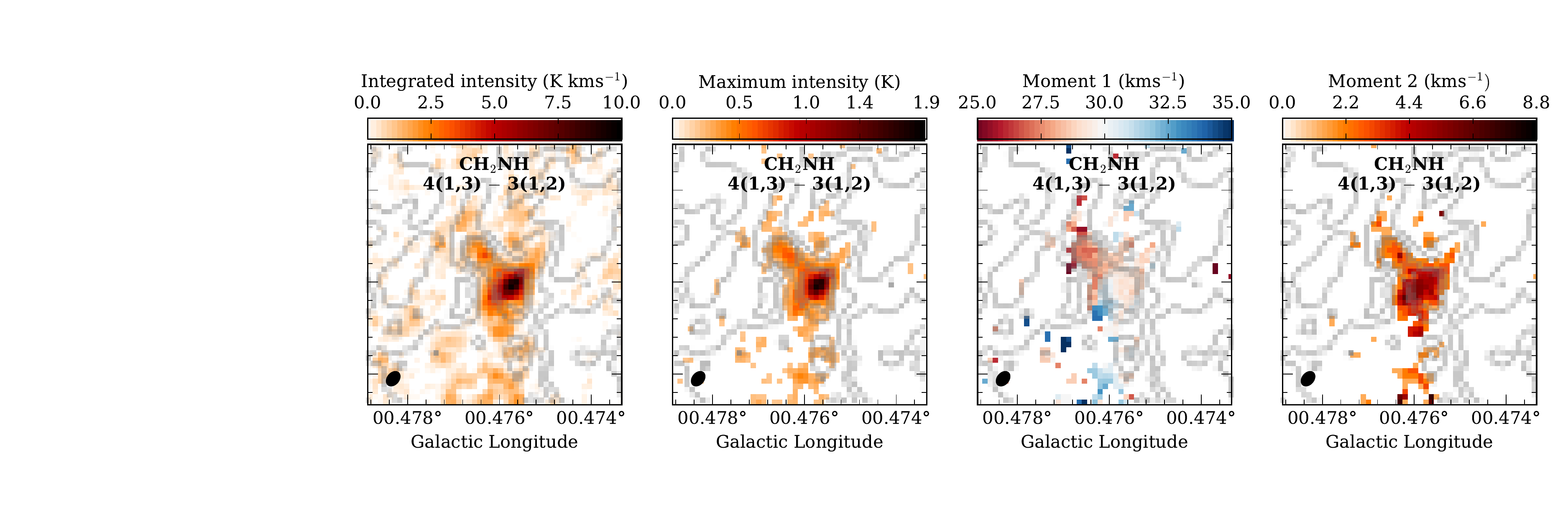}

\caption{Moment maps of the molecular transitions towards Cloud E/F (see table\,\ref{line_idents1}). Show is a zoom-in of the region containing cores e2$_{\rm a}$, e2$_{\rm b}$ (see section\,\ref{virial_cores}). The analysis for the different molecular transitions are presented in each row, with the molecule labeled at the top of each map. Shown in the upper left is the combined 12m, 7m and single dish continuum map, and then from left to right are moment maps of the integrated intensity, peak intensity, intensity weighted centroid velocity, and intensity weighted velocity dispersion for each molecule. Contours on each map are of the continuum shown in levels of [8, 15, 30, 50]\,$\sigma_{\rm rms}$, where $\sigma_{\rm rms}$\,$\sim$\,0.6\,mJy\,beam$^{-1}$.} 
\label{ }
\end{figure*}

\begin{figure*}
\centering

\includegraphics[trim = 3mm 28mm 3mm 15mm, clip,angle=0,width=1\textwidth]{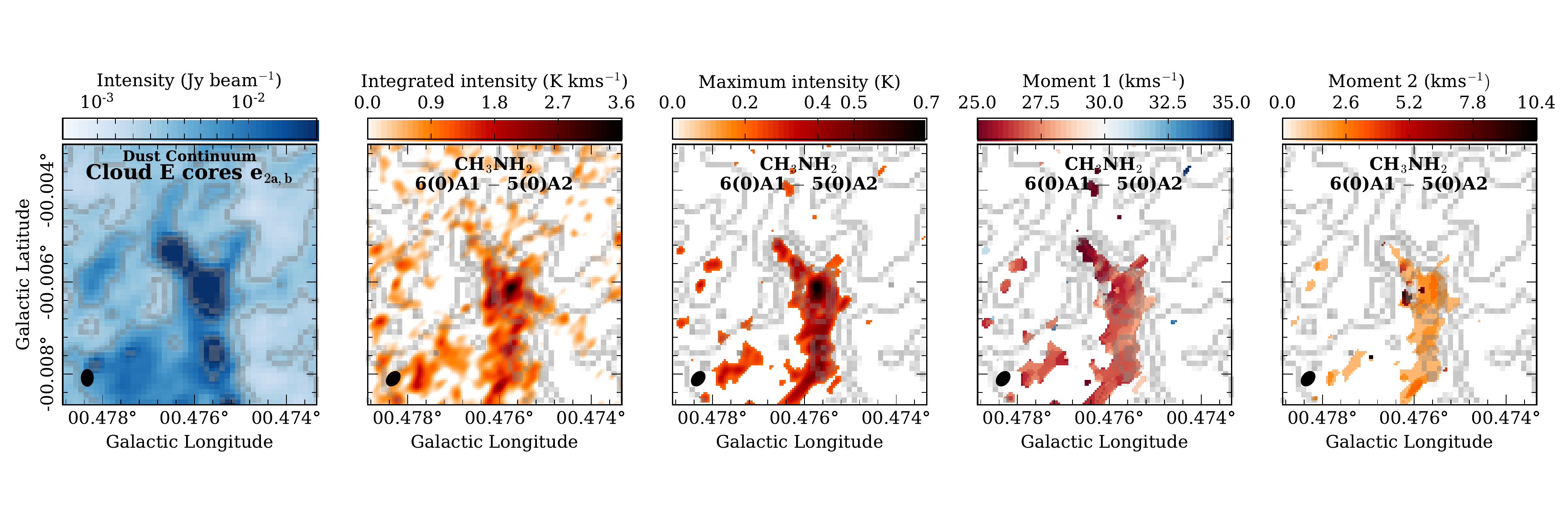}
\includegraphics[trim = 3mm 28mm 3mm 22mm, clip,angle=0,width=1\textwidth]{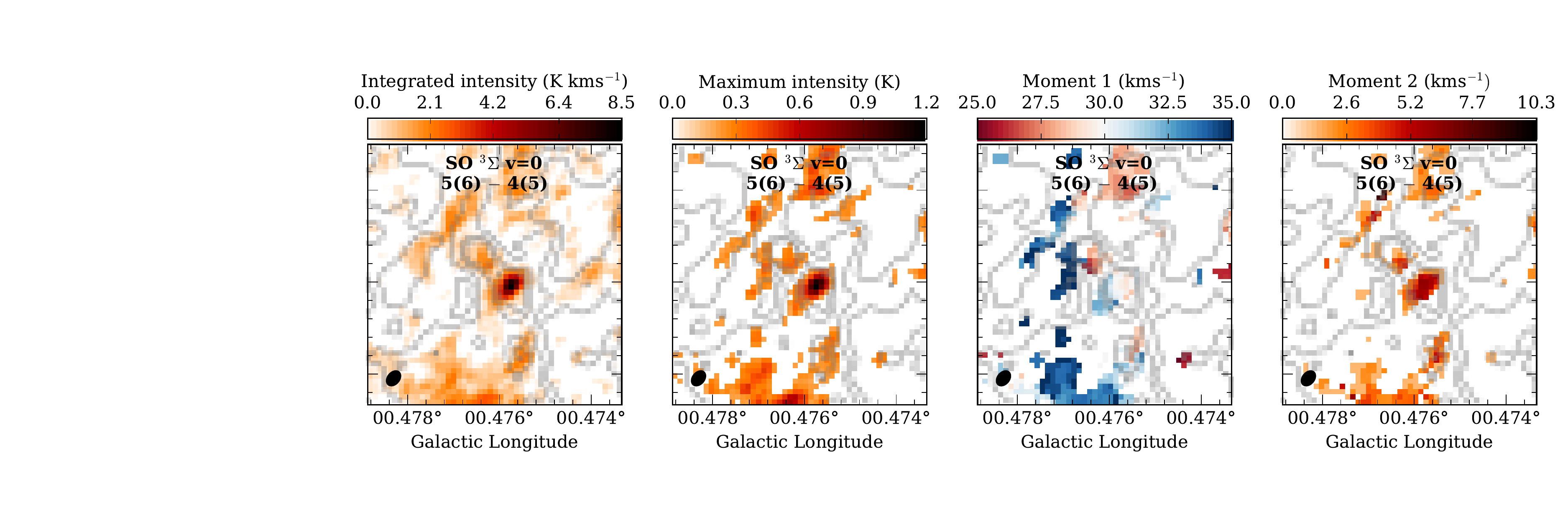}
\includegraphics[trim = 3mm 28mm 3mm 22mm, clip,angle=0,width=1\textwidth]{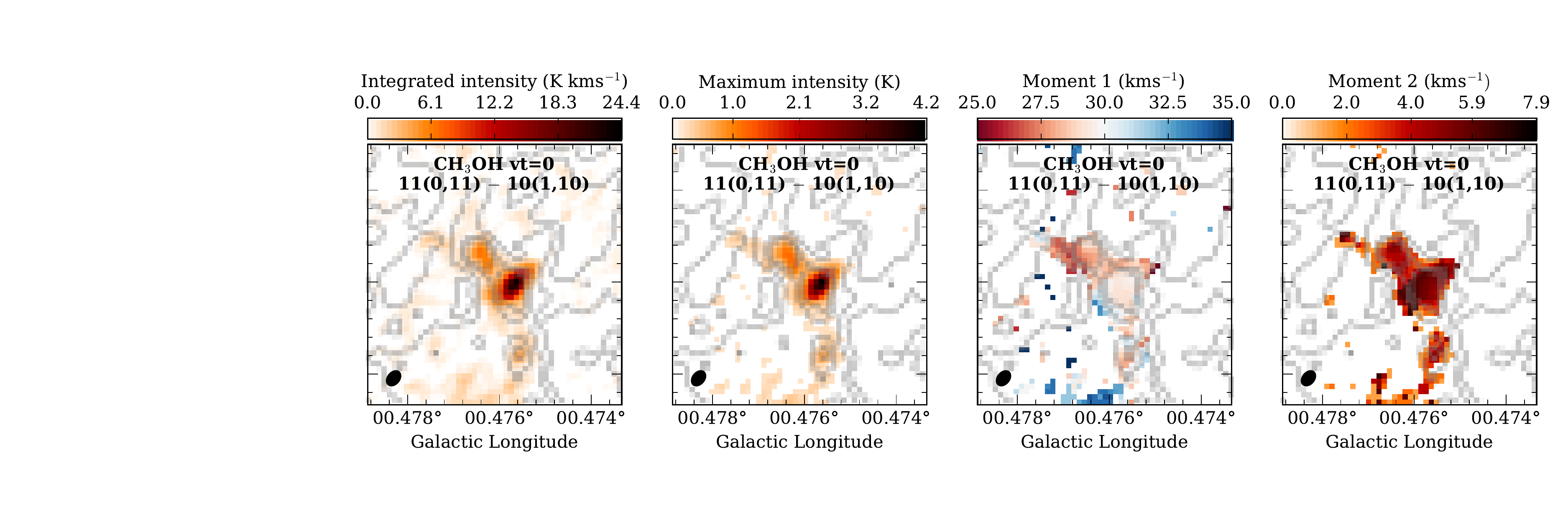}
\includegraphics[trim = 3mm 28mm 3mm 22mm, clip,angle=0,width=1\textwidth]{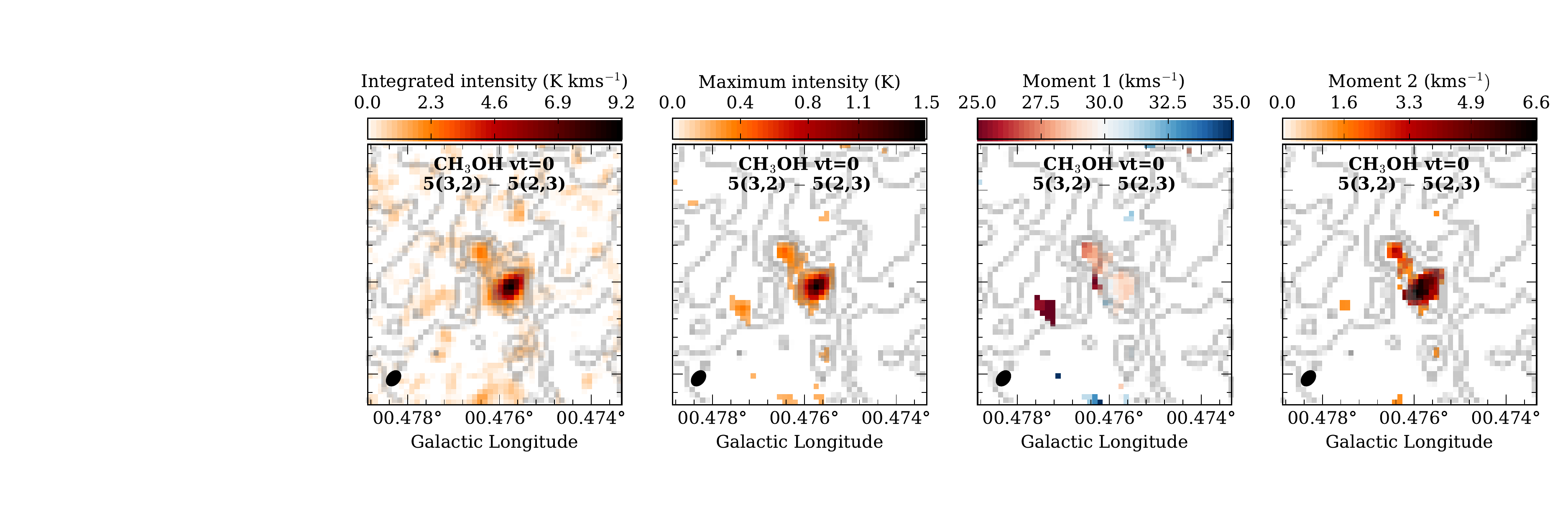}
\includegraphics[trim =  3mm 28mm 3mm 22mm, clip,angle=0,width=1\textwidth]{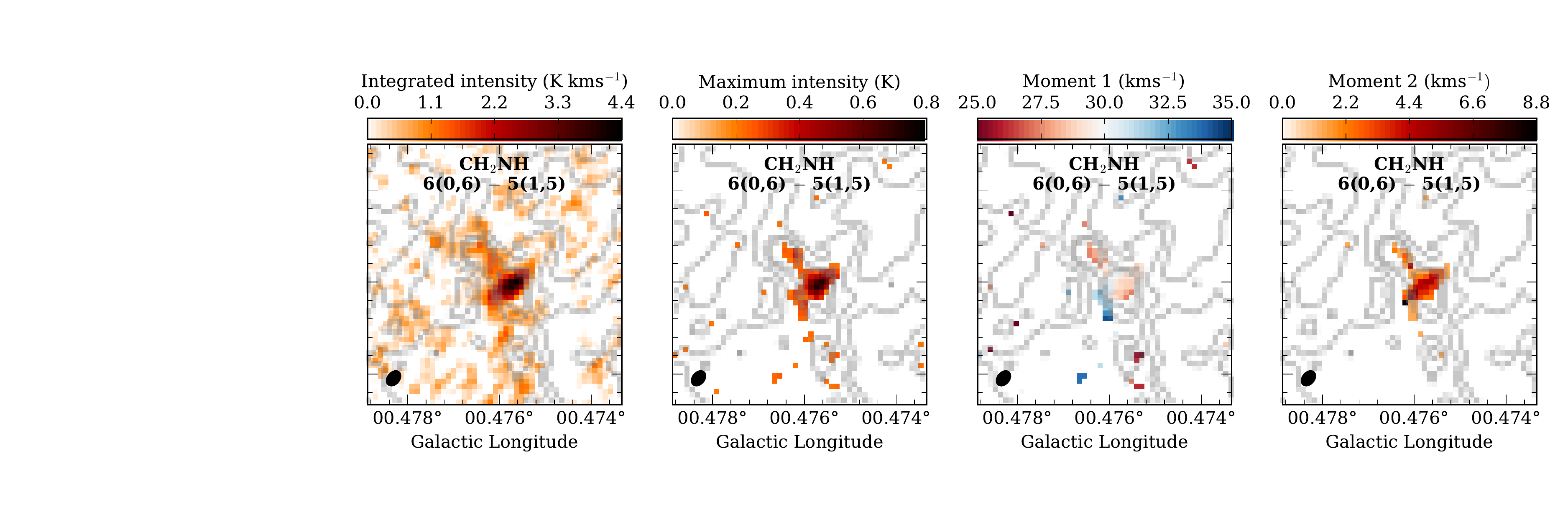}
\includegraphics[trim = 3mm 14mm 3mm 22mm,  clip,angle=0,width=1\textwidth]{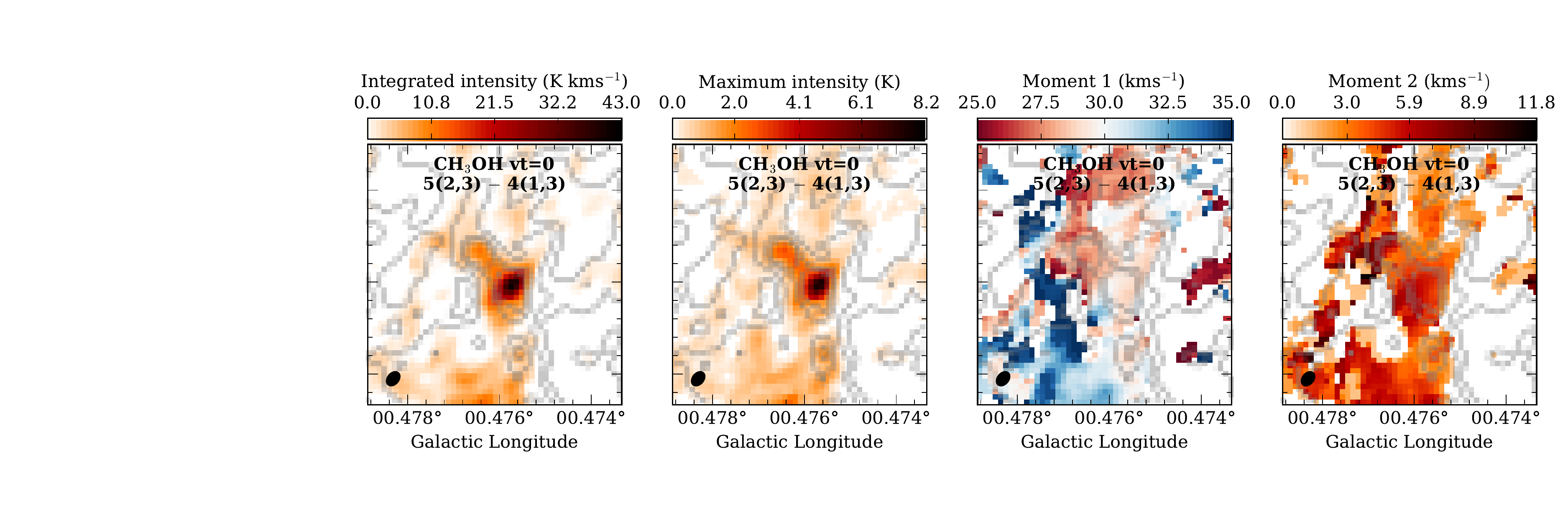}

 \contcaption{}

\label{ }
\end{figure*}

\end{document}